\documentclass[12pt]{article}
\usepackage{eurosym}
\usepackage{amsfonts}
\usepackage{amsmath}
\usepackage{amscd}
\usepackage{amssymb}
\usepackage{graphicx}
\usepackage{wasysym}
\usepackage{pifont}
\usepackage{multirow}
\usepackage{enumitem}
\usepackage{enumitem}
\usepackage[hyphens]{url}
\usepackage[colorlinks=true,urlcolor=black]{hyperref}
\usepackage{hyperref}
\usepackage{boxedminipage}
\usepackage{color}
\usepackage{multirow}
\usepackage[all]{xy}
\usepackage{ulem}
\usepackage{cancel}
\usepackage{bigints}
\newtheorem{theorem}{Theorem}[section]

\newtheorem{corollary}[theorem]{Corollary}

\newtheorem{proposition}[theorem]{Proposition}

\newcommand{\qeed}{\hfill\textrm{QED}\break\null}
\newcommand{\LL}{L^2(\mathrm{SL}(2,\mathbb R))}
\newcommand{\SL}{\mathrm{SL}(2,\mathbb R)}
\newenvironment{demo}{\noindent\textit{Proof.}~}{\qeed}
\usepackage{orcidlink}
\newcommand{\g}{\mathfrak{g}}
\newcommand{\beqa}{\begin{eqnarray}}
\newcommand{\eeqa}{\end{eqnarray}}
\newcommand{\nn}{\nonumber}
\newcommand{\M}{{\cal M}}

\newcommand{\B}{{\cal B}}
\renewcommand{\d}{\text{d}}

\newcommand{\no}{{}_\circ \hskip -.17truecm {}^\circ }
\hypersetup{
backref = true,
pagebackref = true,
hyperindex = true,
colorlinks = true,
breaklinks = true,
urlcolor = blue,
linkcolor = blue,
bookmarks = true,
bookmarksopen = true,
citecolor=red,
}
\numberwithin{equation}{section}

\begin{document}

\title{New perspectives in Kac-Moody algebras associated to higher dimensional manifolds}
\author{}
\maketitle
\
\begin{center}
{\large Rutwig Campoamor-Stursberg$^{1\ast \;\orcidlink{0000-0003-2907-8533}}$, Alessio Marrani $^{2\dagger\; \orcidlink{0000-0002-7597-1050}}$ ,}
\\[0pt]
and {\large Michel Rausch de Traubenberg$^{3\ddagger \; \orcidlink{0000-0001-5045-2353}}$,}\\[5pt]
\bigskip \bigskip

$^1$ Instituto de Matem\'atica Interdisciplinar and Dp.to de Geometr\'\i a y
Topolog\'\i a, \\UCM, E-28040 Madrid, Spain\\[5pt]

$^2$ School of Physics, Engineering and Computer Science, University of Hertfordshire, AL10 9AB Hatfield, UK
\\[5pt]

$^3$ Universit\'e de Strasbourg, CNRS, IPHC UMR7178, F-67037 Strasbourg
Cedex, France\\[5pt]
\end{center}

\bigskip \bigskip \bigskip

\noindent $^\ast$ rutwig@ucm.es

\noindent $^\dagger$ jazzphyzz@gmail.com 

\noindent $^\ddagger$ Michel.Rausch@iphc.cnrs.fr
\abstract{In this review, we present a general framework for the construction of
Kac-Moody (KM) algebras associated to higher-dimensional manifolds. Starting
from the classical case of loop algebras on the circle $\mathbb{S}^{1}$, we
extend the approach to compact and non-compact group manifolds, coset
spaces, and soft deformations thereof. After recalling the necessary
geometric background on Riemannian manifolds, Hilbert bases and Killing
vectors, we present the construction of generalized current algebras $%
\mathfrak{g}(\mathcal{M})$, their semidirect extensions with isometry
algebras, and their central extensions. We show how the resulting algebras
are controlled by the structure of the underlying manifold, and illustrate
the framework through explicit realizations on $SU(2)$, $SU(2)/U(1)$, and
higher-dimensional spheres, highlighting their relation to Virasoro-like
algebras. We also discuss the compatibility conditions for cocycles, the
role of harmonic analysis, and some applications in higher-dimensional field
theory and supergravity compactifications. This provides a unifying
perspective on KM algebras beyond one-dimensional settings, paving the way
for further exploration of their mathematical and physical implications. }

\bigskip

\noindent
{\bf keyword}: Kac-Moody algebras; Virasoro algebra; central extensions; harmonic analysis; higher dimensional manifolds.

\renewcommand{\d}{\text{d}}

\newpage

 \tableofcontents

\newpage

\section{Introduction}

\subsubsection*{\textit{From finite- to infinite- dimensional symmetry}}

For quite a long time, finite-dimensional simple Lie algebras have provided
the backbone of symmetry in mathematics and physics. The classification by
Cartan and the subsequent development of representation theory turned these
structures into essential tools throughout geometry, number theory and
quantum theory. Beginning in the late 1960s, two independent lines of
development---Moody's and Kac's---generalized this landscape to the
infinite-dimensional realm, resulting in what are now called Kac--Moody (KM)
algebras \cite{Moody:1966gf,Kac:1967jr,Kac:1990gs}. A particularly tractable
and physically relevant subclass is formed by the \textit{affine} algebras,
realized as central extensions of loop algebras. Pressley and Segal's
monograph rigorously analyzed the functional-analytic and group-theoretic
underpinnings in terms of loop groups and their projective unitary
representations \cite{Pressley:1988qk}.

The importance of the affine KM symmetry in physics was then amplified by
its tight relation to the Virasoro algebra, via the Sugawara construction
and current algebra methods \cite{Goddard:1986bp}. This connection underlies
the solvability of two-dimensional conformal field theories (CFTs) and the
Wess--Zumino--Witten (WZW) models, and it has been comprehensively discussed
in the CFT literature \cite{DiFrancesco:1997nk,Belavin:1984vu}. At the same
time, the broader spectrum of KM algebras beyond the affine case (e.g.\
indefinite and hyperbolic generalizations) has been invoked in diverse
corners of high-energy theoretical physics, notably in hidden symmetries of
supergravity, cosmology and string/M-theory.

\subsubsection*{\textit{Beyond the circle : why ?}}

The best-known laboratory for KM algebras is the loop algebra over the
circle $\mathbb{S}^{1}$. Physically, the ubiquity of the circle stems from
the role of closed strings and from the reduction of two-dimensional field
theories on a spatial circle. Mathematically, $\mathbb{S}^{1}$ is
distinguished by its Fourier basis, as well as by the resulting algebraic
simplifications that enable a complete representation theory for affine
algebras. However, many problems of current interest in mathematical physics
naturally involve \textit{higher-dimensional} manifolds :

\begin{itemize}
\item Compactifications in Kaluza--Klein (KK) theory and string/M-theory
crucially involve group manifolds and coset spaces; examples including
spheres, tori, and symmetric spaces are e.g. discussed in \cite%
{Salam:1981xd,Bailin:1987jd,KK1,KK3,Castellani:1991et,Harada:2020exi}.

\item Non-compact, Riemannian target spaces (e.g.\ $\mathrm{SL}(2,\mathbb{R})
$ or $\mathrm{SL}(2,\mathbb{R})/U(1)$) arise in (ungauged and gauged)
supergravity theories in diverse space-time dimensions, as well as in the
theory of black hole attractors (see e.g. \cite{Ferrara:2008de} and references therein) .

\item Deformations of group manifolds (\textit{soft} or non-homogeneous
manifolds) appear in the effective descriptions of flux compactifications
and in group-geometric approaches to supergravity \cite{cas}.
\end{itemize}

In all such settings, the natural notion of \textquotedblleft
currents\textquotedblright\ is no longer related to loops on $\mathbb{S}^{1}$%
, but rather it is connected to functions (or sections) on a manifold $%
\mathcal{M}$, with values in a (generally finite-dimensional) Lie algebra $%
\mathfrak{g}$. The resulting current algebra $\mathfrak{g}(\mathcal{M})$
inherits both the algebraic structure from $\mathfrak{g}$ and the
geometric/analytic structure from $\mathcal{M}$. The central question we
address in this review is: \textit{to what extent do the characteristic
features of affine KM algebras---such as central extensions, semidirect
actions by diffeomorphisms or isometries, and rich representation
theory---survive when }$\mathbb{S}^{1}$\textit{\ is replaced by a
higher-dimensional manifold }$\mathcal{M}$\textit{?}

\subsubsection*{\textit{A unifying viewpoint}}

Our viewpoint is twofold : geometric and representation-theoretic. We begin
and consider a Riemannian (or pseudo-Riemannian) manifold $\mathcal{M}$,
equipped with a measure and a Hilbert space $L^{2}(\mathcal{M})$ of
square-integrable functions. A key point is the existence of a \textit{%
Hilbert basis} adapted to the symmetries of $\mathcal{M}$. On the one hand,
for compact Lie groups $G_{c}$, the Peter-Weyl theorem provides such a basis
in terms of matrix elements of unitary irreducible representations \cite{PW}%
. On the other hand, for non-compact Lie groups $G_{nc}$, the Plancherel
theorem organizes $L^{2}(G_{nc})$ functions as a `mixture' of discrete and
continuous series contributions \cite{Hel,bar}. This harmonic-analytic
formalism allows us to expand $\mathfrak{g}$-valued fields on $\mathcal{M}$
and thus to define current algebras $\mathfrak{g}(\mathcal{M}),$ as well as
their (semi-)direct extensions by symmetry algebras generated by Killing
vectors of $\mathcal{M}$ itself. In general, central extensions can be
characterized cohomologically, in terms of closed currents on $\mathcal{M}$,
and they can also be explicitly matched to local Schwinger terms in current
commutators.

All this can be summarized into an algorithmic construction, which provides
a coherent perspective encompassing spheres $\mathbb{S}^{n}$, compact groups
such as $SU(2)$ and their homogeneous spaces, non-compact groups such as $%
\mathrm{SL}(2,\mathbb{R})$, and generalizations with softened group
structure. The steps of such an algorithmic construction list as follows :

\begin{enumerate}
\item Choose $\mathcal{M}$ (compact or non-compact group manifold,
Riemannian or pseudo-Rieman\-nian coset thereof, soft deformations thereof,
\dots ) and its isometry algebra.

\item Select an appropriate orthonormal basis on $L^{2}(\mathcal{M})$ (as
stated by the Peter-Weyl or Plancherel theorems), and then promote $%
\mathfrak{g}$-valued modes to generators.

\item Determine the semidirect actions by isometries/diffeomorphisms, and
describe/classify all compatible 2-cocycles that yield central extensions.

\item Identify structural features (e.g.\ Witt/Virasoro analogues,
area-preserving diffeomorphisms) related to the geometry of $\mathcal{M}$%
.\smallskip
\end{enumerate}

\subsubsection*{\textit{Applications in physics}}

Historically, the amazingly fruitful cross-fertilization between
infinite-dimensional algebras and physics has evolved along several lines of
research. In two-dimensional CFT, the Virasoro and affine Kac--Moody
algebras have provided the dynamical symmetry principles underlying the
theory of exactly solvable models, modular covariance, and the
classification of primary fields \cite%
{Belavin:1984vu,DiFrancesco:1997nk,Goddard:1986bp,Pressley:1988qk}. In
string theory, worldsheet reparameterization and current algebra symmetries
determine crucial consistency conditions (termed as anomaly cancellation
condition, possibly exploiting BRST cohomology), while WZW models provide
exact backgrounds with affine symmetry. The celebrated and surprising
Monstrous moonshine and generalized KM structures (Borcherds--Kac--Moody
algebras) connect CFTs, sporadic groups, and automorphic forms, highlighting
the depth of the algebraic structures which are emergent in quantum theories
\cite{BORCHERDS1991330,Borcherds:1992jjg,monstrous}.

Another remarkable appearance of KM algebras pertains to the large hidden
Lie symmetries arising in supergravity and in cosmological dynamics. For
instance, cosmological billiards uncovered links between the asymptotic
dynamics of gravity near spacelike singularities and hyperbolic KM
symmetries \cite{Belinsky:1970ew,Henneaux:2010ys}. Moreover, extended
supergravities and dimensional reductions have suggested hierarchies of
very-extended algebras, as well as infinite towers of dual potentials; see, for instance, \cite{Kleinschmidt:2006ad,Bergshoeff:2008xv,Bossard:2017wxl,Tumanov:2017whf,Bossard:2019ksx} 
and, for geometric and group-theoretic analyses, \cite%
{Fre:2005gvm,Castellani:1991et,Harada:2020exi}. Within this context, the KK
theory has provided a wealth of examples in which harmonic analysis (on
compact $\mathcal{M}$) describe the spectra of lower-dimensional fields \cite%
{Salam:1981xd,Bailin:1987jd,KK1,KK3}. Finally, current algebras naturally
arise on $\mathcal{M}$, being defined in terms of Noether charges integrated
over the internal spaces, and a possible, subsequent mode
expansion.\smallskip

\subsubsection*{\textit{Scope and plan of this review}}

Expectedly, the generalization from $\mathbb{S}^{1}$ to higher-dimensional
manifolds $\mathcal{M}$ generates both opportunities and challenges. To list
a few :

\begin{description}
\item[Central extensions.] On $\mathbb{S}^{1}$, there is essentially a
\textit{unique} (up to normalization) central extension of the loop algebra,
yielding an affine KM algebra. Typically, for higher-dimensional manifolds
the space of admissible 2-cocycles
   is
\textit{infinite-dimensional}. Cohomological constructions tied to closed $%
(n-1)$-currents on $\mathcal{M}$ and divergence-free vector fields can be
exploited, directly relating to \textit{Schwinger terms} in the current
algebra.

\item[Symmetry by isometries and diffeomorphisms.] The Lie algebra generated
by the Killing vectors (or broader diffeomorphism algebras) acts naturally
on $\mathfrak{g}(\mathcal{M})$. In favorable cases, one can identify
subalgebras which are analogues to the Witt/Virasoro algebras (e.g.\ de~Witt
algebra on $\mathbb{S}^{n}$, or the area-preserving diffeomorphisms on $%
\mathbb{S}^{2}$), yielding to semidirect products which hints to the
affine--Virasoro interplay, familiar from two dimensions.

\item[Harmonic analysis and representation theory.] On compact groups and
homogeneous spaces thereof, the Peter-Weyl theorem and the Clebsch--Gordan
machinery allow to express products of basis functions in terms of
representation-theoretic data. On the other hand, for non-compact groups
such as $\mathrm{SL}(2,\mathbb{R})$, a `mixture' of discrete and continuous
series appears via the Plancherel theory (implemented also through
Bargmann's classification), and new unitary structures emerge.

\item[Applications.] The above constructions provide an algebraic background
for higher-dimensional current algebras in effective field theories,
organize spectra in compactifications, and suggest new Virasoro-like
structures that may control sectors of dynamics in higher-dimensional CFTs
or holographic models.
\end{description}

This review aims to be pedagogical while remaining faithful to the breadth
of the subject. We deliberately separate general principles from case
studies:

\begin{enumerate}
\item We present a construction based on manifolds, that keeps track of
metric, measure and symmetry data, and we adopt a formalism based on the
Hilbert space, in order to emphasize unitarity and completeness of mode
expansions.

\item We use group- and representation- theoretic tools (such as the
Peter-Weyl and Plancherel theorems, or the Clebsch--Gordan coefficients) to
highlight the algebraic structure, and to allow explicit computations of
structure constants in current algebras on $\mathcal{M}$.

\item We discuss central extensions from two perspectives: cohomology of
current algebras on $\mathcal{M}$, and anomalies/Schwinger terms in local
commutators.

\item Throughout our treatment, we strive to highlight links to
well-established areas in physics, such as two-dimensional CFT and string
theory \cite%
{Pressley:1988qk,Goddard:1986bp,DiFrancesco:1997nk,Belavin:1984vu}, or
higher-dimensional supergravity and compactifications \cite%
{Salam:1981xd,Bailin:1987jd,Castellani:1991et,Fre:2005gvm,Harada:2020exi,KK1,KK3,
Kleinschmidt:2006ad,Bergshoeff:2008xv,Bossard:2017wxl,Tumanov:2017whf,Bossard:2019ksx}%
.
\end{enumerate}

There exist several classic reference works on affine algebras and loop
groups \cite{Pressley:1988qk,DiFrancesco:1997nk,Goddard:1986bp}, on group
manifolds and harmonic analysis \cite{PW,Hel}, and on
supergravity/sigma-model applications \cite%
{Salam:1981xd,Bailin:1987jd,Castellani:1991et,Fre:2005gvm,Harada:2020exi}.
In this respect, it is worth emphasizing that our focus is complementary: we
aim at presenting a \textit{single} construction, which applies to wide
classes of manifolds, such as compact groups and cosets, non-compact groups
and their homogeneous spaces, and (soft) deformations of group manifolds. As
mentioned above, we also present a systematic analysis of the space of
compatible central extensions. Various examples are discussed and treated in
some detail, in order to provide a glimpse of the wealth of the subject,
rather than to exhaust all possibilities.\bigskip

The plan of this review is as follows.

\begin{description}
\item[Section 1] \textbf{General framework on manifolds.} We recall the
geometric background on (pseudo-)Riemannian manifolds, measures, and Killing
vectors; we introduce $L^{2}(\mathcal{M})$ and orthonormal bases adapted to
symmetries (exploiting the Peter-Weyl or Plancherel theorems). We then
define the current algebra $\mathfrak{g}(\mathcal{M})$, its semidirect
extension by isometries, and set up the cohomological language for central
extensions.

\item[Section 2] \textbf{Compact group manifolds and cosets.} Invoking the
Peter-Weyl theorem, we construct $\mathfrak{g}(\mathcal{M})$ on $G_{c}$ and
on homogeneous spaces $G_{c}/H$. We explain how products of basis functions
are controlled by Clebsch-Gordan coefficients and how central extensions are
enumerated by closed currents. As examples, we highlight analogues of
Witt/Virasoro subalgebras on spheres and their role in semidirect products.

\item[Section 3] \textbf{Non-compact groups and Plancherel analysis.} We
treat $\mathrm{SL}(2,\mathbb{R})$ and $\mathrm{SL}(2,\mathbb{R})/U(1)$ as
canonical examples, reviewing Bargmann's discrete/continuous series and the
Plancherel decomposition. We construct $\mathfrak{g}(\mathcal{M})$ and
compatible central extensions, discuss unitarity issues and constraints for
the resulting representations.

\item[Section 4] \textbf{Soft (deformed) group manifolds.} We consider the
so-called \textit{soft} deformations of group manifolds motivated by
supergravity and string compactifications, discussing various aspects in
some detail.

\item[Section 5] \textbf{Root systems and representation theory.} We
introduce a system of roots for the classes of Kac-Moody algebras introduced
above. In a second part of this Section we introduce some elements of
representations theory, stressing important differences with respect to
affine Lie algebras. For simplicity's sake, we will here assume that $%
\mathfrak{g}$ is a compact Lie algebra.

\item[Section 6] \textbf{Applications to physics.} We discuss some
applications of the above mathematical machinery to physics, most notably to
: \textit{(i)} two-dimensional current algebra/CFT (WZW, Sugawara,
Virasoro); \textit{(ii)} higher-dimensional compactifications and spectra in
Kaluza--Klein theory; \textit{(iii)} structures emerging within the theory
of cosmological billiards, as well as underlying the hidden symmetries of
supergravity.

\item[Section 7] \textbf{Outlook and open problems.} We outline
classification questions for central extensions on general $\mathcal{M}$,
representation-theoretic challenges beyond compact groups, and possible
applications to holography and integrability.
\end{description}



\section{Kac-Moody (KM) algebras on higher dimensional manifolds: Some generalities}
Let $\g$ be a semisimple (complex or real) Lie algebra. To the simplest (one-dimensional) manifold,
namely the circle $\mathbb S^1$, on can associate two non-isomomorphic  infinite dimensional Lie algebras.
The first is the affine extension of the loop algebra of smooth maps from $\mathbb S^1$ to
$\g$, which  allows to construct KM (or, better, affine) Lie algebras \cite{Kac:1967jr, Kac:1990gs,Moody:1966gf,Pressley:1988qk,Goddard:1986bp}. A second possibility is given by the Virasoro algebra, which is a central extension of the de Witt algebra ({\it i.e.}, of the algebra of polynomial vector fields on $\mathbb S^1$), and is intimately connected to affine Kac-Moody algebras via the Sugawara construction (see e.g. \cite{Belavin:1984vu,Fuks}).

 A natural question concerns the possibility
 to define   infinite-dimensional KM algebras related to higher dimensional
 manifolds, possibly maintaining some of the structural properties observed for the previous relevant cases.
Along the years, a number of infinite-dimensional KM algebras have been constructed.
The first examples are associated to specific manifolds, such as
 the two-sphere, the three-sphere or the $n-$tori \cite{Bars:1983uv,HOEGHKROHN1990106,Frappat:1989gn,MRT,Harada:2020exi, dow, bb}. Then, a general study for more general manifolds were undertaken in \cite{mrm,rmm2,
   Campoamor-Stursberg:2022ane, Campoamor-Stursberg:2022lyx, rm,ram,Campoamor-Strusberg:2024kpl,ram3}.
To this extent, various type of manifolds $\cal M$ were considered.
 This systematic study started with $\cal M$ being a compact $n-$dimensional manifold
\cite{mrm,rmm2, Campoamor-Stursberg:2022ane, Campoamor-Stursberg:2022lyx, rm}
 (essentially, a compact  Lie group $G_c$ --not to be confused with $G$, the Lie group associated to $\g$-- or a coset space
 $G_c/H$ with $H$ a closed subgroup of $G_c$). This construction is ultimately motivated
 by higher-dimensional physical theories, in which harmonic expansions
\textit{\`{a} la Kaluza-Klein} play a crucial role (see e.g. \cite{Salam:1981xd}\nocite{Bailin:1987jd}-\cite{KK3}, with the latter reference
being itself motivated by the supergravity context).
The second series of extensions concerns non-compact manifolds, principally SL$(2,\mathbb R)$ or the
coset SL$(2,\mathbb R)/$U$(1)$ \cite{ram,Campoamor-Strusberg:2024kpl}. These constructions
were motivated by the fact that the non-compact manifold
$SL(2,\mathbb{R})/U(1)$  appears as
the target space of one complex scalar field in the bosonic sector of some
Maxwell-Einstein supergravity theories in $D=3+1$ space-time dimensions. Finally, the last type of algebras
\cite{ram3}
are associated to manifolds which are  deformations of Lie groups, and called  ‘soft’ group manifold
(see e.g. \cite{cas}).

In this section, before considering the  KM algebras associated to the specific types of  manifolds  presented above, we briefly recall some general properties of real manifolds and explain the various steps to construct
the corresponding Kac-Moody algebra.

\subsection{Relevant properties of manifolds} \label{sec:M}
Let $\cal M$ be an $n-$dimensional real Riemannian manifold.
Let ${\cal U} \subset {\cal M}$ be an open set of $\cal M$. We assume that
the manifold is such that ${\cal M} \setminus {\cal U}$ is of zero measure.
Define an homomorphism from $\cal U$ to an open set of $ {\cal O} \subset\mathbb R^n$ by
\beqa
\begin{array}{lccl}
  f:&\cal U&\to&{\cal O} \subset \mathbb R^n\\
  & m&\mapsto&f(m)=(y^1,\cdots,y^n)\ ,\nn
\end{array}
  \eeqa
 {\it i.e.}, we can associate a system of coordinates for points $m$ in $\cal M$
 almost everywhere.  Let $g_{IJ}$ be the metric tensor of $\M$ in the the system of
 coordinates $y^I=(x^1,\cdots,y^n)$:
 \beqa
\text{d}^2 s = \text{d} x^I \text{d} x^J  g_{IJ}\ \nn
 \eeqa
and  define $g= \big| \det g_{IJ} \big|$.
 Consider now  an infinitesimal diffeomorphism:
 \beqa
y^I \to y'^I + \epsilon \xi^I(y) \ . \nn
 \eeqa
 This transformation is an isometry of $\cal M$, {\it i.e.}, it preserves
 the tensor metric $g_{IJ}$, {\it iff} the $\xi^I$ satisfy the Killing equation:
 \beqa
 \label{eq:killing}
\nabla_I \xi_J(y) + \nabla_J \xi_I(y) = 0 \ ,
 \eeqa
 where $\xi_I = g_{IJ} \xi^J$ and $\nabla_I$ is the covariant
 derivative \cite{Golab}. Suppose that the Killing equation has $s-$independent
 solutions $\xi_A^I(y), A=1,\cdots,s$, then the operators
 \beqa
 \label{eq:xiK}
 \g_\xi=\Big\{K_A= i \xi^I_A(y) \partial_I, A=1,\cdots,s\Big\}
\eeqa
 generate the Lie algebra with Lie brackets
 \beqa
 \label{eq:LieK}
\big[K_A,K_B\big]=ic_{AB}{}^C K_C \ ,
\eeqa
since the composition of two isometries is an isometry, and where
\beqa
-\xi^J_A\partial_J\xi^I_B+\xi^J_B\partial_J\xi^I_A= -c_{AB}{}^C \xi^I_C \ , \nn
\eeqa
because the Killing vectors are associated with independent infinitesimal isometries.

We endow the manifold $\cal M$ we an Hermitian scalar product:
\beqa
\label{eq:sp}
(f,g) = \int_{\cal O} \sqrt{g} \;\text{d}^n y\; \overline{f(y)} g(y)
\eeqa
where ${\cal O} = f({\cal U}) \subset \mathbb R^n$.
Hermiticity of the operators $K_A$ with respect to the scalar product \eqref{eq:sp} translates into two conditions:
\beqa
\label{eq:herm}
\begin{array}{ll}
  1:&\partial_I(\sqrt{|g|} \xi^I_A)=0,\\
  2:&\xi^I_A\big|=0\ ,\ \  I=0,\cdots, n
  \end{array}
\eeqa
where $\xi^I_A\big|=0$ means that  the boundary term associated to all directions vanishes. Note in particular that if $\xi^I_A$
are periodic in a direction,  condition
2 is automatically satisfied. Similarly, if a  direction is unbounded, say  $y^{I_0} \in \mathbb R$, condition 2 is also satisfied since
$\xi^{I_0}_a \to 0$ whenever $y^{I_0} \to \pm \infty$. Furthermore, as $\xi^I_a$ is a Killing
vector, we have a Levi-Civita connection {\it satisfying}
$\nabla_I g_{JK}=0$, hence the Killing condition \eqref{eq:killing} implies the
condition 1 above\footnote{Recall that
 $\nabla_I V^I = 1/\sqrt{g} \;\partial_I\big(\sqrt{g}\; V^I\big)$.}.

The last step of this general introduction is to consider the Hilbert space
of square integrable functions on $\cal M$, namely $L^2(\M)$,
or more precisely $L^2({\cal O})$. It is known that
any Hilbert space admits a Hilbert basis \cite{RS, GP}. Let ${\cal B}=\{b_M, M\in \mathbb N\}$ be a Hilbert basis of $L^2({\cal O})$, {\it i.e.},
 a complete orthonormal set
of vectors (for the scalar product \eqref{eq:sp}). In our case, the vectors $b_N$ become functions on $\cal O$.
We also assume that for any generators $K_A$ of the Lie algebra generated by the Killing vectors (see \eqref{eq:LieK}) and for any vector $b_M$ in $\cal B$, the
function $K_A b_N$ is square integrable. Thus,
\beqa
\label{eq:Kb}
K_A b_M(y)= (M_A)_M{}^N\; b_N(y) \
\eeqa
where the symbol of summation is omitted,  where $(M_A)_M{}^N$ is a matrix representing the action of $\g_\xi$ on $b_M$. Indeed,
this equation in fact means that the vectors $b_M$ transform according
to a given representation of $\g_\xi$. In general, this representation is
fully reducible, but it may happen that a subset of vectors does not belong
to an irreducible representation (see Sect. \ref{sec:Gnc}).
Since the
set of vectors $b_M$ is orthonormal and constitutes a Hilbert
basis of $L^2({\cal O})$, these vector also constitute
a unitary representation of $\g_\xi$, hence the
operators $K_A$ are Hermitian and thus satisfy \eqref{eq:herm} (and, in
particular, the second condition).\\

In the following, two types of manifolds will be considered. The first type corresponds to group
manifolds of compact Lie groups $G_c$, while the second corresponds to group manifolds of  non-compact Lie groups $G_{nc}$. In both cases the decomposition of square integrable functions (or the Hilbert basis $\B$) organize within the representation theory of $G_c$ (resp. $G_{nc}$), but it should be observed that these two
cases are structurally quite different. If the group is compact, its unitary representations are finite dimensional and it turns out
that, once correctly normalized, the set of all matrix elements constitutes
an orthonormal Hilbert basis of $L^2(G_c)$. This corresponds to the Peter--Weyl Theorem
\cite{PW}. If the group is non-compact, its unitary representations are infinite dimensional and square integrable functions decompose into a sum over the matrix elements of the discrete series and an integral over the matrix elements of the
continuous series (see below). In this situation, the Plancherel Theorem has to be used \cite{sch}.

Returning to the Hilbert basis $\cal B$, we assume further that for any elements
$b_N, b_M$ of $\cal B$ the product $b_N b_M$ is square integrable, thus in
particular we have
\beqa
\label{eq:bb}
b_M(y) b_M(y)=C_{MN}{}^P b_P(y) \
\eeqa
where the symbol of summation is ommitted.
In general, the coefficients $C_{MN}{}^P$ are difficult to compute explicitly,
except (at least) for two types of manifolds: (i) $\M$ is related to a compact
Lie group $G_c$, or (ii)  $\M$ is related to a non-compact
Lie group $G_{nc}$. In both cases, the coefficients $C_{MN}{}^P$ can be expressed
explicitly by means  of the Clebsch-Gordan coefficients of $G_c$ (resp. $G_{nc}$).

Since $\M$ is a real manifold, one can choose the set of functions $b_M$ to be real functions. It is however useful for
certain manifolds $\M$ to consider complex-valued functions. Reality of $\M$ imposes then that for any $b_M$ in $\B$
we have
\beqa
\label{eq:bc}
\overline{b}^N(y) =\eta^{NM}\; b_M(y)  \ , \ \ \text{or}\ \
b_M(y) = \eta_{MN}\; \overline{b}^N(y)\ ,
\eeqa
where  $\eta_{MN}=\eta^{MN}$ and such that for a given $M\in \mathbb N$, there exists a unique $N \in \mathbb N$ such
that $\eta_{MN}$ is non-vanishing. One can of course chose $\eta_{MN}=1$, albeit for group manifolds we could also have $\eta_{MN}=\pm 1$
(see Section \ref{sec:GM}). Finally, in this section the index of the functions $b_M$ belongs to $\mathbb N$. In turn, other choices will be
more relevant in the next section, in particular, negative integer values for $M$. In this case, conditions \eqref{eq:bc} will be obvious
and $\eta_{MN} \ne 0$ if $N=-M$.

Since the basis $\cal B$ is complete, we have the completness relation:
\beqa
\label{eq:comp}
\sum \limits_{N\in \mathbb N}\; \overline{b}^N(y) b_N(y') =
\frac{\delta^n(y-y')}{\sqrt{g(y)}} \ .
\eeqa
\subsection{A KM algebra associated to $\M$}\label{sec:KMM}
Let $\g=\{T_a, a=1,\cdots,d\}$ be a $d-$dimensional simple Lie algebra with
Lie brackets
\beqa
\big[T_a, T_b\big] = i f_{ab}{}^c T_c \  . \nn
\eeqa
The Lie
algebra can be real or complex. We assume for simplicity that $\g$ is a compact Lie
algebra with Killing form:
\beqa
\big<T_a, T_b \big>_0 = k_{ab} = \text{tr}(\text{ad}(T_a) \text{ad}(T_b))\ , \nn
\eeqa
where ${\rm ad}$ denotes the adjoint representation of $\g$ and $k_{ab}$ is definite positive.

In the next step of our construction we define  the space of smooth mappings from
$\M$ into $\mathfrak{g}$ as
\begin{eqnarray}
\mathfrak{g}(\M) =\Big\{T_{aM} = T_a b_M(y) \ ,\ \ a =
1,\cdots, d \ , \ \ M \in \mathbb N \Big\} \ ,
\end{eqnarray}
which inherits the structure of a Lie algebra
\begin{eqnarray}  \label{eq:hatg}
\big[T_{aM} , T_{bN}\big] = i f_{ab}{}^c \;C_{MN}{}^P T_{cP} \ ,
\end{eqnarray}
where the coefficients $C_{MN}{}^P$ are defined in \eqref{eq:bb}.
The Killing form in $\mathfrak g({\cal M})$ is given by
\beqa
\Big<X,Y \Big>_1 = \int _{\cal O} \sqrt{g} \;\text{d}x^n\; \Big<X,Y \Big>_0 \ , \nn
\eeqa
for $X,Y \in \mathfrak g({\cal M})$.
From   \eqref{eq:bc} it follows  that
\beqa
\label{eq:loop2}
\Big<T_{aM} ,T_{bN}  \Big>_1 = k_{ab} \eta_{MN} \ .
\eeqa
In fact, if we assume
\beqa
\label{eq:del}
\big[T_{a}(y) , T_{b}(y')\big] = i f_{ab}{}^c T_{c}(y) \frac{\delta^n(y-y')}{\sqrt{g}},
\eeqa
where
\beqa
T_a(y)= \sum\limits_{M \in \mathbb N}  T_{aM} \; \overline{b}^M(y)
\eeqa
then integrating both sides of the equation above by
$\int \sqrt{g} \;\text{d}^n y\; b_P(y) \int \sqrt{g} \;\text{d}^n y' \;b_q(y')$,
\eqref{eq:del} reproduce \eqref{eq:hatg} using \eqref{eq:comp}
and \eqref{eq:bb}. Clearly the algebra $\g(\M)$ is the generalization of the
loop algebra corresponding to   the $n-$dimensional manifold $\M$
(the loop algebra is associated to  $\M = \mathbb S^1$,  the circle).\\

Continuing the construction, we recall that $\M$ is such that the operators associated to the Killing
vectors
$\g_\xi=\{K_A, A=1,\cdots, s\}$  are Hermitian with respect to the
scalar product \eqref{eq:sp} (see \eqref{eq:herm}).
Let
$\g(\M) \rtimes \g_\xi$. This algebra admits a semidirect
structure:
\beqa
\label{eq:Kloop}
\big[T_{aM} , T_{bN}\big] &=& i f_{ab}{}^c \;C_{MN}{}^P T_{cP} \ ,\nn\\
\big[K_A, b_M(y)\big] &=& (M_A)_M{}^N \;b_N(y)\ ,\\
\big[K_A,K_B]&=& i \;c_{AB}{}^C K_C \ ,\nn
\eeqa
(see \eqref{eq:Kb}). It may be interesting to identify the maximal set of commuting operators.
Let $r$ be the rank of $\g_\xi$, and
let $\{D_1,\cdots, D_r\}$ be a Cartan subalgebra. Without loss of generality, we
can assume that the elements of the basis $\cal B$ are eigenvectors of the $D$s:
\beqa
\label{eq:Db}
\big[D_i, b_M(y)\big] = M(i) \;b_M(y)
\eeqa
with $M(i) \in \mathbb R$. It is important to emphasize that,
\beqa
\label{eq:bc2}
\big[D_i, b_M(y)\big] = M(i) \;b_M(y)  \ \ \Leftrightarrow \ \ \big[D_i, \overline{b}^M(y)\big] = -M(i) \;\overline{b}^M(y) \ .
\eeqa

The algebra above can even be further  extended. Introduce the
 Lie algebra of vector fields
$\mathfrak{X}({\cal M})=\{V_{IM} = -i\;  b_M(y) \partial_I, I=1,\cdots, n, M \in \mathbb N\}$ on ${\cal M}$,
where  $\partial_I = \frac{\partial}{\partial y^I}$.  The algebra \eqref{eq:Kloop} extends to
 $\g(\M) \rtimes \mathfrak{X}({\cal M})$:
\beqa
\label{eq:KM-vect}
\big[T_{aM},T_{bN}\big]&=& i\; f_{a b}{}^c c_{MN}{}^P  T_{cP}\ , \nn\\
\big[V_{IM},V_ {JN}\big]&=& - i \Big( (\partial_I  b_N) V_{JM} -
(\partial_J b_M) V_{IN}\Big) \ ,\\
\big[V_{IM},T_{a N}\big]&=&  -i b_M \partial_I  b_N T_{a } \equiv-i d_{IM,N}{}^P T_{aP}\ ,  \nn
\eeqa
and has also a semidirect product structure. This algebra admits
\eqref{eq:Kloop} as a subalgebra,  also some analogue of the de Witt algebras. Let
\beqa
\label{eq:Witt}
\text{Witt}_i(\M)=\Big\{\ell_{iM}=-b_M(y)\; D_i\ ,\ \  M\in \mathbb N\Big\}\ ,\ \  i=1,\cdots,s
\eeqa
which
are a subalgebras of the algebra of vector fields on $\M$. The
algebra \eqref{eq:Kloop} extends to $\g(\M) \rtimes \text{Witt}_i({\cal M})$
\beqa
\label{eq:KM-witt}
\big[T_{aM},T_{bN}\big]&=& i\; f_{a b}{}^c c_{MN}{}^P  T_{cP}\ , \nn\\
\big[\ell_{iM},\ell_ {jN}\big]&=& (M(i) -N(j)) C_{MN}{}^P \ell_{iP}\Big) \ ,\\
\big[\ell_{iM},T_{a N}\big]&=&  -M(i)C_{MN}{}^P T_{aP}\ .  \nn
\eeqa
\bigskip

The last step of the construction
is to extend centrally the algebra $\g(\M)$. Central extensions of  $\g(\M)$
were identified   by Pressley and Segal in \cite{ps} (see Proposition 4.28  therein). Given a one-chain $C$ ({\it i.e.}, a closed one-dimensional
 piecewise smooth curve), the central extension is given by the two-cocycle
 \beqa
 \label{eq:cos}
\omega_C(X,Y) = \oint_C \big<X, \text{d} Y\big>_0  \ ,
\eeqa
where $\text{d} Y =\partial_I Y\ \text{d} y^I$ is the exterior derivative of $Y$.
The two-cocycle can be written in alternative form. Indeed, we have  \cite{bt}
\beqa
 \label{eq:cos2}
\omega_C(X,Y) = \int_{{\cal O}}  \big<X, \text{d} Y\big>_0 \wedge \gamma \ ,
\eeqa
where $\gamma$ is a closed $(n-1)-$current (a distribution)  associated to $C$.
We now show that the cocycle \eqref{eq:cos} can also be associated to a vector
field on $L\in \mathfrak{X}(\M)$.
Let $L=f^I(y)\; \partial_I$ a vector field
be such that
\beqa
\label{eq:df=0}
\partial_I(\sqrt{g} f^I)=0 \ .
\eeqa
To the operator $L$
  one can associate  the $1-$form:
  \beqa
  \label{eq:F1f}
F(y)&=&   f_I(y)\; \text{d} y^I
\eeqa
where $f_I = g_{IJ} f^J$,
and  the $(n-1)-$form:
\beqa
\label{eq:gamnf}
\gamma(y)=\sqrt{g}\sum\limits_{I=1}^n (-)^{I-1} f^I(y)\; \d y^1 \wedge \cdots
\d y^{I-1}\wedge \d y^{I+1} \wedge \cdots \wedge \d y^n \ .
\eeqa
Obviously, if $\cal M$ is orientable, $\gamma= {}^\ast F$
is the Hodge dual of $F$. Now the condition \eqref{eq:df=0} implies that
$\gamma$ is closed, equivalently, that $F$ is co-closed. Moreover,
if the cohomology group $H^{n-1}({\cal M}) = 1$ is trivial, then $F$ is
co-exact or $\gamma$ is exact and there exists a one-form $G$ such that
\beqa
F= {}^\ast \d {}^\ast G \ , \ \
\gamma=(-1)^{q+n-1} \d h \ , \nn
\eeqa
where $h={}^\ast G$ and the metric $g$ has signature $(p,q)$ with $q$ minus
signs.
Since the $(n-1)-$form associated to the operator $L$ is closed,
the corresponding two-cocycle \eqref{eq:cos2} reduces to
\beqa
\label{eq:cosL}
\omega(X,Y)_f= \int \limits_{\cal O} \sqrt{g} \;\d^n y \;\big<X,L Y\big>_0
= \int \limits_{\cal O} \big<X, \d Y\big>_o\wedge \gamma\ .
\eeqa
We conclude that there is an infinite number of central extensions associated
to any operator $L$ satisfying \eqref{eq:df=0}.
In \cite{fei}, Feigin  showed
that the number of central extensions is equal to the dimension of
$H^2(\g(\M))$, and proved that when $\dim \M >1$ and $\M$ is compact,
we have an infinite number of central extensions. Central extensions were
systematically computed in the case of the two- and
three-sphere in \cite{Frappat:1989gn,dow,bb}.

The result above can also be obtained differently. Indeed,  one can centrally
  extend the `loop' algebra $\g({\cal M})$, using the current
  algebra and introducing an anomaly or a Schwinger
  term:
  \beqa
  \label{eq:anom}
  \big[T_{a_1}(y_1),T_{a_2}(y_2)\big] =\Big( i f_{a_1 a_2}{}^{b} T_b(y_2)
  + k_{a_1 a_2}\;f^I(y_2)\;\partial^{(2)}_I\Big)
  \frac{\delta^n(y_1-y_2)}{\sqrt{g(y_2)}} \ ,
  \eeqa
  where  $\partial^{(2)}_I$ means that we take the derivative with respect to the second variable $y_2^I$.
  Thus,
\beqa
\Big[\big[T_{a_1}(y_1),T_{a_2}(y_2)\big],T_{a_3}(y_3)\big]
&=&\Big[\Big( i f_{a_1 a_2}{}^{b} T_b(y_2)
  + k_{a_1 a_2}\;f^I(y_2)\;\partial^{(2)}_I\Big)\frac{\delta^n(y_1-y_2)}{\sqrt{g(y_2)}}, T_{a_3}(y_3)\Big]\nn\\
&=&\cdots + i f_{a_1 a_2 a_3} f^I(y_3)
\frac{\delta^n(y_1-y_2)}{\sqrt{g(y_2)}}\;\partial^{(3)}_I
\Big(\frac{\delta^n(y_2-y_3)}{\sqrt{g(y_3)}}\Big)\nn
\eeqa
where $\dots$ represents terms not involving the anomaly and $f_{a_1a_2 a_3}=
f_{a_1 a_2}{}^b k_{b a_3}$.  Integration by parts,
using the symmetry property of the Dirac $\delta$-distribution and
the antisymmetry of the structure constants $f_{a_1a_2 a_3}$ in the computation
$\Big[\big[T_{a_1}(y_1),T_{a_2}(y_2)\big],T_{a_3}(y_3)\big]+$ cyclic terms, the
Jacobi indentity leads to \eqref{eq:df=0},
which is the two-cocycle condition that we have obtained using the Pressley-Segal two-cocycle.
Furthermore, performing integration $\int \sqrt{g(y)} \d^n y\; b_M(y)
\int \sqrt{g(y')} \d^n y'\; b_N(y')$ for the anomaly term in \eqref{eq:anom}  leads, due to
the relation \eqref{eq:df=0} above, to the two-cocycle \eqref{eq:cosL}.
This means that the Pressley-Segal results on central extensions can  equivalently
be given in terms of an appropriate Schwinger term.
\\

We now address the compatibility of central extensions and the differential
operators of $\mathfrak{X}(\M)$. Assume that we have one central extension associated
to the operator $L$ above or the two-cocycle $\omega_L$. The new Lie brackets
read
\beqa
\label{eq:compat}
\big[X,Y\big]_L= \big[X,Y]_0 + \omega_L(X,Y) \ ,
\eeqa
where $[\;,\;]_0$ are the Lie brackets without central extension, {\it i.e.}, the Lie brackets of $\g(\M)$. Let $L'$ be an operator in $\mathfrak{X}(\M)$. The
compatibility condition, {\it i.e.}, the Jacobi identities, between the KM algebra with two-cocycle $\omega_L$ and
$L'$  turn out to be
\beqa
\omega_L(L'X, Y) + \omega_L(X,L'Y) = 0 \  . \nn
\eeqa
Because of \eqref{eq:df=0}, after integration by parts,  we get
that $L'=L$ is an obvious solution.

Let  now $(L,\omega_L)$ be a doublet differential operator/two-cocycle.
On the one hand, the  two-cocycle $\omega_L$ and the differential operator $L$
are  in duality (see \eqref{eq:F1f} and \eqref{eq:gamnf})
and on the other hand, they are  compatible.  One may wonder if for a given manifold
$\M$ one can have more than one couple $(\omega_L,L)$ in duality and compatible
with each other.
In fact, as will be seen in Section \ref{sec:Gc}, in general this is the case.
In particular, if we consider the  commuting  Hermitian operators $D_i$
(see \eqref{eq:Db})
and the corresponding two-cocycle $\omega_i$:
\beqa
\label{eq:cc}
\omega_i(T_{aN},T_{bM}) =   M(i) k_{ab} \eta_{MN}\ ,
\eeqa
the compatibility condition above between $D_j$ and  $\omega_{i}$ reduces to
\beqa
\omega_i(D_j T_{aN} ,T_{bM}) +\omega_i( T_{aN}, , D_jT_{bM})=
 k_{ab} M(i)(N(j)+M(j)) \eta_{MN} = 0 \ . \nn
 \eeqa
 The  condition of compatibility translates into
 \beqa
 \label{eq:comp2}
(N(j)+M(j)) \eta_{MN} = 0 \ , \ \ \forall M,N \ .
\eeqa
Let now $\g_D=\Big\{D_i, i=1,\cdots,r' \le r: \eqref{eq:comp2} ~\text{
  is~satisfied}\Big\}$. We thus define the KM  algebra
$\widetilde{\mathfrak{g}}(\M) = \big\{\mathcal{T}_{aM},a=1\cdots, d, M\in \mathbb N,  k^i, i=1,\cdots r'\}\rtimes \big\{D_i,i=1,\cdots r'\} $.
From \eqref{eq:hatg}, \eqref{eq:Db} and \eqref{eq:cc} the non-vanishing  Lie brackets are:
\beqa
\label{eq:KM-gen}
\big[{\cal T}_{aM},{\cal T}_{bN}\big] &=&i\;  f_{ab}{}^c C_{MN}{}^P {\cal T}_{cP} +    k_{ab}  \;\eta_{MN} \; \sum \limits_{i=1}^{r'} k^i M(i)
 \ , \nn\\
\big[D_i, {\cal T}_{aM}\big] &=&   M(i)  {\cal T}_{aM}\ .
\eeqa
Note that $k^i, i=1,\cdots r'$ are the central charges associated to the two-cocycle $\omega_i$. 

\medskip
In the construction above, we determined the maximal set of compatible cocycles and differentials operators $(\omega_L, L)$. We consider here another possibility which will be relevant for the construction of the analogue of
  the Virasoro algebra (see Sec. \ref{sec:SU2} for ${\cal M}=$ SU$(2)$).  Let $\omega_L$ be a cocycle associated to the differential operator $L$ and let $T$ be any function in $L^2({\cal M })$. The compatibility
  condition of the cocycle $\omega_L$ with
  the operator $L$ reads
  \beqa
  \label{eq:TL}
\omega_L\big(T(m) L X,Y\big)+ \omega_L\big(X,T(m) L Y\big)=0 \ ,
\eeqa
whose second term on the LHS is
\beqa
\omega_L\big(X,T(m) L Y\big) = \int \limits_{\cal M} \big<X, L\big((T(m) L Y\big)\big>_0
 \sqrt{g}\;\d^nm\nn \ \ ,
 \eeqa
 whilst the first one, after an integration by parts, reduces to
 \beqa
 \omega_L\big(T(m) L X,Y\big)&=&\int \limits_{\cal M} \big<T(m) LX, L Y\big>_0
 \sqrt{g}\;\d^nm\nn\\
&=& -\int \limits_{\cal M}\Big( \big<X, L\big((T(m) L Y\big)\big>_0
 \sqrt{g}+ \big<X, (T(m) L Y\big)\big>_0
 \partial_m(\xi^m \sqrt{g})\Big)\;\d^n m\nn\\
 &&+\int \limits_{\cal M}\partial_m\Big(\xi^m \big<X, (T(m) L Y\big)\big>_0
  \sqrt{g}\Big)\d^n m\nn \ ,
 \eeqa
 where we have used $L=\xi^m\partial_m$.
 The second term in the second equality vanishes because of \eqref{eq:df=0}. On the other hand,  the boundary term
 vanishes along any compact direction of $\cal M$, but also vanishes along  non-compact directions of
 $\cal M$, since $X$ and $Y$ go to zero  at  infinity.
 Finally, Eq.[\ref{eq:TL}] is satisfied and
the set of differential operators
\beqa
\label{eq:DT}
D_L=\Big\{L^T := T(m) L\ , T(m) \in L^2({\cal M})\Big\}
\eeqa
is compatible with the cocycle $\omega_L$. Furthermore,  we have the following commutation relations
\beqa
\big[L^{T_1}, L^{T_2}\big] = L^{T_{12}}\ , \ \ T_{12}(m)= T_1(m) \big(L T_2(m)\big)- T_2(m) \big(L T_1(m)\big) \ . \nn
\eeqa
Thus $D_L \subset \mathfrak{X}({\cal M})$ is a subalgebra of the algebra of vector fields on $\cal M$, which should be regarded as
the analogue of the de Witt algebra in the case of ${\cal M}=$U$(1)$.
This algebra was found in \cite{Frappat:1989gn} for ${\cal M}=$ SU$(2)$.
We shall see in Section \ref{sec:SU2} that, in the case of SU$(2)$, a convenient choice of $L$  enables us to
introduce an analogue of the Virasoro algebra on SU$(2)$, {\it i.e.}, to central extend  the algebra $D_L$.
This construction certainly extends to other manifolds, as, for instance, any compact group manifold ${\cal M} =G_c$. In particular,
for a given cocycle associated to $L$,
 further compatible differential
operators  can be added. For instance, if $L=D_i$,  the set $D_{D_i}$ is the de Witt algebra
(see \eqref{eq:Witt} and the second equation of \eqref{eq:KM-witt}). Next, the  algebra $D_{D_i}$  can certainly be centrally extended along the lines of
Sec. \ref{sec:SU2}, yielding to an analogue of the semidirect product of the Virasoro algebra with the Kac-Moody algebra.\\

Before ending  this section, it should be  noted  that the loop algebra $\g({\cal M})$ admits also other central extensions. See
  {\it e.g.} \cite{mick, fad} for the three-dimensional case and \cite{ceder} (and references therein) for higher-dimensional cases.

\section{KM algebras on compact group manifolds}\label{sec:GM}
In the previous section we have given a general strategy to construct a generalized KM algebra associated to a manifold $\M$. We have assumed throughout   \eqref{eq:Kb}  and \eqref{eq:bb}. Let $G_c$ be a compact Lie group manifold and
let $H\subset G_c$ be a closed subgroup of $G_c$. In this section we are  considering the case where $\M$ is either
the compact group manifold $\M=G_c$ or the  coset space $\M=G_c/H$. In both cases, the conditions   \eqref{eq:Kb} and \eqref{eq:bb} (together
with   \eqref{eq:herm}(2)) are
natural.

\subsection{Compact group manifolds}\label{sec:Gc}
Let $G_c$ be an $n-$dimensional compact Lie group.
Let $m^A$ with $i=A,\cdots,n$ be a
parameterization of $G_c$.
Then an element of $G_c$ connected to the identity takes the form:
\begin{eqnarray}  \label{eq:paramGc}
g(m)= e^{i m^A J_A} \ ,
\end{eqnarray}
where $J_A, A=1,\cdots,n$ are the generators of $\mathfrak{g}_c$, the Lie
algebra of $G_c$ (not to be confused with the generators of $\mathfrak{g}$).
Then,  the coordinates of a group element (in a local coordinate
chart) are
\begin{eqnarray}
  \label{eq:virb}
g(m)^M \equiv m^M \ , \ \ M=1,\cdots, n \ .
\end{eqnarray}
The indices $A, B, \cdots$ are tangent space indices, \textit{i.e.}, flat
indices, whilst the indices $M,N,\cdots$ are world indices, \textit{i.e.},
curved indices.
It should be mentionned that for specific parameterizations,  the variables
$m^A$ decompose into $m^A=(\varphi^i,\theta^r), i=1,\cdots,p, r=1,\cdots,q, p+q=n$ where the  matrix elements of $m^M$ are periodic in $\varphi$ and non-periodic in $\theta$ (see Section \ref{sec:GM}).

The Vielbein one-form associated to the parameterization
\eqref{eq:paramGc} is thus
\begin{eqnarray}  \label{eq:e}
e(m)= g(m)^ {-1} \;\mathrm{d} g(m) \ ,
\end{eqnarray}
where $\text{d} g(m)$ is the exterior derivative of $g$. The Vielbein  satisfies the Maurer-Cartan equation
\begin{eqnarray}
  \label{eq:MC}
\text{d} e + e \wedge e = 0 \ .
\end{eqnarray}
Expanding the Vielbein in the basis $J_A$: $e=i e^A J_A$ we have for the one-forms $e^A$:
\begin{eqnarray}
e^A(m) = e_M{}^A(m)\; \text{d} m^M \ .  \notag
\end{eqnarray}
The one-forms are left-invariant by construction, {\it i.e.}, invariant by the left multiplication $L_h(g)=hg$ with $h\in G_c$.
Note that defining the alternative Vielbein $e'(m) =\text{d} g(m) g(m)^{-1}$, we obtain a right-invariant one-forms, {\it i.e.}, invariant
by the right multiplication  $R_h(g)=gh$.
The metric tensor on $G_c$ can be defined by the left or right-invariant one-forms
\begin{eqnarray}  \label{eq:gG}
  g_{MN} = e_M{}^A(m) e_N{}^B(m) \delta_{AB}
  = e'_M{}^A(m) e'_N{}^B(m) \delta_{AB}\ .
\end{eqnarray}

We endow the manifold $G_c$ with the scalar product
\beqa
\label{eq:spGc}
(f,g) = \frac 1 V \int \limits_{G_c} \sqrt{g}\; \text{d} \varphi^p \text{d} \theta^q  \overline{f(\varphi,\theta)} f(\varphi,\theta) \ ,
\eeqa
where $V$ is the volume of $G_c$. \\

To identify a Hilbert basis of $L^2(G_c)$, we first recall that since $G_c$ is a compact Lie group, its unitary irreducible
representations are finite dimensional. Let  $\hat {\cal R} =\{{\cal R}_k, k \in \hat G_c\}$ be the set of all irreducible unitary representations of $G_c$,
and let $\hat G_c$ be the set of labels of such representations (see below). Let $d_k$ be the dimension of the representation ${\cal R}_k$
and let  $D_{(k)}{}^i{}_j(g)$ be its matrix elements.
With these notations, have the following theorem:

\begin{theorem}[Peter-Weyl \cite{PW}]\label{theo:PW}
Let  $\hat {\cal R}= \{{\cal R}_k, k \in \hat G_c\}$ be the set of all unitary irreducible representations of $G_c$, and  let
$D_{(k)}(g)\in {\cal R}_k$ for $g\in G_c$. Then the set of functions on $G_c$,
\beqa
\psi_{(k)}{}^i{}_{j}(g) = \sqrt{d_k} D_{(k)}{}^i{}_j(g), \ \ k \in \hat G_c\ , i,j=1,\cdots,d_k, g \in G_c,
\eeqa
forms a complete Hilbert  basis of $L^2(G_c)$ with  inner product
\beqa
\int _{G_c} \sqrt{g}\; \text{d} \varphi^p \text{d} \theta^q  \; \overline{\psi}^{(k)}{}_{i}{}^{j} (g) \psi_{(k')}{}^{ i'}{}_{j'}(g) =
\delta^{k}_{k'}
\delta_{i}^{i'}\delta^{j}_{j'} \ . \nn
\eeqa
\end{theorem}

Since the notations of Theorem \ref{theo:PW} are not appropriate for our purpose, we introduce more convenient  notations. Assume that the Lie
algebra $\g_c$ if of rank $\ell$. Then $\g_c$ admits $\ell$ independent Casimir operator $\{C_1,\cdots, C_{\ell}\}$.  Let
$\cal R$ be a unitary representation of $G_c$, $\cal R$
is uniquely specified by the eigenvalues of the $\ell$ Casimir operators (alternatively the representation can also be specified by the
highest weight with respect to a given Cartan subalgebra $\mathfrak{h} \in \g_c$). Denote ${Q}=(c_1,\cdots,c_\ell)$ the eigenvalues
$\{C_1,\cdots, C_{\ell}\}$. Racah showed in \cite{Ra} that
a number of $1/2(\dim g -\ell)$  internal labels are required to separate the states within the multiplet $\cal R$.
Given a Cartan subalgebra, its eigenvalues constitute an appropriate set of $\ell$ labels.
The choice of the remaining additional internal labels is far from being unique, and  usually depends on a specific chain of proper subalgebras
\begin{equation*}
\frak{g}_1\subset \frak{g}_2 \subset \dots \subset \frak{g}_c
\end{equation*}
such that, in each step, the Casimir operators of the subalgebra  are used to separate states \cite{Bdh,Lou}.
Let $D_{(k)}$ and $D'_{(k)}$ be two matrix elements of the representation ${\cal R}_k$. Since
the product  $D_{(k)}D'_{(k)}$
 is still a matrix element of ${\cal R}_k$, the lines (resp. the column) of $D_{(k)}$ are in the representation
 ${\cal R}_k$ (lines are in left representations and column in right representations).
 Consequently, we denote now any matrix element
 appearing in Theorem \ref{theo:PW} by $\Psi_{LQR}$,
 where $Q$ denotes the $\ell$ eigenvalues of the Casimir operators that identify the representation,
 and $L$  (resp. $R$) the  $1/2(\dim g -\ell)$  labels to separate the states within the left (resp. right) action of $G_c$.
 Let $\cal I$ be the range for the variables $L,Q$ and $R$; with these notations, the Hilbert
 basis of $L^2(G_c)$ associated to the Peter-Weyl theorem is given by:
 \beqa
 \label{eq:HilGc}
{\cal B}_{G_c} = \Big\{\Psi_{LQR}\ , \ \ (L,Q,R) \in {\cal I} \Big\} \ ,
 \eeqa
 and the matrix elements are normalized:
 \beqa
(\Psi_{LQR}, \Psi_{L'Q'R'})= \delta^Q_{Q'} \delta^L_{L'} \delta^R_{R'} \ . \nn
 \eeqa\\

 The parameterisation $m^A=(\varphi^i,\theta^r)$ leads naturally to a
differential realization of the Lie algebra $\mathfrak{g}_c$ for the
generators of the left and right action. Indeed, since
the manifold $G_c$ has a natural left and right action, the Killing equation \eqref{eq:killing}
automatically admits solutions, namely the generators of the left (resp. right) action
and $\g_\xi = (\g_c)_L \oplus (\g_c)_R$.
Let $L_A$ (resp. $R_A$) $A=1,\cdots
n$, be the generators of the left (resp. right) action. We have
\begin{eqnarray}
  \label{eq:LR}
\big[L_A,L_B\big]=i c_{AB}{}^C L_C \ , \ \ \big[R_A,R_B\big]=i c_{AB}{}^C
R_C \ , \ \ \big[L_A,R_B\big]=0 \ ,
\end{eqnarray}
where $c_{AB}{}^C$ are the structure constants of $\mathfrak{g}_c$.
Furthermore, the operators $L_A$ and $R_A$ are Hermitian with respect to the scalar product \eqref{eq:spGc} and
act naturally on the matrix elements $\Psi_{LQR}$:
\begin{eqnarray}
L_A \Psi_{LQR}(\varphi,\theta)&=& (M^Q_A)_L{}^{L^{\prime }}\Psi_{L^{\prime
}QR}(\varphi,\theta)  \notag \\
R_A \Psi_{LQR}(\varphi,\theta)&=& (M^Q_A)_R{}^{R^{\prime }}\Psi_{LQR^{\prime
}}(\varphi,\theta) \   \notag
\end{eqnarray}
where $(M^Q_A)_{L}{}^{L^{\prime }}, (M^Q_A)_{R}{}^{R^{\prime }}$  are the
matrix elements of the left or right action for the representation specified
by $Q$.

Let ${\cal D}_Q, {\cal D}_{Q'}$ be two representations of $G_c$ specified by the eigenvalues
of the Casimir operators. Consider the tensor product:
\beqa
    {\cal D}_Q\otimes {\cal D}_{Q'}= \bigoplus_{\hskip .25truecm Q''}\;  {\cal D}_{Q''}\nn \ .
    \eeqa
    Introducing the Clebsch-Gordan coefficient
    $\begin{pmatrix} Q&Q'&Q''\\L&L'&L''\end{pmatrix}$,  we have
      \beqa
      \big|Q'',L''\big>= \sum \begin{pmatrix} Q&Q'&Q''\\L&L'&L''\end{pmatrix}
     \big|Q,L\big>    \otimes  \big|Q',L'\big>
      \eeqa
      Thus, observing that
      \beqa
\Psi_{LQR}(0)=\sqrt{d} \delta_{LR} \ ,\nn
\eeqa
where $d$ is the dimension of the representation ${\cal D}_Q$, we have
\beqa
\label{eq:psipsi}
\Psi_{LQR}(m)\;\Psi_{L'Q'R'}(m) = C_{LQR;L'Q'R'}^{L''Q''R''}\; \Psi_{L''Q''R''}(m) \ ,
\eeqa
where the summation is implicit, and with
\beqa
\label{eq:Coeff}
C_{LQR;L'Q'R'}^{L''Q''R''}=
\sqrt{\frac{dd'}{d''}}\;
\begin{pmatrix} Q&Q'&Q''\\L&L'&L''\end{pmatrix}\;
 \overline{\begin{pmatrix} Q&Q'&Q''\\R&R'&R''\end{pmatrix}},
 \eeqa
 where $d$ (resp. $d',d''$) are the dimension of ${\cal D}_Q$ (resp. ${\cal D}_{Q'}, {\cal D}_{Q''}$).

\subsection{Coset spaces of compact group manifolds}\label{sec:coset}

Now let $H$ be a closed subgroup of $G_c$, $\pi: G_c\rightarrow G_c/H$ be the quotient map and let $n=\dim G_c, s=\dim H$. Then we can always find a chart $\left(U,\varphi=(y^1,\dots ,y^n)\right)$ around the identity in $G_c$ such that
\begin{enumerate}
\item $\varphi(U)=\left\{ (\xi^1,\dots ,\xi^n)\; |\; |\xi^p|< \varepsilon,\; p=1,\dots ,n\right\}$ for some $\varepsilon>0$.
\item Each slice with $y^{s+1}=\xi^{s+1},\dots ,y^n=\xi^n$ is a relatively open set in some coset $gH$, and these cosets are all distinct.
\item The restriction of $\pi$ to the slice $y^q=0,\; q=1,\dots s$ is an open homeomorphim, and hence determines a chart around the identity element in $G_c/H$.
\end{enumerate}
Using these charts, the coset $G_c/H$ can be endowed with the structure of a differential manifold, such that the $\pi$ and the action of $G_c$ are differentiable. If in addition $H$ is a normal subgroup, then $G_c/H$ inherits the structure of a Lie group, and the differential of the $G_c\rightarrow G_c/H$ corresponds to the to the quotient map $\mathfrak g_c\rightarrow \mathfrak g_c/\mathfrak h$, where $\mathfrak h$ is the Lie algebra of $H$.

In the general case (i.e., there is some $g\in G_c$ with $g^{-1}Hg\neq H$), the coset space $G_c/H$ is not a Lie group, and hence the factor space $\mathfrak g_c/\mathfrak h$ is not a Lie algebra. We write the generators of $\mathfrak g_c$,  namely $T_A$ (with $A=1, \cdots ,\dim \mathfrak g_c$), as follows:
$U_i$ with $i=1,\cdots ,\dim \mathfrak h$,  and $V_p$  with  $p=1,\cdots ,\dim \mathfrak g_c -\dim \mathfrak h$, where $V_p$ denotes the complementary space of $\mathfrak h$ in $\mathfrak g_c$.
The commutations relations take the generic form
\beqa
\begin{array}{ll}
(a)&[U_j,U_k]=i\;g_{jk}{}^\ell U_\ell \ ,\\
(b)&[U_j,V_p]=i\; N_jp^\ell U_\ell+ i\; (R_j)_p{}^q V_q \ ,\\
(c)&[V_p,V_q]= i \;g_{pq}{}^j U_j + i \;g_{pq}{}^r V_r \ .
\end{array}\nn
\eeqa
The relations (a) are trivially satisfied, as $\mathfrak h$ is a Lie subalgebra of $\mathfrak g_{c}$, whereas the
relations (b)  imply that $\mathfrak g_{c}/\mathfrak h$ is a representation of $\mathfrak g_{c}$, whenever the condition $N_{jp}^{\ell}=0$ is satisfied. In particular, the coset space $G_c/H$ is called reductive if there exists an $Ad(H)$-invariant subspace $\mathfrak{m}$ of $\mathfrak{g}_c$ that is complementary to $\mathfrak{h}$ in $\mathfrak{g}_c$.\footnote{The space $\mathfrak{m}$ is sometimes called the Lie subspace for $G_c/H$.} It follows from the invariance that $\left[\mathfrak{h},\mathfrak{m}\right]\subset \mathfrak{m}$. A sufficient condition for the coset space to be reductive is that either $H$ or ${\rm Ad}(H)$ is compact \cite{Hel}. If, in addition $g_{pq}{}^r=0$ holds in (c), then the manifold $G_c/H$ is related to a specially relevant class of homogeneous spaces (endowed with a Riemannian structure), the so-called symmetric spaces \cite{ANG}.

In the following, we shortly summarize the construction steps of the previous section, adapted to (generic) homogeneous spaces. A generic
element of $G_c/H$ is of the form:
\beqa
L=e^{im^p V_p} \ , \nn
\eeqa
and the Vielbein one-form given by
\beqa
e= L^{-1}\text{d} L \ ,
\eeqa
which decomposes as
\beqa
\label{eq:ecoset}
e(m)= i e^p U_p + i e^i V_i \ .
\eeqa
Denoting
\beqa
e^p(m)= e^p{}_{\tilde p}(m) \text{d} m^{\tilde p} \ , \nn
\eeqa
where $p,q,\cdots$ represent the indices in the flat tangent space
and $\tilde p, \tilde q, \cdots$ the curved indices of the manifold $G_c/H$,
the metric tensor on the coset space is given by
\beqa
\label{eq:gGH}
g_{\tilde p \tilde q}= e^p{}_{\tilde p}(m) e^q{}_{\tilde q}(m) \delta_{pq} \ .
\eeqa
We identify the Killing vectors and the corresponding symmetry algebra of the manifold $G_c/H$
\beqa
\label{eq:KGH}
\g_\xi = \Big\{K_A = i \xi_A^p \partial_p: \ \  \xi_A^p\ \ \text{Killing~vector~of} \ \ g_{\tilde p \tilde q}
\Big\} \subset (\g_c)_L \oplus (\g_c)_R \ .
\eeqa

We endow the manifold $G_c/H$ with the scalar product
\beqa
\label{eq:spGcH}
(f,g) = \frac 1 {V'} \int \limits_{G_c/H} \sqrt{g'}\; \text{d} m^N
\overline{f(m)}\; g(m) \ ,
\eeqa
where $V'$ is the volume of $G_c/H$, which is the restriction of \eqref{eq:spGc}
to $G_c/H$ and $N=\dim \g_c -\dim {\mathfrak h}$. \\

We now identify a Hilbert basis of $G_c/H$.
Given a representation  ${\cal R}_k$  of $G_c$, which is such that
${\cal R}_k$ admits  a scalar
representation with respect to the embedding   $H\subset G_c$. Let
${\cal R}_k|_{H}$ be the set of representations with this property.
The set of matrix elements $\Psi_{LQR}$ of a representation  ${\cal R}_k|_{H}$ decomposes into
  \beqa
\Big\{\Psi_{LQR} \ , (LQR) \in {\cal I} \Big\} =
\Big\{\Psi_{LQR_0} \ , (LQR) \in {\cal I}: \Psi_{LQR_O} \text{trival~under~the~right~action~of~}
H\Big\} \oplus \cdots
\nn
\eeqa
Stated differently,
$R_0$ corresponds to the set of
indices associated to the trivial representation  of the right  $H-$action.
The matrix elements $\Psi_{LQR_0}$
contribute to the harmonic analysis
 in $G_c/H$. With the notation of Section \ref{sec:Gc} we denote
 the corresponding normalized matrix elements
 \beqa
 \label{eq:psiGH}
 \phi_{LQ}=\psi_{LQR_0} \ ,
 \eeqa
 where the index $R_0$ means that that  $\phi_{LQR_0}$ transform trivially under the right action
 of $H$  \cite{Salam:1981xd, mrm}.
 The number of internal labels
   is $\ell$ (the rank of $\g_c$) to separate the representations,  and $1/2(\dim \g_c-\ell)$
internal labels to distinguish states within representations \cite{mrm}.
Solving the Killing equation for the coset leads to the generators of
the left action $G_c$, and possibly additional generators associated to
the right action which survive the coset process.
Finally, computing the product $\phi_{LQ}\phi_{L'Q'}$, we obtain coefficients
that can be deduced from \eqref{eq:Coeff} (see \ref{sec:KMGc}).

\subsection{KM  algebras associated to compact Lie groups}\label{sec:KMGc}

Consider first $\M=G_c$.
The first step to associate a KM algebra to $G_c$, is to define the Lie algebra
\beqa
\g(G_c)=
\Big\{T_{aLQR}(m)= T_a\psi_{LQR}(m)\ ,\ \  A=1,\cdots ,\  d,\  (L,  Q, R) \in {\cal I}\Big\} \nn
\eeqa
(see \eqref{eq:HilGc}) with Lie brackets
\beqa
\label{eq:LoopGc}
\big[T_{aLQR}(m), T_{a' L'Q'R'}(m)\big]= if_{aa'}{}^{a''}
C_{LQR;L'Q'R'}^{L''Q''R''} \;T_{a'' L''Q''R''} \ .
\eeqa
This algebra can be obtained in a different way. Let
\beqa
T_a(m)= \sum \limits_{(L,Q,R)\in {\cal I}} T_{aLQR} \;\overline{\psi^{LQR}}(m) \ ,\nn
\eeqa
then
\beqa
\label{eq:LoopGc2}
\big[T_a(m), T_b(m')\big] = i f_{ab}{}^C T_c(m') \frac{\delta(m-m')}{\sqrt{g}}
\eeqa
lead to \eqref{eq:LoopGc} upon integration by $\int \text{d}^n m \sqrt{g} \;\Psi_{LQR}(m)
\int \text{d}^n m'\sqrt{g}\; \Psi_{L'Q'R'}(m')$.\\

This algebra can indeed be obtained easily in the context of Kaluza-Klein theories. Consider the
spacetime $K=\mathbb R^{1,D-1} \times G_c$, {\it i.e.}, a $D-$dimensional Minkowski spacetime with a compact manifolfd $G_c$ of dimension $\dim \g_c=n$
as internal space.
Denote $X^I=(x^\mu, m^M)$ the components in $K$ where $\mu=0,\cdots,D-1$
and $M=1,\cdots, n$. Assume  further the metric on $K$ (in this simple
analysis we don't endow $\mathbb R^{1,d-1}$ with a Riemannian structure, thus
the metric is Minkowskian)
\beqa
\text{d} s^2 = \text{d} x^\mu \text{d} x^\nu \eta_{\mu \nu} -\text{d} m^M \text{d} m^N g_{MN}:=
g_{IJ} \text{d} X^I \text{d} X^J \ ,\nn
\eeqa
where $\eta_{\mu \nu}=$diag$(1,-1,\cdots,-1)$ is the Minkowski metric on $\mathbb R^{1,d-1}$ and $g_{MN}$ is the metric \eqref{eq:gG} on $G_c$.

Consider
a massless free complex  scalar field, with action
\beqa
    {\cal S} = \int \limits_{\mathbb R^{1,D-1}} \text{d} x^D\; \int \limits _{G_c}\;
    \text{d} m^{n}
      \sqrt{g} \;
      g^{IJ} \;\partial_I \Phi^\dag(x,m) \;\partial_J \Phi(x,m)\  . \nn
      \eeqa
      The equations of the motion read
      \beqa
(\Box-\nabla^2) \Phi(x,m) =0 \ ,\nn
      \eeqa
where $\Box= \partial_\mu \partial_\nu \eta^{\mu \nu}$ and
$\nabla^2= \frac 1{\sqrt{g}} \partial_M(g^{MN} \partial_N)$ are respectively the
$D-$dimensional d'Alembertian and the Laplace-Beltrami operator of $G_c$.
Next, decomposing the scalar field in the compact manifold $G_c$ as
\beqa
\Phi(x,m)= \sum \limits_{(L, Q, R) \in {\cal I}} \phi_{LQR}(x)\; \overline{\psi}^{LQR}(m),\nn
\eeqa
and observing that  the action of the  Laplace-Beltrami operator on the scalar functions is related to the quadratic Casimir
operator   \cite{Hel,Gil},
\beqa
\nabla^2= -k C_2\ , \ \ k>0\ \ , \nn
\eeqa
(where $C_2$ is the quadratic Casimir operator of $\g_c$ with eigenvalue $c_2$)  we obtain
\beqa
\nabla^2 \overline{\psi}^{LQR}(m)= -k c_2   \overline{\psi}^{LQR}(m)\ ,\ \  c_2\ge 0 \ , \nn
\eeqa
with the  mass of the field $\phi_{LQR}$ given by $\sqrt{kc_2}$.
Thus the solution of the relativistic wave equation takes the form:
\beqa
\Phi(x,m)=\sum\limits_{(L,Q,R) \in {\cal I}} \int \frac{\text{d}^{D-1}x}{\sqrt{2E_{c_2}}}
\Big(a_{\vec p, QLR} \;e^{-i {\bf p}\cdot {\bf x}}
+b^\dag_{\vec p, QLR} \;e^{i {\bf p}\cdot {\bf x}}\Big)\;\overline{\psi}^{LQR}(m) \ ,\nn
\eeqa
where ${\bf p}\cdot {\bf x}$ is the scalar product in the Minkowski spacetime, and ${\bf p}\cdot {\bf p} = E^2-\vec p \cdot \vec p =k c_2$, or $E_{c_2}\equiv \sqrt{\vec p \cdot \vec p+
  kc_2}$.

Let $\Pi(x,m) = \dot \Phi^\dag(x,m)$ be the conjugate momentum of $\Phi$. If we now quantize the scalar field by
the usual equal-time commutation relations, an elementary computation leads to
\beqa
&&\big[\Phi(t,\vec x,m),\Pi(t,\vec x',m')\big]=i\delta^{D-1}(\vec x -\vec x')
\frac{\delta^n(m-m')}{\sqrt{g}} \nn\\
&\Leftrightarrow&
\left\{\begin{array}{lll}
\big[a_{\vec p, LQR}, a^\dag_{\vec p',L'Q'R'}\big]&=&(2\pi)^{D-1}\;\delta^{D-1}(\vec p - \vec p')\; \delta_{LL'}\; \delta_{QQ'}\; \delta_{RR'}\\
\big[b_{\vec p, LQR}, b^\dag_{\vec p',L'Q'R'}\big]&=&(2\pi)^{D-1}\;
\delta^{D-1}(\vec p - \vec p')\; \delta_{LL'}\; \delta_{QQ'} \;\delta_{RR'} \ .
\end{array} \right.\nn
\eeqa

Consider now, that the scalar field is in a representation ${\cal R}$ of
the compact Lie algebra $\g$. Let $M_a$ be the corresponding matrix
representation. Then the N\oe ther theorem leads to the conserved current
\beqa
j^\mu_a(t,\vec x ,m)=i\big(\Pi(t,\vec x,m) M_a \Phi(t,\vec x,m)-
\Phi^\dag(t,\vec x,m) M_a \Pi^\dag(t,\vec x,m) \big)\nn
\eeqa
and the equal-time commutation relations imply
\beqa
\big[j^0_a(t,\vec x ,m),j^0_b(t',\vec x' ,m')\big]=i f_{ab}{}^c
j^0_c(t,\vec x ,m)\delta^{D-1}(\vec x -\vec x')
\frac{\delta^n(m-m')}{\sqrt{g}} \ . \nn
\eeqa
Upon space integration $\int \text{d} x^{D-1}$ we reproduce \eqref{eq:LoopGc2}. A second integration $\int \sqrt{g}\;\text{d} m^{d}$ leads to \eqref{eq:LoopGc}. Thus
the conserved charges $Q_a= \int \text{d} x^{D-1} \int \sqrt{g}\;\text{d} m^{d}\; j_a^0(t,\vec x,m)$
generate the Lie algebra \eqref{eq:LoopGc}, which appears  naturally upon
compactification. This fact was already observed in \cite{dd}. for the case $G_c =U(1)$.
\\

The construction of the KM algebras associated to $G_c$ follows naturally from the description given in Section \ref{sec:KMM}.
We now introduce the generators of the Cartan subalgebra of $\g_c$ for the left (resp. right action)
 $D^L_i, i=1,\cdots, \ell'\le \ell$ (resp. $D_i^R, i=1\cdots, \ell'$)
 satisfying \eqref{eq:comp2}, together with their associated two-cocycles
\beqa
\label{eq:cosGc}
\begin{split}
\omega^L_i(T_{aLQR},  T_{a'L'Q'R'}) &=&
\int \text{d} m^d \;\sqrt{g} \;\Big<T_{aLQR},  D^L_i T_{a'L'Q'R'}\Big>_0
=k_{aa'} L'(i) \eta_{LQ  R ,L' Q'R'}\;
 ,  \\
\omega^R_i(T_{aLQR},  T_{a'L'Q'R'}) &=&\int \text{d} m^d \;\sqrt{g}\; \Big<T_{aLQR},
D^R_i T_{a'L'Q'R'}\Big>_0 =
k_{aa'} R'(i) \eta_{L Q R,L'Q' R'}\; \ ,
\end{split}
\eeqa
(see \eqref{eq:gamnf}).  
As the detailed steps where given in Sect. \ref{sec:KMM}, we merely indicate the results
without further details. The KM algebra associated to $G_c$ can be defined (with the notations
of Sect. \ref{sec:KMM}) by
$\widetilde{\mathfrak{g}}($G$_c) = \big\{\mathcal{T}_{aLQR},a=1\cdots, d, (LQR) \in {\cal I}, D^L_i, D^R_i, k_L^i, k_R^j, i=1,\cdots \ell'\}$,
with non-vanishing Lie brackets:
\beqa
\begin{split}
\label{eq:KM-GC1}
\big[{\cal T}_{aLQR},{\cal T}_{a'L'Q'R'}\big] &=
i\;  f_{aa'}{}^{a''} C_{LQR;L'Q'R'}^{L''Q''R''} {\cal T}_{a'' L''Q''R''}\nn\\
& +    k_{aa'}  \;\eta_{LQR,L'Q'R'} \; \sum \limits_{i=1}^{\ell'} \big(k_L^i L(i) +k_R^i R(i)\Big)
 \ , \nn\\
 \big[D^L_i, {\cal T}_{aLRQ}\big] &=   L(i)  {\cal T}_{aLQM}\ ,\\
 \big[D^R_i, {\cal T}_{aLRQ}\big] &=   R(i)  {\cal T}_{aLQP}.\nn
 \end{split}
\eeqa
 Note that here, $L(i), R(i)$ are the components of the weight
vectors of the corresponding representation.

Consider now briefly the construction of a KM algebra associated to a coset $G_c/H$ where
$H\subset G_c$ is a closed subgroup of $G_c$.
The construction of a KM algebra associated to
the manifold $G_c/H$ is entirely analogous to the construction of the KM algebra
associated to $G_c$, {\it mutas mutandis} the following replacements:
\begin{enumerate}[noitemsep]
\item  the metric  on $G/H$ is given by \eqref{eq:gGH};
\item the scalar product reduces to \eqref{eq:spGcH};
\item the Hilbert basis of $L^2($G$_c/$H$)$  is given by the set of $H-$invariant
  vectors of the right-action \eqref{eq:psiGH};
\item the Killing vectors and the corresponding symmetry algebra of $G/H$ are given in
  \eqref{eq:KGH}. Denote $D_i, i=1,\cdots,\ell'\le \ell$ the generators of the Cartan subalgebra satisfying \eqref{eq:comp2}.
\item the two-cocycles are obtained through  \eqref{eq:cc} (with the commuting
  Hermitian operators satisfying \eqref{eq:comp2} $D_i, i=1,\cdots,\ell'\le \ell$).

\end{enumerate}
The Lie brackets of the algebra  $\widetilde{\g}($G$_c/$H) are given by the Lie
brackets \eqref{eq:KM-GC1} {\it modulo} the substitutions above.

\subsection{KM algebra associated to SU$(2)$ and SU$(2)/$U$(1)$} \label{sec:SU2}
In this section we give  explicit examples of  KM algebras associated to
 compact group manifolds or  cosets. A detailed account on their construction was given
in \cite{mrm}, for this reason we merely present two examples here.

We first consider  $G_c=$ SU$(2)$. We give here the principal steps, the details of which can be found also in \cite{mrm,ram3}.
We have
\beqa
\text{SU}(2) =\Big\{z_1, z_2 \in \mathbb C^2: |z_1|^2+|z_2|^2\Big\}\cong {\mathbb S}^3 \ ,\nn
\eeqa
where  ${\mathbb S}^3$ is the three-sphere.
A parameterisation of $\mathbb{S}^{3}$ is given by
\begin{eqnarray}  \label{eq:pSU2}
\begin{split}
z_1 = \cos \frac \theta 2 e^{i\frac{\varphi+\psi}2} \ , \ \
z_2 = \sin \frac \theta 2 e^{i\frac{\varphi-\psi}2} \ ,\ \  0\le \theta \le \pi\ , \ \ 0\le \varphi <2\pi \ , \ \ 0\le \psi <4 \pi \ .
\end{split}%
\end{eqnarray}
From
\beqa
m=\begin{pmatrix} z_1&-\bar z_2\\
 z_2&\phantom{-}\bar z_1
\end{pmatrix} \in \text{SU}(2)
\eeqa
equation \eqref{eq:e} leads to the left-invariant one-forms
\begin{eqnarray}  \label{eq:LI1F}
\lambda_1&=& \sin \varphi \;\text{d} \theta -\cos\varphi\; \sin\theta\; \text{d}
\psi  \notag \\
\lambda_2&=& \cos \varphi\; \text{d} \theta +\sin\varphi\; \sin\theta \;\text{d}
\psi \\
\lambda_3&=&\cos \theta \;\text{d} \psi + \text{d} \varphi  \notag
\end{eqnarray}
and to the metric tensor
\beqa
\text{d} s^2 = \lambda_1^2 + \lambda_2^2 + \lambda_3^2 =
\text{d} \theta^2 + \text{d} \varphi^2 + \text{d} \psi^2 + 2 \cos\theta \;\text{d} \phi \;\text{d} \psi\ .
\eeqa
The left/right invariant vectors fields  obtained by solving the Killing equation \eqref{eq:killing} are
\begin{eqnarray}
\begin{array}{llllll}
L_\pm & = & e^{\pm i \psi}\Big( -\frac i {\sin \theta} \partial_\varphi +
i\cot \theta \partial_\psi \pm \partial_\theta \Big) \ , & L_0 & = &
-i\partial_\psi \nonumber \\
R_\pm & = & e^{\pm i \varphi}\Big(\phantom{-}\; \frac i {\sin \theta}
\partial_\psi - i\cot \theta \partial_\varphi \mp \partial_\theta \Big)\ , &
R_0 & = & -i\partial_\varphi \nonumber%
\end{array}
\notag
\end{eqnarray}
and satisfy the commutation relations
\begin{eqnarray}
\begin{array}{llllll}
\big[L_0, L_\pm\big] & = & \pm L_\pm \ , & \big[L_+, L_-\big] & = & 2 L_0 \ , \\%
[2pt]
\big[R_0, R_\pm\big] & = & \pm R_\pm \ , & \big[R_+, R_-\big] & = & 2 R_0 \ ,
\end{array}\ \ \big[L_a,R_b\big]=0 \ .
\notag
\end{eqnarray}
The quadratic Casimir operator reduces to:
\begin{eqnarray}
C_2= -\partial^2_\theta -\cot \theta \partial_\theta -\frac 1 {\sin^2 \theta}%
\Big(\partial_\varphi^2 +\partial_\psi^2\Big)+2 \frac{\cos \theta}{%
\sin^2\theta} \partial_\varphi\partial_\psi \ .  \notag
\end{eqnarray}

Finally, the SU$(2)-$scalar product is given by
\beqa
(f,g)= \frac 1 {16 \pi^2} \int \limits_{{\mathbb S}^3} \sin \theta \;\text{d} \theta\; \text{d} \psi \;\text{d} \phi\;
  \overline{f(\theta,\varphi,\psi)}\; g (\theta,\varphi,\psi) \ . \nn
  \eeqa

  We now identify the matrix elements. Since $\dim \mathfrak{su}(2)=3$ and $\text{rk}\; \mathfrak{su}(2)=1$, we need
  two labels to identify the matrix elements of $\mathfrak{su}(2)$, one external corresponding to the eigenvalue of the quadratic Casimir operator, and one internal label determined by the Cartan subalgebra eigenvalues of the
  left/right action of $\mathfrak{su}(2)$.
  We thus obtain \cite{ram3} $(s \in \frac 12\mathbb N, -s\le n,m\le s$):
  {\small
  \begin{eqnarray}
    \label{eq:matsu2}
\psi_{nsm}(\theta,\varphi,\psi) = \left\{
\begin{array}{lc}
\frac {(-1)^{m-n} \sqrt{(2s+1)}} {(n-m)!}\sqrt{ \frac{(s+n)!}{(s-n)!} \frac{%
(s-m)!}{(s+m)!}} e^{im\varphi + i n \psi} \cos^{-n-m} \frac \theta 2
\sin^{n-m} \frac \theta 2 & n\ge m\  \\
{}_2F_1(-m-s,-m+s+1;1-m+n; \; \sin^2 \frac \theta 2) & -n-m\ge 0 \\[6pt]
\frac { \sqrt{(2s+1)}} {(m-n)!} \sqrt{ \frac{(s+m)!}{(s-m)!} \frac{(s-n)!}{%
(s+n)!}} e^{im\varphi + i n \psi} \cos^{-n-m} \frac \theta 2 \sin^{m-n}
\frac \theta 2 & m\ge n\  \\
{}_2F_1(-n-s,-n+s+1;1-n+m; \; \sin^2 \frac \theta 2) & -n-m\ge 0 \\[6pt]
\frac {(-1)^{m-n} \sqrt{2s+1}} {(n-m)!} \sqrt{ \frac{(s+n)!}{(s-n)!} \frac{%
(s-m)!}{(s+m)!}} e^{im\varphi + i n \psi} \cos^{n+m} \frac \theta 2
\sin^{n-m} \frac \theta 2 & n\ge m\  \\
{}_2F_1(n-s,n+s+1;1+n-m; \; \sin^2 \frac \theta 2) & n+m\ge 0 \\[6pt]
\frac{\sqrt{2s+1}} {(m-n)!} \sqrt{ \frac{(s+m)!}{(s-m)!} \frac{(s-n)!}{(s+n)!%
}} e^{im\varphi + i n \psi} \cos^{n+m} \frac \theta 2 \sin^{m-n} \frac
\theta 2 & m\ge n\  \\
{}_2F_1(m-s,m+s+1;1+m-n; \; \sin^2 \frac \theta 2) & n+m\ge 0%
\end{array}
\right.
  \end{eqnarray}
  }
where ${}_2F_1$ denotes the Euler hypergeometric polynomial
(see e.g. \cite{ram} for definitions).
A direct  inspection gives rise to
\beqa
\overline{\psi^{nsm}}(\theta,\varphi,\psi)=(-1)^{n-m} \psi_{-ns-m}(\theta,\varphi,\psi) \ ,\nn
  \eeqa
  leading to $\eta_{nm,n'm'}=(-1)^{n-m}\delta_{n,-n'}\delta_{m,-m'}$ (see Sect. \ref{sec:M}).
These matrix elements were obtained solving the differential equations
\beqa
L_0 \psi_{nsm}(\theta,\varphi,\psi)&=&n \psi_{nsm}(\theta,\varphi,\psi)\ , \nn \\
R_0 \psi_{nsm}(\theta,\varphi,\psi)&=&m \psi_{nsm}(\theta,\varphi,\psi)\ , \nn \\
C_2 \psi_{nsm}(\theta,\varphi,\psi)&=&s(s+1) \psi_{nsm}(\theta,\varphi,\psi)\nn
\eeqa
where $C_2$ is the quadratic Casimir operator \cite{ram3}, and we have
\beqa
\begin{split}
L_\pm \psi_{nsm}(\theta,\varphi,\psi) =&\sqrt{(s\mp n)(s\pm n +1)} \psi_{n\pm 1sm}\ ,\\
L_0 \psi_{nsm}(\theta,\varphi,\psi)=& n  \psi_{nsm}(\theta,\varphi,\psi)\\
R_\pm \psi_{nsm}(\theta,\varphi,\psi) =&\sqrt{(s\mp m)(s\pm m +1)} \psi_{nsm \pm 1}\ ,\\
R_0 \psi_{nsm}(\theta,\varphi,\psi)=& m  \psi_{nsm}(\theta,\varphi,\psi) \ .
\end{split}\nn
\eeqa
The set
\begin{eqnarray}  \label{eq:HS3}
\mathcal{B} = \Big\{\psi_{nsm}, s \in \frac 12 \mathbb{N}, -s\le n,m \le s %
\Big\}
\end{eqnarray}
constitutes a Hilbert basis of $L^2(\mathrm{SU}(2))$, and the following is satisfied :
\begin{eqnarray}
  \label{eq:HSU2}
(\psi_{nsm},\psi_{n^{\prime },s^{\prime },m^{\prime }}) = \delta_{ss^{\prime }} \delta_{n n^{\prime }}
\delta_{mm^{\prime }} \ .
\end{eqnarray}

\bigskip
We have now all the ingredients to define the
KM algebra associated to SU$(2)$. First observe that the differential operators
$L_0, R_0$ satisfy both \eqref{eq:comp2}. Let $\omega_L, \omega_R$
be the corresponding two-cocycles.
One may now wonder whether there exists additional differential operators compatible with  the two cocycles $\omega_L, \omega_R$. This will be studied in the next
section.
Thus in turn we are able to define the possible extensions of the algebra
$\g(\M)=\Big\{T_a \psi_{nsm}(\theta, \varphi, \psi),\ a=1,\cdots, d \ , \ell \in \frac 12 \mathbb{N}, -\ell\le n,m \le \ell\Big\}$  (see \eqref{eq:LoopGc}) by
\begin{eqnarray}
\widetilde{\mathfrak{g}}(\text{SU}(2))\rtimes \{L_0,R_0\}=\Big\{\mathcal{T}_{a nsm},
k_L,k_R, \ \ a=1,\cdots, d \ , \ell \in \frac 12 \mathbb{N}, -\ell\le n,m
\le \ell\Big\} \rtimes\Big\{L_0,R_0\Big\} \notag
\end{eqnarray}
with Lie brackets (see \cite{mrm})
{\small
\begin{eqnarray}
  \label{eq:KMSU2}
\big[\mathcal{T}_{a nsm}, \mathcal{T}_{a^{\prime }n^{\prime }s^{\prime
}m^{\prime }}\big]&=& if_{a a^{\prime }}{}^{a^{\prime \prime }} C^{s^{\prime
\prime }}_{ss^{\prime }nn^{\prime }mm^{\prime }} \mathcal{T}_{a^{\prime
    \prime }n+n^{\prime }s^{\prime \prime }m+m^{\prime }}\nn\\
&&+
k_{ab} (-1)^{m-n} \delta_{ss^{\prime }}
\delta_{n,-n^{\prime }}\delta_{m,-m^{\prime }}\big(k_Ln^{\prime
}+k_Rm^{\prime })
\notag \\
\big[L_0,\mathcal{T}_{a nsm}\big]&=&n \mathcal{T}_{a nsm}   \\
\big[R_0,\mathcal{T}_{a nsm}\big]&=&m \mathcal{T}_{a nsm} \ .  \notag
\end{eqnarray}}
Here  we have
\beqa
  \label{eq:CCSU2}
C_{s_1 s_2}^{S}{}_{n_1n_2 m_1m_2} =\sqrt{ \frac{(2s_1+1)(2s_2+1)}{2S+1}}
\begin{pmatrix}
s_1 & s_2 & S \\
n_1 & n_2 & n_1+n_2%
\end{pmatrix}%
\; \overline{
\begin{pmatrix}
s_1 & s_2 & S \\
m_1 & m_2 & m_1+m_2%
\end{pmatrix}
}
\eeqa
with $\begin{pmatrix}
s_1 & s_2 & S \\
n_1 & n_2 & n_1+n_2%
\end{pmatrix}$ the Clebsch-Gordan coefficients.
Note that we recover the algebra obtained in \cite{mrm,ram3}. There exists
a second algebra associated to $\g(\text{SU}(2))$, that  will be studied in the next
section.
\\

The KM algebra associated to SU$(2)/$U$(1)\cong {\mathbb S}^2$ follows directly from the construction above. A point in the coset is parameterized by
\beqa
m= \begin{pmatrix} e^{i \varphi} \cos \frac \theta 2& -\sin \frac \theta 2\\
  \sin \frac \theta 2 & e^{-i\varphi}\cos \frac \theta 2
  \end{pmatrix}\nn
\eeqa
obtained by the substitution $\psi=-\varphi$ in \eqref{eq:pSU2}.
From \eqref{eq:ecoset},
we obtain the left-invariant vector fields
\beqa
\omega_1 = \sin\theta\; \sin \varphi \;\text{d} \varphi - \sin \varphi \;\text{d} \theta  , \ \
\omega_2 = \sin\theta \;\cos \varphi \;\text{d} \varphi + \cos \varphi\; \text{d} \theta \ , \nn
\eeqa
such that the metric on the two-sphere reduces to
\beqa
\text{d} s^2= \text{d} \theta^2 + \sin^2 \theta \;\text{d} \varphi^2 \ . \nn
\eeqa
The invariant vector fields obtained by solving the Killing equation \eqref{eq:killing} are
\beqa
L_\pm= e^{\pm i \varphi}\Big( i\cot \theta \partial_\varphi \pm \partial_\theta\Big) \ , \ \
L_0=-i\partial_\varphi \ , \nn
\eeqa
and generate the $\mathfrak{so}(3)-$Lie algebra.
We finally define the scalar product
\beqa
(f,g) = \frac 1 {4 \pi} \int \limits_{\mathbb S^2} \sin \theta \;\text{d} \theta\; \text{d} \varphi\; \overline{f(\theta,\varphi)} \;g(\theta,\varphi) \ . \nn
\eeqa

Following Sec. \ref{sec:coset}, the only representations to be considered for harmonic analysis on the two-sphere
are those having matrix elements with eigenvalues $R_0$ equal to zero
(see \eqref{eq:psiGH}). This is possible {\it iff} $s \in \mathbb N$ and for $m=0$.
Substituting $m=0$ in \eqref{eq:matsu2}, we obtain $(\ell\in \mathbb N$):
\begin{eqnarray}
&Y_{\ell n}(\theta,\varphi) =\frac{(-1)^{{\ell}+\frac{|n|+n}2}\sqrt{2\ell+1} }{|n|!} \sqrt{
\frac{(\ell+|n|)!}{(\ell-|n|)!}} e^{ i n \varphi} \cos^{|n|} \frac \theta 2
\sin^{|n|} \frac \theta 2\nn\\
    &{}_2F_1(|n|-\ell,|n|+\ell+1;1+|n|; \; \sin^2 \frac
\theta 2),%
\nn
\end{eqnarray}
where
\beqa
\overline{Y^{\ell n}}(\theta,\varphi) = (-1)^n Y^{\ell-n}(\theta,\varphi) \ . \nn
\eeqa

The KM algebras associated to $\mathbb S^2$ follows immediately:
 \begin{eqnarray}
\widetilde{\mathfrak{g}}(\text{SU}(2)/\text{U}(1))=\Big\{\mathcal{T}_{a \ell m },
k, \ \ a=1,\cdots, d \ , \ell \in  \mathbb{N}, -\ell\le m
\le \ell\Big\} \rtimes \{L_0\} \notag
\end{eqnarray}
with Lie brackets (see \cite{mrm})
\begin{eqnarray}
\big[\mathcal{T}_{a \ell m}, \mathcal{T}_{a^{\prime }\ell'm^{\prime }}\big]&=& if_{a a^{\prime }}{}^{a^{\prime \prime }} C^{\ell^{\prime
\prime }}_{\ell \ell^{\prime }mm^{\prime }} \mathcal{T}_{a^{\prime
\prime }m+m^{\prime }\ell^{\prime \prime }}  + k k_{ab} (-1)^{m} \delta_{\ell \ell^{\prime }}
\delta_{m,-m^{\prime }} \notag \\
\big[L_0,\mathcal{T}_{a \ell m}\big]&=&m \mathcal{T}_{a \ell m}  \notag
\end{eqnarray}
Since in the coset SU$(2)/$U$(1)$ the matrix elements of the left action are obtained
from the matrix elements of SU$(2)$, $\psi_{n\ell m}$ for $m=0$, the coefficients
\eqref{eq:CCSU2} reduce to
\beqa
C_{\ell_1 \ell_2}^{L}{}_{n_1n_2} =\sqrt{ \frac{(2\ell_1+1)(2\ell_2+1)}{2L+1}}
\begin{pmatrix}
\ell_1 & \ell_2 & L \\
n_1 & n_2 & n_1+n_2%
\end{pmatrix}%
\; \overline{
\begin{pmatrix}
\ell_1 & \ell_2 & L \\
0 & 0 & 0%
\end{pmatrix}
}\nn
\eeqa
This algebra was obtained for the first time in \cite{Bars:1983uv}, see also \cite{mrm, ram3}.

\subsection{Virasoro algebra associated to the two- and three- sphere}
\label{sec:virS}

In this paragraph we show that, for two- and the three-spheres, an analogue of the Virasoro algebra can be constructed.

Consider first the case of $\mathbb S^2$.
Let $L$ be a vector field on $\mathbb S^2$:
\beqa
L= A_\varphi(\theta,\varphi) \partial_\varphi + A_\theta(\theta,\varphi) \partial_\theta \ \nn
\eeqa
As $H^1(\mathbb S^2)=1$, every closed one-form on the sphere is exact.
Let $\gamma=\d h$ be a closed one-form associated to the cocycle
\beqa
\omega_h(X,Y) = \frac 1 {4\pi} \int \limits_{\mathbb S^2} \big<X,\d Y\big>\wedge \d h \ . \nn
\eeqa
The compatibility condition between the KM algebra with two-cocycle $\omega_h$
\beqa
\big[X,Y\big]_h= \big[X,Y]_0 + \omega_h(X,Y) \ , \nn
\eeqa
where $[\;,\;]_0$ are the Lie brackets without central extension, {\it i.e.}, the Lie brackets
of $\g(\mathbb S^2)$,
turns out to be
\beqa
L h=c \ ,\nn
\eeqa
with $c$ a constant whose solutions are given by (see \cite{Frappat:1989gn})
\beqa
\label{eq:diff-cos}
L^T_h(\theta,\varphi)&=&T(\theta,\varphi)\big(\partial_\theta h \partial_\varphi - \partial _\varphi h \partial_\theta)\ , \\\
\Lambda(\theta,\varphi)&=&\frac{\partial_\theta h \partial_\varphi + \partial _\varphi h \partial_\theta}{\partial_ \theta h \partial_\varphi h} \ . \nn
\eeqa
The generators $L^T_h$ generate the area preserving diffeomorphism algebra
\cite{FI, bps}.
The second solution is possible if the denominator does not vanish.
In particular, if we define the cocycle associated to $L_0$
the generators of the de Witt algebra
 \beqa
 \label{eq:llm}
\ell_{\ell m} = - Y_{\ell m}(\theta,\varphi) L_0 \ ,
\eeqa
are compatible with the two-cocycle $\omega_0$. We are thus able to
define the algebra
\beqa
&\widetilde{\mathfrak{g}}'(\text{SU}(2)/\text{U}(1))\rtimes \text{Witt}\nn\\
&=
  \Big\{\mathcal{T}_{a \ell m},
k \ \ a=1,\cdots, d \ , \ell \in  \mathbb{N}, -\ell\le m
\le \ell\Big\} \rtimes\Big\{\ell_{\ell m} \ ,  \ell \in  \mathbb{N}, -\ell\le m
\le \ell\Big\} \notag
\end{eqnarray}
with the following Lie brackets:
\begin{eqnarray}
  \label{eq:KMSO3-2}
  \big[\mathcal{T}_{a \ell m}, \mathcal{T}_{a^{\prime } \ell' m^{\prime }}\big]&=& if_{a a^{\prime }}{}^{a^{\prime \prime }} C^{\ell^{\prime
\prime }}_{\ell \ell^{\prime }mm^{\prime }} \mathcal{T}_{a^{\prime
\prime }\ell'' m+m^{\prime }}  + k_0 m'k_{ab} (-1)^{m} \delta_{\ell \ell^{\prime }}
\delta_{m,-m^{\prime }} \notag \\
\big[L_0,\mathcal{T}_{a \ell m}\big]&=&m \mathcal{T}_{a \ell m}
\end{eqnarray}

\bigskip
A similar analysis holds for $\mathbb S^3$.
As $H^2(\mathbb S^3)=1$, a closed  one-form is exact.
Let $(u_1,u_2,u_3)= (\varphi,\psi,\cos \theta)$ and let $h(u)= h_i(u) \d u^i$.
The corresponding two-cocycle associated to the closed form $\d h$ is given by
\beqa
\omega_h(X,Y)=\int \limits_{\mathbb S^3} \big<X,\d Y\Big> \wedge \d h \ . \nn
\eeqa
From
$
\d(X\d Y\wedge h)=\d X \wedge \d Y \wedge h -X \d Y\wedge \d h\nn
$
and since $\mathbb S^3$ has no boundary, we have
\beqa
\omega_h(X,Y)&=&\int \limits_{\mathbb S^3} \Big< X\wedge \d Y\Big>\wedge \d h
=  \int \limits_{\mathbb S^3} \Big<\d X\wedge \d Y\Big>\wedge h \nn\\
&=&\int \limits_{\mathbb S^3}\Big( h_1 \{X,Y\}_{23} +h_2 \{X,Y\}_{31} + h_3 \{X,Y\}_{12}\Big)\d u_1  \d u_2 \d u_3 \ , \nn
\eeqa
where
\beqa
\{X,Y\}_{ij}=\partial_i X \partial_j Y - \partial_j X \partial_i Y \ . \nn
\eeqa
It can be shown that the vector field (see also Eq.[\ref{eq:DT}])
\beqa
L_h=A(u_1,u_2,u_3)\Big(\big(\partial_1 h_2 \partial_3 - \partial_3 h_2 \partial_1\big) +
\big(\partial_2 h_3 \partial_1 - \partial_1 h_3 \partial_3\big)+
\big(\partial_3 h_1 \partial_2 - \partial_2 h_1 \partial_3\big)\Big)\ , \nn
\eeqa
is compatible with the two-cocycle $\omega_h$ \cite{Frappat:1989gn}.
The algebra generated by the operators of the form $L_h$ above are
called area preserving diffeomorphisms in \cite{FI, bps}.
We are thus able to define
 (pay attention to the fact that, in this case, we only
have one central charge, say $k_L=k$)
\beqa
&\widetilde{\mathfrak{g}'}(\text{SU}(2))\rtimes \text{Witt}_{L_0}=\nn\\
&\Big\{\mathcal{T}_{a nsm},
k, \ \ a=1,\cdots, d \ , s \in \frac 12 \mathbb{N}, -s\le n,m
\le s\Big\} \rtimes\Big\{\ell_{msn}, s \in \frac 12 \mathbb{N}, -s\le n,m
\le s \Big\} \notag
\eeqa
where
\beqa
\ell_{msn}=-\psi_{msn}(\varphi,\psi,\theta) L_0 \nn
\eeqa
and with Lie brackets:
\beqa
\big[\mathcal{T}_{a nsm}, \mathcal{T}_{a^{\prime }n^{\prime }s^{\prime
}m^{\prime }}\big]&=& if_{a a^{\prime }}{}^{a^{\prime \prime }} C^{s^{\prime
\prime }}_{ss^{\prime }nn^{\prime }mm^{\prime }} \mathcal{T}_{a^{\prime
\prime }n+n^{\prime }s^{\prime \prime }m+m^{\prime }} 
+n' k\; k_{aa'} (-1)^{m+n} \delta_{ss^{\prime }}
\delta_{n,-n^{\prime }}\delta_{m,-m^{\prime }}  \notag \\
\big[\ell_{nsm}, {\cal T}_{a'n's'm'}\big]&=&
-n' C^{s^{\prime
    \prime }}_{ss^{\prime }nn^{\prime }mm^{\prime }}  {\cal T}_{a n+n's''m+m'}\nn\\
\big[\ell_{nsm}, \ell_{n's'm'}\big]&=&(n-n') C^{s^{\prime
    \prime }}_{ss^{\prime }nn^{\prime }mm^{\prime }}  \ell_{n+n's''m+m'}\
\eeqa

\bigskip
We now show that  the de  Witt algebra of the two- and three-sphere
can be centrally extended. Before analysing possible central extensions, let us make
several observations. A basis of the algebra of vector fields $\mathfrak{X}(\mathbb S^2)$ is given
by $\{X^\varphi_{\ell m} = Y_{\ell m} \partial_\varphi, X^\theta_{\ell m} = Y_{\ell m} \partial_\theta,
\ell \in \mathbb N, -\ell \le m\le \ell\}$. Several subalgebras of $\mathfrak{X}(\mathbb S^2)$  with different properties may
be considered. For instance, Floratos and Iliopoulos studied the area preserving diffeomorphism algebra which appears in the analysis of bosonic membranes. If the membrane has the topology
of the two-sphere, this algebra is generated by the vector fields of the form \cite{FI}
$L_h=\partial_{\cos \theta} h(\theta,\varphi)\;\partial_\varphi -\partial_\varphi h(\theta,\varphi)\;\partial_{\cos\theta}$ and satisfy
\beqa
\label{eq:alg1}
\big[V_{h_1}, V_{h_2}\big]= V_{\{h_2,h_1\}} \ ,
\eeqa
where $\{h_1,h_2\}= \partial_{\cos \theta} h_1(\theta,\varphi) \partial_\varphi h_2(\theta,\varphi)-
\partial_{\cos \theta} h_2(\theta,\varphi) \partial_\varphi h_1(\theta,\varphi)$ is the Poisson bracket.
In particular, we have for $V_{\ell m}=\partial_{\cos \theta} Y_{\ell m}(\theta,\varphi)\partial_\varphi -\partial_\varphi  Y_{\ell m}(\theta,\varphi)\partial_{\cos\theta}$
\beqa
\big[V_{\ell_1 m_1}, V_{\ell_2 m_2}\big]= - g _{\ell_1 m_1 \ell_2 m_2}^{\ell_3 m_3}\; V_{\ell_3 m_3} \ ,
\eeqa
where the structure constants
\beqa
\{Y_{\ell_1 m_1}, Y_{\ell_2 m_2}\}= g _{\ell_1 m_1 \ell_2 m_2}^{\ell_3 m_3}\; Y_{\ell_3 m_3} \ , \nn
\eeqa
are given in \cite{Hop}.
The second relevant algebra is that associated  to the compatibility
of the cocycle $\omega_h$  \cite{Frappat:1989gn}. This algebra is generated by the vector fields of the
form $L_h^T=T(\theta,\varphi)\Big(\partial_{\cos \theta} h(\theta,\varphi)\partial_\varphi -\partial_\varphi h(\theta,\varphi)\partial_{\cos\theta}\Big)$:
\beqa
\label{eq:alg2}
\big[L_h^{T_1}, L_h^{T_2}\big]= L_h^{T_3} \ ,\ \  T_3= T_1\{T_2,h\}- T_2 \{T_1,h\} \ .
\eeqa
The difference between the algebras given in \eqref{eq:alg1} and \eqref{eq:alg2} is that, for the former, the function
$h$ in $V_h$ varies, whilst in the latter, the function $h$ in $V_h^T$ is fixed. It has been shown in
\cite{bps} that the algebra \eqref{eq:alg1}, or the area preserving diffeomorphism algebra, does not admit central extensions.
Differently, we now show that the algebra \eqref{eq:alg2} for $h(\theta,\varphi)= Y_{\ell m}(\theta,\varphi)$ admits
a central extension.
Introduce   the spherical harmonics in the form:
\beqa
\label{eq:Y}
Y_{\ell m}(\theta,\varphi)= \sqrt{\frac{(\ell -m)!}{(\ell+m)!}}P_{\ell m}(\cos \theta)e^{i m \varphi} = Q_{\ell m}(\cos \theta) e^{i m \varphi},
\eeqa
where $P_{\ell m}$ are the associated Legendre  functions,
satisfying the orthonormality  property  for $ \ell_1,\ell_2 \ge |m| $:
\beqa
\label{eq:orth}
(Q_{\ell_1 m}, Q_{\ell_2 m}) = \frac 12 \int \limits_0^\pi \sin \theta \text{d} \theta\; Q_{\ell_2 m}(\cos\theta) \;Q_{\ell_1 m}(\cos\theta)=
\delta_{\ell_1 \ell_2} \ .   \
\eeqa
We also have
\beqa
\label{eq:QQ}
Q_{\ell_1 m_1}(\cos\theta) Q_{\ell_2 m_2}(\cos\theta) = C_{\ell_1 \ell_2 m_1 m_2}^\ell Q_{\ell m_1+m_2}(\cos\theta) \ ,
\eeqa
and
\beqa
\label{eq:mQ}
Q_{\ell -m}(\cos\theta) = (-1)^m Q_{\ell m}(\cos \theta) \ ,
\eeqa
which are a direct consequence of analogous relations for $Y_{\ell m}$ and from the definition of $Q_{\ell m}$
(see \eqref{eq:Y}).
Let
\beqa
L(\theta,\varphi)&=&   \sum\limits_{m=-\infty}^{+ \infty} \Big( \sum \limits_{\ell = |m|}^{+ \infty} L_{\ell m}  Q_{\ell m}(\cos\theta)\Big)
                         e^{i m\varphi}
                         = \sum\limits_{m=-\infty}^{+ \infty} L_m(\theta) e^{i m \varphi} \ . \nn
                         \eeqa
   We further assume that the $L_m(\theta)$ satisfy the relations

   \beqa
   \label{eq:KMS2}
\big[L_m(\theta),L_n(\theta')\big]&=&\Big((m-n) L_{m+n} + \frac c{12}m(m^2-1) \delta_{m,-n}\Big)\delta(\cos \theta -\cos \theta')
\eeqa
Now,  using  the orthonormality relation \eqref{eq:orth}, \eqref{eq:QQ} and \eqref{eq:mQ}, integration by
\beqa
\frac 12 \int \d u Q_{\ell m}(u) \frac 12 \int \d u' Q_{\ell'm'}(u')  \ , \  \ u=\cos\theta, u'=\cos\theta'\nn
\eeqa
leads to:
\beqa
\label{eq:KV-S2}
\big[L_{\ell_1 m_1},L_{\ell_2 m_2}\big]&=&(m-n) C_{\ell_1 \ell_2 m_1 m_2}^{\ell_3}  L_{\ell_3 m_1+m_2} +(-1)^{m_1} \frac c{12}m_1(m_1^2-1)
\delta_{m_1,-m_2}\delta_{\ell_1\ell_2} \nn
\eeqa

We now check that the Jacobi identities are  satisfied using \eqref{eq:KMS2}. Indeed,
\beqa
&[[L_{m_1}(\theta_1),L_{m_2}(\theta_2)],L_{m_3}(\theta_3)]\nn\\
&+[[L_{m_2}(\theta_2),L_{m_3}(\theta_3)],L_{m_1}(\theta_1)]\nn\\
&+[[L_{m_3}(\theta_3),L_{m_1}(\theta_1)],L_{m_2}(\theta_2)]\nn\\
&=-\frac{c}{12}(m_2 - m_3)(m_1 - m_3)(m_1 - m_2)(m_1 + m_3 + m_2)\nn\\
&\delta_{m_1+m_2+m_3,0}
\delta(\cos\theta_1 -\cos\theta_2) \delta(\cos \theta_2-\cos\theta_3)=0
\nn
\eeqa
where we have used $(m_1-m_2)(m_3^3-m_3)+(m_2-m_3)(m_1^3-m_1) +(m_3-m_3)(m_1^3-m_1)=
(m_2 - m_3)(m_1 - m_3)(m_1 - m_2)(m_1 + m_3 + m_2)$.

Thus altogether, the algebra
\beqa
&\widetilde{\g}'(\mathbb S^2) \rtimes \text{Vir}(\mathbb S^2)=\nn\\
&
\Big\{\mathcal{T}_{a \ell m},
k, \ \ a=1,\cdots, d \ , \ell  \in  \mathbb{N}, -\ell \le m
\le \ell \Big\} \rtimes\Big\{L_{\ell m } \ , \ell \in \mathbb N \ ,  -\ell \le m
\le \ell \Big\}
 \notag
\eeqa
has Lie brackets:
\begin{eqnarray}
  \label{eq:KMV}
  \big[\mathcal{T}_{a \ell m}, \mathcal{T}_{a^{\prime } \ell' m^{\prime }}\big]&=& if_{a a^{\prime }}{}^{a^{\prime \prime }} C^{\ell^{\prime
\prime }}_{\ell \ell^{\prime }mm^{\prime }} \mathcal{T}_{a^{\prime
\prime }\ell'' m+m^{\prime }}  + k_0 m'k_{ab} (-1)^{m} \delta_{\ell \ell^{\prime }}
  \delta_{m,-m^{\prime }} \notag \\
   \big[L_{\ell_1 m_1},L_{\ell_2 m_2}\big]&=&(m-n) C_{\ell_1 \ell_2 m_1 m_2}^{\ell_3}  L_{\ell_3 m_1+m_2} +(-1)^{m_1} \frac c{12}m_1(m_1^2-1)
  \delta_{m_1+m_2}\delta_{\ell_1\ell_2} \nn\\
  \big[L_{\ell m}, {\cal T}_{a_1 \ell_1 m_1}\big]&=& -m_1 C_{\ell \ell_1 m m_1}^{\ell'} {\cal T}_{a_1 \ell' m+m_1} \ , 
\end{eqnarray}
where now $L_0=-L_{00}$.

It remains to be shown that the algebra above is the {\it only} centrally extended Lie algebra obtained from
$\g(\mathbb S^2)$ and $\mathfrak{X}(\mathbb S^2)$  that one can
consistently define.
Recall that $\g(\mathbb S^2) \rtimes \mathfrak{X}(\mathbb S^2)
=\big\{T_{a \ell m}, a =1,\cdots,\dim \g, \ell \in N, -\ell \le m \le \ell\big\}\rtimes\big\{
X^\theta_{\ell m}= Y_{\ell m}(\theta,\varphi) \partial_\theta, X^\varphi _{\ell m}= Y_{\ell m}(\theta,\varphi) \partial_\varphi,
\ell \in \mathbb N,  -\ell \le m \le \ell \big\}$, which contains as a subalgebra
$\g(\mathbb S^2) \rtimes \{L_0, L_+, L_-\}$ (where  $\{L_0, L_+, L_-\}$ are the algebra
associated with the isometry of $\mathbb S^2$), has Lie brackets:
\beqa
\big[T_{a_1\ell_1 m_1}, T_{a_2 \ell_2 m_2}\big] &=& i f_{a_1 a_2}{}^{a_3} C_{\ell_1 \ell_2 m_1 m_2}^{\ell_3}\;
T_{a_3 \ell_3 m_1 +m_2} \nn\\
\big[L_0,T_{a_1\ell_1 m_1}]&=& m_1 T_{a_1\ell_1 m_1}\nn\\
\big[L_\pm,T_{a_1\ell_1 m_1}]&=& \sqrt{(\ell_1\mp m_1)(\ell_1\pm m_1+1)} T_{a_1\ell_1 m_1 +\pm 1}\nn
\eeqa
The algebra $\g(\mathbb S^2)$ admits a infinite number of central extensions
\cite{ps}. Explicit expressions can be found in  \cite{Frappat:1989gn,dow,bb}.
Now the compatibility of the algebra which centrally extends $\g(\mathbb S^2)$ and $\mathfrak{X}(\mathbb S^2)$ is two-fold:
on the one hand only one cocycle can be considered, say $\omega_h$. On the other hand, only the vector fields
\eqref{eq:diff-cos} are compatible with $\omega_h$. 
If we now centrally extend $\g(\mathbb S^2)$ introducing the two-cocyle $\omega_0$, the only generators of the first algebra
which are compatible with $\omega_0$ are the operators $\ell_{\ell m}$ (see \eqref{eq:llm}) and the only operator
of the second algebra which is compatible with $\omega_0$ is $L_0$ \footnote{Strictly speaking the operator $R_0$ is also compatible with $\omega_0$.}:
\beqa
&&\big\{ {\cal T}_{a\ell m},   a =1,\cdots,\dim \g, \ell \in N, -\ell \le m, k_0 \le \ell\big\}\rtimes \{L_0\}\nn\\
&\subset&\big\{ {\cal T}_{a\ell m},   a =1,\cdots,\dim \g, \ell \in N, -\ell \le m, k_0 \le \ell\big\}\rtimes\big\{\ell_{\ell m},  \ell \in N, -\ell \le m, k_0 \le \ell\big\} \nn\ .
\eeqa
Finally if we centrally extend the de Witt algebra $\big\{\ell_{\ell m},  \ell \in N, -\ell \le m \le \ell\big\}$,
we obtain the Virasoro algebra of the two-sphere.
Thus, in conclusion, the algebra $\widetilde{\g}'(\mathbb S^2) \rtimes \text{Vir}(\mathbb S^2)$
with brackets given in \eqref{eq:KMV} is the only centrally
extended algebra that one can define along these lines. \\

The analysis for the three-sphere is similar. Starting from
\beqa
 L(\theta,\varphi,\psi)&=& \sum \limits _{\epsilon=0,1/2}\;
 \sum \limits_{\ell \in \mathbb N + \epsilon}\;
 \sum \limits_{m,n =-\ell}^\ell\;
 L_{n\ell m} \; \psi_{n\ell m} (\theta,\varphi,\psi)\nn\\
 &=& \sum \limits _{\epsilon=0,1/2}\; \sum \limits_{n \in \mathbb Z + \epsilon}
 \Bigg(
 \sum \limits_{\ell\ge |n|}\;
 \sum \limits_{m =-\ell}^\ell\;
 L_{n\ell m} \; F_{\ell m} (\theta,\varphi) \Bigg) e^{in\psi}\nn\\
 &\equiv&  \sum \limits _{\epsilon=0,1/2}\; \sum \limits_{n \in \mathbb Z + \epsilon}
 L_n(\theta,\varphi)\; e^{in\psi}
 \nn
 \eeqa
 We have introduced the functions $F_{n\ell m}$ defined by
 $\psi_{ns m}(\theta,\varphi,\psi)=F_{n s m} (\theta,\varphi) \; e^{i n\psi}$.
 Thus, if we assume:
 \beqa
\big[L_n(\theta,\varphi),L_{n'}(\theta',\varphi')\big]&=&
\Bigg((n-n') L_{n+n'}(\theta,\varphi) + \frac c{12}(n^3-n)\; \delta_{n,-n'}\Bigg)
\frac{\delta(\theta-\theta')\;\delta(\varphi-\varphi')}
{\sin \theta'}
\nn
 \eeqa
as for the two-sphere, we obtain the algebra
\beqa
&\widetilde{g}'(\text{SU}(2)) \rtimes \text{Vir}(\text{SU}(2))=\nn\\
&\Big\{\mathcal{T}_{a nsm},
k, \ \ a=1,\cdots, d \ , s \in \frac 12 \mathbb{N}, -s\le n,m
\le s\Big\} \rtimes\Big\{L_{msn}, s \in \frac 12 \mathbb{N}, -s\le n,m
\le s \Big\} \notag
\eeqa
where
\beqa
\ell_{msn}=-\psi_{msn}(\varphi,\psi,\theta) L_0 \nn
\eeqa
and with Lie brackets:
\beqa
\label{eq:VirKM}
\big[\mathcal{T}_{a nsm}, \mathcal{T}_{a^{\prime }n^{\prime }s^{\prime
}m^{\prime }}\big]&=& if_{a a^{\prime }}{}^{a^{\prime \prime }} C^{s^{\prime
\prime }}_{ss^{\prime }nn^{\prime }mm^{\prime }} \mathcal{T}_{a^{\prime
\prime }n+n^{\prime }s^{\prime \prime }m+m^{\prime }} 
+n' k\; k_{aa'} (-1)^{m+n} \delta_{ss^{\prime }}
\delta_{n,-n^{\prime }}\delta_{m,-m^{\prime }}  \notag \\
\big[L_{nsm}, {\cal T}_{a'n's'm'}\big]&=&
-n' C^{s^{\prime
    \prime }}_{ss^{\prime }nn^{\prime }mm^{\prime }}  {\cal T}_{a n+n's''m+m'}\nn\\
\big[L_{nsm}, L_{n's'm'}\big]&=&(n-n') C^{s^{\prime
    \prime }}_{ss^{\prime }nn^{\prime }mm^{\prime }}  \L_{n+n's''m+m'}
+\frac c {12} (n^3-n)\delta_{ss'} \delta_{n,-n'}\delta_{m,-m'} \ . \nn\
\eeqa
We  now have $L_0=- L_{000}$.  This algebra has been defined in \cite{rm}.

\section{KM algebras on non-compact Lie groups}\label{sec:Gnc}

The purpose of this section is to extend the results of the previous section to the case of non-compact manifolds.

   \subsection{Some generalities}
Let $G_{nc}$ be a non-compact Lie group and let $\g_{nc}=\{J_A, A=1,\cdots, n\}$ be its corresponding
Lie algebra. In contrast to compact Lie algebras, where the Killing form is positive definite\footnote{We use the physicists' notation, in which
 the Lie brackets have  an additional  $i-$factor. Thus,  for unitary representations the generators of
    the Lie algebra $\g$ are Hermitian, and the Killing form is definite positive for a compact semisimple Lie algebra.}, for non-compact Lie algebras $\g_{nc}$, the Killing form has signature $(n_+,n_-)$ with $n=n_++n_-$.
In spite of this
difference, the metric $g_{MN}$ on $G_{nc}$  is obtained in a similar way than the metric
on $G_c$ (see  Eqs.[\ref{eq:paramGc}-\ref{eq:gG}]). Since in this case the manifold $G_{nc}$ is
non-compact,  its volume is infinite (the volume is finite for a compact Lie group $G_c$).
Thus we define the scalar product
for $f,g\in L^2(G_{nc})$ by:
\beqa
(f,g) =\int \limits_{G_{nc}} \sqrt{g}\;\text{d} m^n  \;\overline{f(m)}
\;g(m)
\eeqa
for the variables $m$, with the same notations as in Sect. \ref{sec:Gc}.
As it holds for compact manifolds, when solving the Killing equations, the
generators of the left (resp. right) action and $\g_\xi = (\g_{nc})_L \oplus
(g_{nc})_R$ corresponding of the left/right action of $G_{nc}$ automatically appear as solutions. For the corresponding generators, we adopt here the same notation used for $G_c$ (see \eqref{eq:LR}).\\

The next step in the construction of a KM algebra associated to $G_{nc}$ is to decompose square
integrable functions on $L^2(G_{nc})$. However, the situation for non-compact Lie groups is very different
than the situation for compact Lie groups. The first difference resides in the unitary representations
of $G_{nc}$. Due to non-compactness, unitary representations are infinite
dimensional (recall that for a compact Lie group, all unitary representations are finite dimensional).
Next, there exist two types of unitary representations: the discrete series and the continuous series \cite{Kli}.
A non-compact Lie group admits always continuous series as unitary representation, but the former
exist {\it iff} the rank of the non-compact group is equal to the rank of its maximal compact subgroup.
The discrete series are characterized by discrete eigenvalues of the Casimir operators, whereas the
continuous series have continuous eigenvalues of the Casimir operators. This in particular means that
if we realize the Lie algebra on the manifold $G_{nc}$, the continuous series is non normalizable,
whilst the discrete series is normalizable. Hence, the harmonic analysis on $L^2(G_{nc})$ is more involved
in this case, and is summarized in the Plancherel theorem \cite{sch}. This theorem basically states
that any square integrable functions on $G_{nc}$ decomposes as a sum over the matrix
elements of the discrete series and an integral over the matrix elements of the continuous series.
However, since we are considering
Hermitian operators with continuous spectra
 their  eigenfunctions are not normalizable. This means that
some care must be taken. We thus consider
 a Gel'fand triple
of a Hilbert space ${\cal H}$ (see for instance \cite{Ba-Ra}). Let   ${\cal S}$ be
a dense subspace  of smooth functions of  ${\cal H}$ and its dual  ${\cal S}'$
\beqa
\label{eq:triple}
{\cal S} \stackrel{{\tiny\mbox{J}}}{\subset} {\cal H} \stackrel{{\tiny\mbox{K}}}{\subset} {\cal S}' \ ,
\eeqa
where $J:{\cal S} \rightarrow {\cal H}$ is an injective bounded operator with dense image and $K$ is the composition of the canonical isomorphism
${\cal H}\simeq {\cal H}^{\prime}$ determined by the inner product (given by the Riesz theorem) and the dual $J^{\prime}:{\cal H}^{\prime}\rightarrow {\cal S}' $ of $J$.
The set ${\cal S}$ we are considering here is the space of rapidly decreasing functions
in any non-compact directions of $G_{nc}$ or, for short, Schwartz functions.
Schwartz functions (which generalize the well-known set of Schwartz functions for $\mathbb R$)
were defined for any semisimple Lie group by Harish-Chandra (see also \cite{Va} and references therein).
With these definitions, it turns out that the matrix elements of the continuous  series
belong to ${\cal S}'$,
and a closed expression of the Plancherel Theorem can be given
(see for instance \cite{Ba-Ra}, Theorem 1,  p. 426, and Eqs.[34-37] p. 429,
or \cite{Va} Chap. 8).

We are considering also a second way to expand any square integrable functions on $G_{nc}$, by identifying
a Hilbert basis of $L^2(G_{nc})$. This is always possible, since any Hilbert space admits
a Hilbert basis ({\it i.e.} a countable, complete set of orthornoral vectors) \cite{RS}.

Then, following the steps of section \ref{sec:KMGc}, one can naturally associate to $G_{nc}$ a corresponding
KM algebra. In this review we don't consider a generic non-compact manifold $G_{nc}$, but only focus on specific $G_{nc}$, namely $\SL$. There are two isomorphic presentations of the KM algebra associated to $\SL$. The first one
is based on the Plancherel theorem and involves integrals and Dirac $\delta-$distributions in its Lie brackets.
The second is based on the identification of a Hilbert basis of $L^2(G_{nc})$, and involves only sums
and Kronecker symbols in its Lie brackets. Since we are only considering Schwartz functions throughout, both
presentations are isomorphic.

\subsection{A KM algebra associated to SL$(2,\mathbb R)$}
In this section we give with some details about the construction of the KM algebra
associated to $\SL$. At the end of the section we briefly comment the construction
associated to $\SL/\text{U}(1)$.

\subsubsection{The group SL$(2,\mathbb R)$}
The group  $\SL \cong \text{SU}(1,1)$ is defined by the set of $2\times 2$ complex matrices
\beqa
\text{SL}(2,\mathbb R) &=& \Bigg\{U = \begin{pmatrix} z_1 & \bar z_2 \\
   z_2   & \bar z_1 \end{pmatrix}  \ , \ \ z_1, z_2 \in \mathbb C:  \ \ |z_1|^2-|z_2|^2 =1 \Bigg\}  \nn\\
&\cong& \Big\{ z_1, z_2 \in \mathbb C^2: |z_1|^2-|z_2|^2=1\Big\}\equiv \mathbb H_{2,2}\ ,
\eeqa
where $\mathbb H_{2,2}$ is the hyperboloid which
can be parameterized as follows:
\begin{equation}
z_{1}=\cosh \rho e^{i\varphi _{1}}\ ,\ \ z_{2}=\sinh \rho e^{i\varphi _{2}}\
,\ \ \rho \geq 0\ ,0\leq \varphi _{1},\varphi _{2}<2\pi \ . \nn
\end{equation}
From this parameterization
we obtain the  left-invariant one-forms (see \eqref{eq:e})
\begin{eqnarray}
\lambda_0&=& \cosh^2 \rho \;\text{d} \varphi_1 -\sinh^2 \rho\; \text{d} \varphi_2\nn\\
\lambda_1&=& \sin (\varphi_1+\varphi_2) \; \text{d} \rho -\sinh \rho \;\cosh \rho\; \cos (\varphi_1+\varphi_2)
\;(\text{d} \varphi_1 -\text{d} \varphi_2)\nn\\
\lambda_2&=& -\cos (\varphi_1+\varphi_2) \; \text{d} \rho -\sinh \rho \;\cosh \rho\; \sin (\varphi_1+\varphi_2)
\;(\text{d} \varphi_1 -\text{d} \varphi_2)\ . \nn
\end{eqnarray}
Thus the metric tensor is:
\beqa
\text{d} s^2 = \lambda_0^2 - \lambda_2^2 - \lambda_3^2 =
-\text{d} \rho^2   + \cosh^2 \rho \; \text{d} \varphi_1^2 -  \sinh^2 \rho  \;\text{d} \varphi^2 \ . \nn
\eeqa
The generators of left/right action of $\mathfrak{sl}(2,\mathbb R)$,  obtained by solving the Killing equation \eqref{eq:killing}, are
\begin{eqnarray}
  \begin{array}{llcclll}
L_{\pm} &=&\frac{1}{2}e^{i(\varphi _{1}\mp\varphi _{2})}&\Big[i\tanh \rho
\;\partial _{1}\pm\partial _{\rho }-i\coth \rho \;\partial _{2}\Big]\ ,&
L_{0} &=&\frac{i}{2}\big(\partial _{2}-\partial _{1})\ ,
\nonumber \\
R_{\pm} &=&\frac{1}{2}e^{\pm i(\varphi _{1}+\varphi _{2})}&\Big[-i\tanh
\rho \;\partial _{1}\mp\partial _{\rho }-i\coth \rho \;\partial _{2}\Big]\ , &
R_{0} &=&-\frac{i}{2}\big(\partial _{2}+\partial _{1})\ ,
\end{array}
\notag
\end{eqnarray}
and satisfy the commutation relations
\begin{eqnarray}
\begin{array}{llllll}
\big[L_0, L_\pm\big] & = & \pm L_\pm \ , & \big[L_+, L_-\big] & = & -2 L_0 \ , \\%
[2pt]
\big[R_0, R_\pm\big] & = & \pm R_\pm \ , & \big[R_+, R_-\big] & = & -2 R_0 \ ,
\end{array} \ \ \big[L_a,R_b\big]=0 \ .
\notag
\end{eqnarray}
The Casimir operator is given by
\beqa
C_2
  &=& \frac{1}{2}\coth(2 \rho)\partial_{\rho }+\frac{1}{4}\partial _{\rho }^{2}-\frac{1-\tanh^{2}\rho}{4}\partial _{1}^{2}+\frac{\coth^2\rho-1}{4}\partial _{2}^{2} \ .
\notag
\eeqa
and the scalar product on $\mathbb H_{2,2}$ reduces to:
\begin{equation}
\label{eq:spSL}
(f,g)=\frac{1}{ 4\pi^{2}}\int\limits_{\mathbb H_{2,2}}\cosh \rho \;\sinh
\rho \;\text{d}\rho \; \text{d}\varphi_{1}\; \text{d}\varphi _{2}\;\overline{{f}(\rho ,\varphi
_{1},\varphi _{2})}\; g(\rho ,\varphi _{1},\varphi _{2})\ .
\end{equation}%

We recall that unitary representations ${\cal D}_\Lambda = \big\{\big|\Lambda,n\big>, \ n \in I_\Lambda\big\}$ of $\mathfrak{sl}(2,\mathbb R)$ were classified by Bargmann
\cite{bar} (see also \cite{ram}):
\beqa
\label{eq:actsl2}
L_0 \big|\Lambda,n\big>&=&n  \big|\Lambda,n\big> \ , \nn\\
L_+ \big|\Lambda,n\big>&=& \vartheta \; \sqrt{(n+\Lambda)(n-\Lambda+1)}\big|\Lambda,n+1\big>\ , \\
L_- \big|\Lambda,n\big>&=& \vartheta  \;\sqrt{(n-\Lambda)(n+\Lambda-1)}\big|\Lambda,n-1\big>\ , \nn\\
C_2 \big|\Lambda,n\big>&=& \Lambda(\Lambda+1) |L,n\big>  \ ,\nn
\eeqa
with the sign $\vartheta$ being conveniently chosen for each representation
(see Proposition \ref{prop:unit}).
Unitarity restricts $\Lambda$ and  $I_\Lambda$:
\begin{proposition}\label{prop:unit}
  Unitary representations are:
\begin{enumerate}[noitemsep]
\item ${\cal D}_\lambda^+$: discrete series bounded from below $\Lambda=\lambda, \lambda \in
  \mathbb N\setminus \{0\}$ or $\frac 12+
  \mathbb N$ and $I_\lambda=\{n\ge \lambda\}$;
\item   ${\cal D}_\lambda^-$:  discrete series bounded from above
 $\Lambda=\lambda, \lambda \in
  \mathbb N\setminus \{0\}$ or  $\frac 12+
  \mathbb N$ and
  $I_\lambda=\{n\le -\lambda\}$;
\item  ${\cal C}_{i\sigma}^\epsilon$: principal continuous series  $\Lambda=\frac 12 + \frac i2\sigma, \sigma>0$  and  $I_{i \sigma}= \mathbb Z$ or
  $\frac 12 +\mathbb Z$ ($n \in \frac 12 +\mathbb Z$ or $n \in \mathbb Z$);
\item ${\cal C}_\sigma$: supplementary continuous series  $\Lambda=\frac 12 + \frac \sigma2, 0<\sigma^2<1$ and $I_\sigma= \mathbb Z$ ($n\in \mathbb Z$).
\end{enumerate}
The sign $\vartheta$ can be taken equal to 1, however, conveniently
  we take $\vartheta =1$ for the discrete series bounded from below, as well as for the
  (principal and supplementary) continuous series, whilst $\vartheta =-1$
  for the discrete series bounded from above.
\end{proposition}
The discrete series ${\cal D}^+_\lambda$  is bounded from below whilst
the discrete series ${\cal D}^-_\lambda$ is bounded from above.
The continuous series ${\cal C}_{i\sigma}^\epsilon$ and ${\cal C}_\sigma$ are unbounded.
The eigenvalue of $C_2$ is discrete for the two-discrete series, and continuous for the two
continuous series. Note that $\mathfrak{sl}(2,\mathbb R)$ admits discrete series because $\mathfrak{u}(1)\subset
\mathfrak{sl}(2,\mathbb R)$ and rk $\mathfrak{u}(1)=$ rk $\mathfrak{sl}(2,\mathbb R)$.

\subsubsection{Matrix elements of $\mathfrak{sl}(2,\mathbb R)$ and Plancherel Theorem}
 The Plancherel Theorem only involves the two discrete series and the principal
continuous series. Since the supplementary continuous series plays no role (see \cite{sch, vil, hc}), we hereafter only consider discrete and continuous principal series. The corresponding matrix matrix elements
are denoted $\psi_{m\bar \Lambda n}$  or, more specifically:
for the discrete series $\psi_{m\lambda n}^\eta$ where  $\bar \Lambda=(\lambda,\eta)$,
$\eta=\pm, \lambda>1/2,\eta m, \eta n \ge \lambda$, for the discrete series
bounded
from below ($\eta=1$) above ($\eta=-1$)\footnote{Unitarity
of  $\mathfrak{sl}(2,\mathbb R)-$representations
implies $\lambda>0$, but in order to have normalizable matrix elements we now
have $\lambda>1/2$ \cite{bar,ram}.}, and  for the continuous principal series $\psi_{m i\sigma n}^\epsilon$
where $\bar \Lambda= (\frac12 + \frac i 2 \sigma,\epsilon)$,
$\sigma>0, \epsilon=0,1/2, n,m \in \mathbb Z + \epsilon$,
for the continuous principal bosonic (resp. fermionic) $\epsilon=0$ (resp.
$\epsilon=1/2$). These matrix elements
are obtained solving
the differential equations
\beqa
L_0 \psi_{m\Lambda n}(\rho,\varphi_1,\varphi_2)&=& m\;\psi_{m\Lambda n}(\rho,\varphi_1,\varphi_2)\nn\\
R_0 \psi_{m\Lambda n}(\rho,\varphi_1,\varphi_2)&=& n\;\psi_{m\Lambda n}(\rho,\varphi_1,\varphi_2)\nn\\
C_2\psi_{m\Lambda n}(\rho,\varphi_1,\varphi_2)&=& \Lambda(\Lambda+1)\;\psi_{m\Lambda n}(\rho,\varphi_1,\varphi_2)\ .\nn
\eeqa
{\small
We can unify all matrix elements in the following form (for $m\ge n$)
\beqa
\psi_{n \bar \Lambda m}(\rho,\varphi_1,\varphi_2)&=&\frac{{\cal N}}{(m-n)!}
  \sqrt{\frac {\Gamma(\vartheta  m+1 -\Lambda)\;\Gamma(\vartheta m+\Lambda)}
       {\Gamma(\vartheta  n+1 -\Lambda)\;\Gamma(\vartheta  n+\Lambda)}}\;
  e^{i(m+n)\varphi_1 + i (m-n)\varphi_2}
  \nn\\
&&
\hskip -2.truecm\cosh^{-\vartheta (m+n)} \rho \sinh^{\vartheta (m-n)} \rho\;{}_2 F_1(-\vartheta  n + \Lambda, -\vartheta  n -\lambda+1; 1 +\vartheta  (m-n); -\sinh^2 \rho) \nn
\eeqa
}
\hskip -.1truecm where $\bar \Lambda=(\Lambda=\lambda, +),  (\Lambda=\lambda, -)$ with $\lambda>1/2$
for the discrete series and $\bar \Lambda=(\Lambda=\frac 12 + \frac i 2 \sigma, 0),
(\Lambda=\frac 12 + \frac i 2 \sigma, \frac 12)$ with $\sigma>0$ for the continuous series.
We have defined $\vartheta =1$ for the discrete series bounded from below and the
continuous series and $\vartheta =-1$ for the discrete series bounded from above, and
${\cal N}=\sqrt{2(2\lambda -1)}$ (resp. ${\cal N}=1$) for the discrete (resp.
continuous series). The matrix elements for $n\ge m$ are obtained with the substitution $m\leftrightarrow n$ everywhere,
except in the exponential factor, which is unaffected,  and are multiplied by an overall factor $(-1)^{n-m}$.
Recall that ${}_2F_1$ is a hypergeometric function (see \cite{ram} for precise definition). The matrix elements
of discrete series are thus expressed in terms of hypergeometric polynomials, whereas the matrix elements of
continuous series are expressed in terms of hypergeometric functions \cite{ram}. They are normalized such that
\beqa
\label{eq:Psi_N}
\begin{array}{rlll}
(\psi^{\eta_1}_{m_1 \lambda_1 n_1},\psi^{\eta_2}_{m_2 \lambda_2 n_2})&=&\delta^{\eta_1\eta_2} \delta_{\lambda_1 \lambda_2}\delta_{m_1m_2} \delta_{n_1n_2}&
\text{discrete~series}\\
\psi^\epsilon_{mi\sigma n}(0)&=&\delta_{mn}& \text{principal~continuous~series}
\end{array}
\eeqa
and satisfy \eqref{eq:actsl2} for the left and right action. As stated previously, the matrix elements
of the discrete series are normalizable (see \eqref{eq:Psi_N}), but satisfy
\beqa
\psi^{\eta}_{m \lambda n}(0)=\sqrt{2(2\lambda -1)} \;\delta_{mn} \ ,
\eeqa
whilst the matrix elements of continuous principal series are not normalizable and satisfy
\beqa
\label{eq:Non-norm}
(\psi^{\epsilon }_{ni \sigma m}, \psi^{\epsilon' }_{n'i \sigma' m'})=\frac{1}{\sigma\tanh\pi(\sigma+i\epsilon)} \delta_{\epsilon \epsilon'}
\delta_{mm'} \delta_{n n'} \delta(\sigma-\sigma')\ ,
\eeqa
(see \cite{ram}).
Finally, we have
\beqa
\label{eq:Pconj}
\overline{\psi^{m\lambda n}_{\eta}}(\rho,\varphi_1,\varphi_2) &=& \psi^{-\eta}_{-n\lambda -m}(\rho,\varphi_1,\varphi_2) \ ,\nn\\
\overline{\psi^{m i\sigma n}_{\epsilon}}(\rho,\varphi_1,\varphi_2) &=&\psi_{-m i\sigma -n}^{\epsilon}(\rho,\varphi_1,\varphi_2) \ .
\eeqa

Let $\cal S$  be the set of Schwartz (or rapidly decreasing) functions  in the $\rho-$direction and let ${\cal S}'$
be its dual (see \eqref{eq:triple}).
The asymptotic (when $\rho \to + \infty$) expansion of the matrix elements of the discrete series was studied in
\cite{bar,ram},  and it turns out that the matrix elements $\psi^{\eta}_{m \lambda n}$ belong to $\cal S$. Moreover,
 the matrix elements of the continuous principal series
 $\psi^\epsilon_{m i\sigma n}$ belong to ${\cal S}'$; see for instance \cite{Ba-Ra}, Theorem 1,  p. 426, and Eqs.[34-37] p. 429, or \cite{Va} Chap. 8.
 Then, considering a function $f \in {\cal S}$, the Plancherel Theorem
 enables us to expand $f$ as follows (see \cite{hc} and \cite{vil}, pp. 336-337):
\beqa
\label{eq:Plan}
f(\rho,\varphi_1,\varphi_2) &=&\sum \limits_{\lambda>\frac12}\;
\sum \limits_{m,n\ge \lambda}\; f_{+  }^{n\lambda m}\; \psi_{n\lambda m}^{+ }(\rho,\varphi_1,\varphi_2) +
  \sum \limits_{\lambda>\frac 12}\;
 \sum \limits_{m,n\le - \lambda}\; f_{-}^{n\lambda m}\; \psi_{n\lambda m}^{- }(\rho,\varphi_1,\varphi_2)\nn\\
 && + \int \limits_0^{+\infty} \text{d} \sigma\; \sigma \tanh \pi \sigma \sum \limits_{m,n \in \mathbb Z}
 f_0^{nm}(\sigma) \psi^{0 }_{n i \sigma m}(\rho,\varphi_1,\varphi_2)\\
&&  +\int \limits_0^{+\infty} \text{d} \sigma\; \sigma \coth \pi \sigma \sum \limits_{m,n \in \mathbb Z + \frac 12}
 f_\frac 12^{nm}(\sigma) \psi^{\frac 12 }_{n i \sigma m}(\rho,\varphi_1,\varphi_2)\nn
 \eeqa
  {\it i.e.}, as a sum over the matrix elements of the discrete series and an integral over the matrix elements
     of the principal continuous series. This is the Plancherel Theorem for $\SL$.
 The components of $f$ are given by the scalar products \eqref{eq:spSL}
 \beqa
 \label{eq:PC}
 f_{\pm }^{n \lambda m}&=&(\psi^{\pm  }_{n\lambda m},f) \\
 f^{\epsilon}_{nm}(\sigma)&=& (\psi^{\epsilon }_{ni\sigma m},f) \ . \nn
 \eeqa
     Introducing  the symbol
 $\sum \hskip -.3 truecm \int{}$ to denote the summation over all discrete and continuous series, we can rewrite  \eqref{eq:Plan}  as
\beqa
\label{eq:IP}
f(\rho,\varphi_1,\varphi_2) = \sum_{\Lambda,m,n} \hskip -.6 truecm \int{} f^{n \Lambda m} \psi_{n\Lambda,m}(\rho,\varphi_1,\varphi_2)
\eeqa
with $\Lambda=(+,\lambda), (-,\lambda), (0,i\sigma), (1/2,i\sigma)$. \\

\subsubsection{Hilbert basis of $\LL$} \label{sec:HilbSL}

The Plancherel Theorem gives rise to an expansion of Schwartz functions as a sum over the matrix elements of the  discrete series and an integral
over the matrix elements of the continuous series (see \eqref{eq:Plan}).
As any Hilbert space is known to admit a Hilbert basis \cite{RS}, we would like now to identify
a Hilbert basis of $\LL$. Let $\LL= \LL^{\text{d}} \oplus \LL^{\text{d}^\perp}$, where
$\LL^{\text{d}}$ constitutes  the set of square-integrable functions expanded within the discrete series (first line
of \eqref{eq:Plan}). By definition, the set of matrix elements of discrete series is  a Hilbert
basis of $\LL^{\text{d}}$. Differently, since the matrix elements of the continuous series are not
normalizable, they don't constitute
a Hilbert basis of $\LL^{\text{d}^\perp}$.  V. Losert identified for us a Hilbert basis for
$\LL$ (see \cite{ram}).
Let $W_{mn}$ be the eigenspaces of the operators $L_0$ and $R_0$
\beqa
W_{nm} = \Bigg\{ F \in \LL\ , \ \ F(\rho,\varphi_1,\varphi_2)=e^{i(m+n)\varphi_1 + i(m-n)\varphi_2} f(\rho) \Bigg\} \ .\nn
\eeqa
For   $F\in W_{nm}$ we get
\beqa
\label{eq:Los-com}
L_0 F(\rho,\varphi_1,\varphi_2)&=&n F(\rho,\varphi_1,\varphi_2) \ , \nn\\
R_0 F(\rho,\varphi_1,\varphi_2)&=&m F(\rho,\varphi_1,\varphi_2) \ .
\eeqa
The main idea of V. Losert is to identify a Hilbert basis of $W_{nm}$ for each $n$ and $m$.
Let
\beqa
    {\cal B}_{nm} = \Bigg\{\Phi_{nnk}(\rho,\varphi_1,\varphi_2) = e^{i(m+n)\varphi_1 +i(m-n)\varphi_2} e_{nmk}(\cosh 2 \rho)\ ,\ \  k \in \mathbb N \Bigg\} \nn
    \eeqa
    be a Hilbert basis of  $W_{nm}$.
    Observing that for the discrete series  bounded from below (resp. above) we have $nm >0, n,m\ge \lambda>1/2$
    (resp.   $mn>0,n,m\le -\lambda<-1/2$) (see Proposition \ref{prop:unit}), three cases have to be considered  for $W_{nm}$ \cite{ram}:
    \begin{enumerate}[noitemsep]
    \item $mn >0, m,n>1/2$:  ${\cal B}_{nm}$ contains matrix elements of the discrete series bounded from below, together with elements
      of $\LL^{\text{d}^\perp}$.
   \item $mn >0, m,n<-1/2$: ${\cal B}_{nm}$ contains matrix elements of the discrete series bounded from above, together with elements
     of $\LL^{\text{d}^\perp}$.
     \item $mn<0$ or $m=0,1/2$ or $n=0,1/2$: ${\cal B}_{nm}$ contains only elements
     of $\LL^{\text{d}^\perp}$.
    \end{enumerate}
    Let $\epsilon=0$ if $m,n \in \mathbb Z$ and
    $\epsilon=1/2$ if $m,n \in \mathbb Z +\frac 12$,
    and define the condition
    \beqa
    \label{eq:cond}
{\cal C}: \ \ mn >0\ , |m|, |n|>1/2
\eeqa
which ensures that $W_{mn}\cap \LL^{\text{d}} \ne \emptyset$. Then set
\beqa
k_{\text{min}}=\left\{
\begin{array}{clll}\text{min}(|m|,|n|) -\epsilon&\text{if}&{\cal C}&\text{is~satisfied}\\
  0&\text{if}& {\cal C}&\text{is~not~satisfied}
  \end{array}\right.
    \eeqa
    such that if
    \beqa
    \label{eq:kmin}
k< k_{\text{min}}& \Rightarrow&\Phi_{nmk} \in \LL^{\text{d}}\nn \\
k\ge  k_{\text{min}}& \Rightarrow&\Phi_{nmk} \in \LL^{\text{d}^\perp} \ ,
\eeqa
showing that $W_{mn}\cap \LL^{\text{d}}= \emptyset$ when the condition \eqref{eq:cond} does not hold.
For $x = \cosh 2 \rho$, we have for $n\ge m$ \cite{ram}
\beqa
\label{eq:ed}
e_{nmk}(x)&=& \sqrt{2^{2m-1}\frac{(2m-2k-1)k!(m+n-k-1)!}{(2m-k-1)!(n-m+k)!} } \times\\
&&(x-1)^\frac{n-m}2 (x+1)^{-\frac{m+n}2} P_k^{(n-m,-n-m)}(x) \in \LL^{\text{d}}
\ ,\nn\\&& \hskip 3.5truecm   0\le k< k_{\text{min}},\nn\\[10pt]
 e_{nmk}(x) &=&  \sqrt{2^{2k + 2 \epsilon+1}\frac{(2k +n-m + 2\epsilon+1) (k +n-m + 2\epsilon) !k! }{(k+2\epsilon)! (n-m+k)!} } \times\nn\\
  &&  (x-1)^{\frac{n-m}2} (x+1)^{\frac{m-n}2 -k-\epsilon -1}
 P_k^{(n-m,m-n-2k-2\epsilon-1)}(x)\in \LL^{\text{d}^\perp} \ \ ,\nn\\
 &&  \hskip 3.5truecm  k\ge k_{\text{min}} \nn \ ,
    \eeqa
    where $P_k^{(a,b)}$ are Jacobi polynomials \footnote{
    The matrix elements of the discrete series are expressed in terms of hypergeomertric
    polynomials. However, one may show that these hypergeometric polynomials can be easily
    related to Jacobi polynomials \cite{ram}.}. Recall that $\deg P_k^{(a,b)}=k$. Thus, $\LL^{\text{d}}$ involves Jacobi polynomials of
    degree $<k_{\text{min}}$ and  $\LL^{\text{d}^\perp}$ involves Jacobi polynomials of
    degree $\ge k_{\text{min}}$ \cite{ram}.
    We have similar expressions for $n\ge m$ and the conjugation relations:
    \beqa
    \label{eq:Lconj}
    \overline{\Phi^{nmk}}(\rho,\varphi_1,\varphi_2)=
    \Phi_{-n-mk}(\rho,\varphi_1,\varphi_2) \ .
    \eeqa

    The set $\cup_{n,m \in \mathbb Z} {\cal B}_{nm} \cup_{n,m \in \mathbb Z+1/2} {\cal B}_{nm}$ is a complete orthonormal Hilbert basis of
    $\LL$ called the Losert basis:
    \beqa
    \label{eq:orthL}
(\Phi_{nmk},\Phi_{n'm'k'})= \delta_{nn'} \delta_{mm'} \delta_{kk'} \ .
    \eeqa
   Therefore, if
    $f \in \LL$, we have
    \beqa
    \label{eq:los}
    f(\rho,\varphi_1,\varphi_2) &=& \sum \limits_{\epsilon=0,1/2} \sum \limits_{n,m \in \mathbb Z +\epsilon}
    \sum \limits_{k=0}^{+\infty} f^{nmk} \Phi_{nmk}(\rho,\varphi_1,\varphi_2)\ , \\
    f_{nmk}&=&(\Phi_{mnk},f) \ .\nn
    \eeqa

    Since the   $\mathfrak{sl}(2,\mathbb R)$ generators act on $\LL$, and since $\LL^{\text{d}} \subset \LL$  is a sub-representation, {\it i.e.}, an invariant subspace of $\LL$, it follows that $\LL^{\text{d}^\perp}$ is also a representation of $\SL$. The action of $L_\pm, R_\pm$ on $\Phi_{nmk} \in \LL^{\text{d}^\perp}$ is given by
    {\small
    \beqa
 \label{eq:Los-herm}
    \begin{array}{llll}
L_+(\Phi_{nmk})&=&
\alpha^{+L}_{nmk} \Phi_{n+1mk+1} + \beta^{+L}_{nmk}  \Phi_{n+1mk} + \gamma^{+L}_{nmk}
\Phi_{n+1nk-1}
& \left\{\begin{array}{ll}
\gamma^{+L}_{nmk}=0&m>n\\
\alpha^{+L}_{nmk}=0&n\ge m
\end{array}\right.
\\
L_-(\Phi_{nmk})&=&\alpha^{-L}_{nmk} \Phi_{n-1mk+1} + \beta^{-L}_{nmk}
\Phi_{n-1mk}+\gamma^{-L}_{nmk}\Phi_{n-1mk-1}& \left\{\begin{array}{ll}
\alpha^{-L}_{nmk}=0&m\ge n\\
\gamma^{+L}_{nmk}=0&n> m
\end{array}\right.\\
R_+(\Phi_{nmk})&=&
\alpha^{+R}_{nmk} \Phi_{nm+1k+1} + \beta^{+R}_{nmk}  \Phi_{nm+1k} + \gamma^{+R}_{nmk}
\Phi_{nm+1k-1}
& \left\{\begin{array}{ll}
\alpha^{+L}_{nmk}=0&m\ge n\\
\gamma^{+L}_{nmk}=0&n>m
\end{array}\right.
\\
R_-(\Phi_{nmk})&=&\alpha^{-R}_{nmk} \Phi_{nm-1k+1} + \beta^{-R}_{nmk}
\Phi_{nm-1k}+\gamma^{-R}_{nmk}\Phi_{nm-1k-1}& \left\{\begin{array}{ll}
\gamma^{-L}_{nmk}=0&m> n\\
\alpha^{+L}_{nmk}=0&n \ge m
\end{array}\right.
\end{array}
    \eeqa
    }
  The action of the Casimir operator reduces to
  {\small
    \beqa
  C_2\Phi_{nmk}(\rho,\varphi_1,\varphi_2) = a_{nmk} \Phi_{nmk-1}(\rho,\varphi_1,\varphi_2)
  + b_{nmk} \Phi_{nmk}(\rho,\varphi_1,\varphi_2)+
  c_{nmk} \Phi_{nmk+1}(\rho,\varphi_1,\varphi_2)\ . \nn
  \eeqa
  }
   \hskip -.2truecm  This action clearly shows that this representation is unitary but
   {\it is not irreducible.} Explicit expressions of the coefficients $\alpha, \beta$ and $a,b,c$ are given in \cite{ram}.

   The asymptotic   behavior of the functions $e_{nmk} \in W_{nm} $   were studied in \cite{ram}, and it
   turns out that the functions $\Phi_{nmk}$ are actually Schwartz functions.
   Thus,  for $\Phi_{nmk} \in W_{nm} \cap \LL^{\text{d}^\perp}$,  one can use the Plancherel Theorem and write \eqref{eq:Plan}
\beqa
\label{eq:PL}
\Phi_{nmk}(\rho,\varphi_1,\varphi_2) = \int \limits_0^{+\infty} \text{d} \sigma \; \sigma\tanh \pi(\sigma+i\epsilon) f_{nmk}(\sigma) \psi_{n i\sigma \epsilon m}(\rho)\ ,
\eeqa
where $\epsilon=0$ if $n,m$ are integers and $\epsilon=1/2$ if $n,m$ are half-integers. It should be observed
that there is no summation on $n$ and $m$.
Furthermore,  using \eqref{eq:PC}, we have (with the scalar product \eqref{eq:spSL})
\beqa
f^{nmk}(\sigma)= (\psi_{n i\sigma \epsilon m},e_{nmk}) \ .\nn
\eeqa
This implies that the relation \eqref{eq:PL} can be inverted
\beqa
\label{eq:LP}
\psi_{n i\sigma \epsilon m}(\rho,\varphi_1,\varphi_2) = \sum \limits_{k\ge k_{\text{min}}} \overline{ f^{mnk}}(\sigma)\Phi_{nmk}(\rho,\varphi_1,\varphi_2) \ ,
\eeqa
with $k_{\text{min}}$ defined  in  \eqref{eq:kmin}.
 We are thus able
 to express the matrix elements of the  (principal) continuous series in terms of the Losert basis, and conversely. This observation is important for the
 construction of the algebra below.\\

 \subsubsection{Clebsch-Gordan coefficients}

 The next step in the construction of a KM algebra associated to SL$(2,\mathbb R)$ is the computation of
the Clebsch-Gordan coefficients corresponding to the decomposition ${\cal D}_\Lambda\otimes {\cal D}_{\Lambda'}$. This has been studied in
\cite{hb1, hb2}.
 The coupling of two discrete series was studied by means of a bosonic realization of the Lie algebra
$\mathfrak{sl}(2,\mathbb R)$; for the case in which at least one
continuous series is involved, the result was obtained by a cumbersome analytic continuation.
An heuristic decomposition can be deduced from a back and forth.
More precisely let ${\cal D}_\Lambda, {\cal D}_{\Lambda'}$ be two representations of $\SL$ in the Plancherel basis, expend the matrix elements
of ${\cal D}_\Lambda$ and  ${\cal D}_{\Lambda'}$
in the Losert basis
using \eqref{eq:LP}, subsequently
decompose the product $\Psi_{m\Lambda n} \Psi_{m'\Lambda'n'}$ in the Losert basis and then express the results back in the Plancherel basis using
\eqref{eq:PL}.
For instance, consider two discrete series bounded from above: ${\cal D}_\lambda^+\otimes {\cal D}_{\lambda'}^+$. The product of two matrix
elements $\psi_{n\lambda m}^+ \psi^+_{n'\lambda' m'}$ is such that $nm>0, n'm'>0, n,m,n',m'>1/2$ (see Condition \ref{eq:cond}).
So we have
$(n+n')(m+m')>0, (n+n'), (m+m')>1/2$. Thus the product ${\cal D}^+_\lambda \otimes {\cal D}_{\lambda'}^+$ decomposes
only in the discrete series bounded from above. The situation is very different for
the product ${\cal D}_{\lambda_+}^+ \otimes {\cal D}_{\lambda_-}^-$ which involves both matrix elements of discrete series
and continuous series. Indeed, the product of two matrix elements $\psi_{m_+\lambda_+ n_+}^+ \psi_{m_-\lambda_- n_-}^-$
 is such
 that $m_\pm n_\pm >0$ and $\pm m_\pm, \pm n_\pm \ge \lambda_\pm$.
 So it may happen that  $(m_++m_-) (n_++n_-)<0$ (depending of
 the respective value of $m_\pm$ and $n_\pm$), and thus
$\psi_{m_+\lambda_+ n_+}^+ \psi_{m_-\lambda_- n_-}^-$
 belongs to $\LL^{\text{d}^\perp}$. So, in this case, by \eqref{eq:PL}
$\psi_{m_+\lambda_+ n_+}^+ \psi_{m_-\lambda_- n_-}^-$
decomposes as an integral over
matrix elements of continuous representations.

We thus generically write:
\beqa
\label{eq:matmat}
\psi_{m_1\bar \Lambda_1  m'_1}(\rho,\varphi_1,\varphi_2) \psi_{m_2\bar \Lambda_ 2m'_2}(\rho,\varphi_1,\varphi_2) &=&   \sum_{\bar \Lambda} \hskip -.5 truecm \int{}\;
{ C}_{\bar \Lambda_1, \bar \Lambda_2}^{\bar \Lambda}{}_{m_1,m_2, m'_1,m'_2}
\psi_{m_1+m_2 \bar\Lambda m'_1+m'_2}(\rho,\varphi_1,\varphi_2)\nn\\
\eeqa
with the notations \eqref{eq:IP}, and
where $\bar \Lambda_1, \bar \Lambda_2= (\lambda,+), (\lambda,-), (i\sigma,0)$ or $(i\sigma,1/2)$ and  $\bar \Lambda$ takes one of the allowed  values
occurring in the tensor product decomposition (see \cite{hb1, hb2} and \cite{ram} for a case by case treatment with the
same notations). The coefficients are given by the corresponding product of Clebsch-Gordan coefficients
(see \cite{ram, Campoamor-Strusberg:2024kpl}).

Similarly, if we proceed with the Losert basis, as the product $e_{nmk}(x) e_{n'm'k'}(x)$ is square integrable (even better, it is a Schwartz function) \cite{ram}
we have
\beqa
\label{eq:ee}
\Phi_{nmk}(\rho,\varphi_1,\varphi_2) \Phi_{n'm'k'}(\rho,\varphi_1,\varphi_2) = C^{k''}_{kk'nn'mm'} \Phi_{n+n'm+m'k''}(\rho,\varphi_1,\varphi_2) \ .
\eeqa

From now on, we call the Plancherel basis the set of  matrix elements of the discrete and principal continuous series, {\it i.e.},
the $\psi$s
and the Losert basis the $\Phi$s.

\subsubsection{A KM algebra associated to SL$(2,\mathbb R)$}
The KM algebra associated to SL$(2,\mathbb R)$ is obtained in an analogous way to the construction of
the KM algebra associated to a compact Lie group $G_c$. It is a  central extension of the algebra
$\g(\SL)$. In the
Plancherel basis (PB), we have
\beqa
{\g}(\SL)&=& \Bigg\{T^a \psi^{+}_{n\lambda m}(\rho,\varphi_1,\varphi_2)\ ,\ T^a \psi^{-}_{n\lambda m}(\rho,\varphi_1,\varphi_2), \ \ \lambda>\frac12, \ \  mn>0,\ \  |m|,|n|>\lambda , \nn\\
&&\hskip .35truecm T^a \psi^{\epsilon}_{ni\sigma m}(\rho,\varphi_1,\varphi_2), \epsilon=0,\frac12, \sigma>0, m,n\in \mathbb Z + \epsilon, \ \
a=1,\cdots,\dim \g\Bigg\}\nn\\
&=&\Bigg\{T^{a}_{m\bar \Lambda n}=T^a \psi_{m\bar\Lambda n}, \ \ \bar\Lambda=(\lambda, +), (\lambda,-), (i\sigma,0), (i\sigma,\frac12)\ , a=1,\cdots,\dim \g\ \
\Bigg\}\nn
\eeqa
and in the Losert basis (LB) we get
\beqa
{\g}(\SL)&=& \Bigg\{T^a_{m n k}=T^a \phi_{m,n,k}(\rho,\varphi_1,\varphi_2), \ \ m,n \in \mathbb Z  + \epsilon, k \in \mathbb N\ , \epsilon=0,\frac12 \Bigg\}\ . \nn
\eeqa
  The Lie brackets take the form
  \beqa
  \label{eq:loopSL}
\begin{array}{llll}
  \big[T^{a}_{m\bar \Lambda n}, T^{a'}_{m'\bar \Lambda'n'}\big] &=&i f^{a a'}{}_{a''}
\sum\limits_{\bar\Lambda''} \hskip -.45 truecm \bigintsss{}\;
   C_{\bar \Lambda, \bar\Lambda'}^{\bar \Lambda''}{}_{m,m',n, n'} T^{a''}_{m+m'\bar \Lambda''n+n'}& (\text{PB})\\[10pt]
\big[T^a_{m n k}, T^{a'}_{m'n'k'}\big]&=&if^{aa'}{}_{a''} \sum \limits_{k''} C_{kk'}^{k''}{}_{m,m',n, n'} T^{a''}_{m+m'n+n'k''}& (\text{LB})
\end{array}
\eeqa
by \eqref{eq:matmat} and \eqref{eq:ee}.
Of course, since we can express the $T^a_{m n k}$ in terms of the $T^{a}_{m\bar\Lambda  n}$ and {\it vice versa}, because of
\eqref{eq:LP} and \eqref{eq:PL}, the two presentations of the algebra are isomorphic.
In \eqref{eq:loopSL} we explicitly write the symbol $\sum$ in the LB ({\it i.e.}, we do not use the Einstein summation convention) in order to emphasize the
different presentation of the algebra in the PB and in the LB bases. Indeed, in  the former basis the algebra involves an integral, while in the latter, it involves sum.  \\

The next step in the construction is to introduce Hermitian operators and
compatible two-cocycles. The construction is similar to the
SU$(2)$ case. We first introduce the two commuting operators $L_0,R_0$.
In the Losert basis. The action of the commuting Hermitian operators is given in \eqref{eq:Los-com},  and in the Plancherel basis the
action of Hermitian operators is given in \eqref{eq:actsl2} with the corresponding value
of $\Lambda$.
We then associate to $L_0,R_0$ compatible two-cocycles $\omega_L,\omega_R$:
\beqa
\omega_{L}(X,Y)&=&-\frac {k_L} {4\pi^2}  \int\limits _{{\mathbb H}_{2,2}} \text{d}\rho \sinh \rho \cosh \rho \;\text{d} \varphi_1  \;\text{d} \varphi_2
\;\Big<X,L_0 Y\Big>_0
\nn\\
\omega_{R}(X,Y)&=& -\frac {k_R} {4\pi^2}\int\limits _{\mathbb H_{2,2}} \text{d}\rho \sinh \rho \cosh \rho \; \text{d} \varphi_1 \; \text{d} \varphi_2
\;\Big<X,R_0 Y\Big>_0\ . \nn
\eeqa
Thus, using the orthogonality conditions \eqref{eq:Psi_N} and \eqref{eq:Non-norm}
(resp. \eqref{eq:orthL}) and
the conjugation property \eqref{eq:Pconj}  (resp. \eqref{eq:Lconj}) in the Planchel (resp. the Losert) basis,
we obtain
\beqa
\label{eq:cosSL}
\omega_L(T^{a}_{n\lambda\eta m}, T^{a' }_{n'\lambda '\eta'm'})&=&  n \;k_L\; k^{aa'}\; \delta_{\lambda,\lambda'}\; \delta_{\eta,-\eta'}\; \delta_{m,-m'}\; \delta_{n,-n'} \\
\omega_L(T^{a}_{n i\sigma\epsilon m}, T^{a'}_{n' i\sigma'\epsilon'm'})&=&n\;k_L\;k^{aa'} \;\frac{\delta(\sigma-\sigma')}{\sigma\tanh \pi(\sigma+i\epsilon)}
\;\delta_{\epsilon,\epsilon'}\; \delta_{m,-m'}\; \delta_{n,-n'}\nn
\eeqa
in the PB basis, and
\beqa
\label{eq:cosSL2}
\omega_L(T^{a}_{n mk}, T^{a'}_{n'm'k'}) &=& \;n k_L\;k^{aa'} \;\delta_{kk'}\;  \delta_{m,-m'} \;\delta_{n,-n'}
\eeqa
in the LB basis.
Defining
\beqa
\delta(\bar \Lambda,\bar \Lambda') =
\left\{
\begin{array}{cc}
 \delta_{\lambda,\lambda'} \delta_{\eta,-\eta'} & \bar \Lambda=(\lambda,\eta)\ ,\ \  \bar\Lambda'=(\lambda',\eta')\\
 \frac{\delta(\sigma-\sigma')}{\sigma\tanh \pi(\sigma+i\epsilon)} \delta_{\epsilon,\epsilon'} &
 \bar \Lambda=(i\sigma,\epsilon)\ , \ \
 \bar \Lambda'=(i\sigma', \epsilon')\\
 0&\text{elsewhere}
\end{array}
\right.\nn
\eeqa
we obtain
\beqa
\omega_L(T^{a}_{n\bar\Lambda m}, T^{a'}_{n'\bar\Lambda'm'}) &=& \;n k_L\;k^{aa'}
\delta(\bar\Lambda,\bar\Lambda')  \delta_{m,-m'} \delta_{n,-n'}\nn
\eeqa
in the PB basis. We
have a similar expression for $\omega_R$.

With the same notations as before, we define the KM algebra associated to
SL$(2,\mathbb R)$:
\beqa
\widetilde{{\g}}(\SL) =\left\{
\begin{array}{ll}
  \Big\{{\cal T}_{m \bar \Lambda n}^a\ , L_0, R_0, k_L, k_L\Big\}&(\text{PB})\\[10pt]
  \Big\{{\cal T}_{mnk}^a\ , L_0, R_0, k_L, k_L\Big\}&(\text{LB})
\end{array}
\right.
\eeqa
From   \eqref{eq:loopSL}. \eqref{eq:cosSL} and  \eqref{eq:Psi_N} and \eqref{eq:Non-norm}, the Lie brackets are

\beqa
\label{eq:LieP}
 \big[{\cal T}_{a m \bar \Lambda n}, {\cal T}_{a'm'\bar \Lambda'n'}\big] &=&i f_{a a'}{}^{a''}
 \sum\limits_{\bar\Lambda''} \hskip -.4 truecm \bigintssss{}\;
 C_{\bar \Lambda, \bar \Lambda'}^{\bar \Lambda''}{}_{m,m',n, n'} {\cal T}_{a''m+m'\bar \Lambda''n+n'} \nn\\
 &&\hskip 1.5truecm
 + (mk_L+nk_R) h_{aa'}\delta(\bar \Lambda,\bar \Lambda') \delta_{m,-m'} \delta_{n,-n'} , \nn \\
 \big[L_0,{\cal T}_{am\bar \Lambda n}\big]&=& m {\cal T}_{am\bar \Lambda n}\ , \\
 \big[R_0,{\cal T}_{am\bar \Lambda n}\big]&=& n {\cal T}_{am\bar \Lambda n} \ ,
 \nn
 \eeqa
 in the Plancherel basis, and from  \eqref{eq:loopSL}. \eqref{eq:cosSL} and \eqref{eq:Non-norm}
 the Lie brackets are
 
 \beqa
 \label{eq:LieL}
 \big[{\cal T}_{am n k}, {\cal T}_{a'm'n' k'}\big] &=&i f_{a a'}{}^{a''}
 \sum \limits_{k''} C_{k k'}^{k''}{}_{m,m',n,n'} {\cal T}_{a''m+m'n+n' k''}\nn\\
&& + (mk_L+nk_R)h_{aa'} \delta_{kk'} \delta_{m,-m'} \delta_{n,-n'}\ , \nn \\
 \big[L_0,{\cal T}_{amn  k}\big]&=& m {\cal T}_{amn k}\ , \\
  \big[R_0,{\cal T}_{amn k}\big]&=& n {\cal T}_{amn k}\ , \nn
 \eeqa
 in the Losert basis.
This algebra was obtained in \cite{ram}.

Since we can expand  the Losert basis in the Plancherel basis \eqref{eq:PL} and the Plancherel basis in
the Losert basis \eqref{eq:LP}, the two presentations of the algebra $\widetilde{\g}(\SL)$ \eqref{eq:LieP} and \eqref{eq:LieL} are equivalent.
Each presentation has its own
advantages.  In the former case, elements of $\widetilde{\g}(\SL)$ are expressed by means of
unitary irreducible representations of $\mathfrak{sl}(2,\mathbb R)$, but the Lie brackets
involve  integrals and Dirac $\delta-$distributions, as a consequence  of the continuous basis. In the latter case, the
Lie brackets  involve only sums and Kronecker symbols, but the elements
of $\widetilde{\g}(\SL)$ are expressed by means of a reducible representation of
$\mathfrak{sl}(2,\mathbb R)$ associated to the Hilbert basis of $\LL^{\text{d}^\perp}$.

This algebra is very different to its analogue of KM algebra of compact Lie groups. Indeed, for
a compact Lie group we have $\g \subset \widetilde{\g}(G_c)$, because the trivial representation
is a unitary representation of $G_c$. However, in the non-compact case $G_{nc}$,
$\g$ {\it is not included} in    $\widetilde{\g}(G_{nc})$, because the trivial
representation is non-normalizable on the $\SL-$manifold.
\\

We conclude this section mentioning that one can associate a Virasoro algebra to
$\SL$. This construction follows the same lines as the corresponding
construction in Section \ref{sec:virS}. We introduce
\beqa
\begin{array}{llll}
\ell_{m\bar \Lambda n}&=& - \psi_{m\bar \Lambda n}(\theta,\varphi_1,\varphi_2) \; L_0 & \text{(PB)}
\\
\ell_{m  nk}&=&- \Phi_{mnk}(\theta,\varphi_1,\varphi_2) \; L_0
&\text{(LB)}
\end{array}\nn
\eeqa
which are compatible with $\omega_L$. The  commutation relations read:
\beqa
\begin{array}{llll}
\big[\ell_{m\bar \Lambda n}, \ell_{m'\bar \Lambda' n'}\big]&=&
(m-m') \sum\limits_{\bar\Lambda''} \hskip -.4 truecm \bigintssss{}\;
C_{\bar \Lambda, \bar \Lambda'}^{\bar \Lambda''}{}_{m,m',n, n'} \;\ell_{m+m'\bar \Lambda'' n+n'}
&\text{(PB)}\\[6pt]
\big[\ell_{m  nk},\ell_{m'  n'k'}\big]&=&(m-m') \sum \limits_{k''} C_{k k'}^{k''}{}_{m,m',n,n'} \;\ell_{m+m'n+n' k''}&\text{(LB)}
  \end{array}\nn
\eeqa
This algebra admits a central extension, exactly along the lines of \eqref{sec:SU2}, and with the
same notations, we introduce the generators:
\beqa
\text{Vir}(\SL)=\left\{
\begin{array}{ll}
  \big\{L_{m\bar \Lambda n}\ , \ \ c \big\}&(\text{PB})\\ \\
  \big\{L_{m nk} \ , \ \ c\big\}&(\text{LB})
  \end{array}\right.\nn
\eeqa
with Lie brackets in the PB basis:
{\small
  \beqa
  \label{eq:VP}
\big[L_{m\bar \Lambda n}, L_{m'\bar \Lambda' n'}\big]&=&
(m-m') \sum\limits_{\bar\Lambda''} \hskip -.5 truecm \bigintssss{}\;
C_{\bar \Lambda, \bar \Lambda'}^{\bar \Lambda''}{}_{m,m',n, n'} \;L_{m+m'\bar \Lambda'' n+n'}\nn\\&&
+\frac c{12}(m^3-m) \delta_{m,-m'}\delta_{n,-n'}\delta(\bar \Lambda,\bar \Lambda')\nn\\
\eeqa
and in the LB basis:
\beqa
\label{eq:VL}
\big[L_{m  nk},\ell_{m'  n'k'}\big]&=&(m-m') \sum \limits_{k''} C_{k k'}^{k''}{}_{m,m',n,n'} \;L_{m+m'n+n' k''}
+\frac c{12}(m^3-m) \delta_{m,-m'}\delta_{n,-n'}\delta_{kk'}\ .\nn\\
\eeqa
}

Thus, $\widetilde{\g}'(\SL) \rtimes \text{Vir}(\SL)$ (similarly to SU$(2)$, we only introduce one
central charge, say $k_L$) has a semidirect structure with
the action of the Virasoro algebra on the KM algebra, the remaining part being:
\beqa
\label{eq:VKM}
\begin{array}{llll}
\big[L_{m\bar \Lambda n}, {\cal T}^a_{m'\bar \Lambda' n'}\big]&=&
-m' \sum\limits_{\bar\Lambda''} \hskip -.5 truecm \bigintssss{}\;
C_{\bar \Lambda, \bar \Lambda'}^{\bar \Lambda''}{}_{m,m',n, n'} \;{\cal T}^a_{m+m'\bar \Lambda'' n+n'}
&\text{(PB)}\\[6pt]
\big[L_{m  nk},{\cal T}^a_{m'  n'k'}\big]&=&-m' \sum \limits_{k''} C_{k k'}^{k''}{}_{m,m',n,n'} \;
{\cal T}^a_{m+m'n+n' k''}&\text{(LB)}
  \end{array} \ . \nn
\eeqa
The algebra $\widetilde{\g}'(\SL) \rtimes \text{Vir}(\SL)$ is thus defined by \eqref{eq:LieP}/\eqref{eq:LieL}
(with the central charge $k_R=0$), \eqref{eq:VP}/\eqref{eq:VL} and \eqref{eq:VKM}.
Again, the Virasoro algebra of $\SL$ is very different to the Virasoro algebra of SU$(2)$.
Recall that for the Virasoro algebra of SU$(2)$ we have $L_0=-L_{000}$ (see \eqref{eq:VirKM}), but
here, as the trivial representation is not normalizable, we cannot relate $L_0$ to a
specific element of Vir$(\SL)$. However, one can extend the Virasoro algebra to Vir$'(\SL)=$Vir$(\SL)\times
\{L_0\}$, where the action of $L_0$ is given by
\beqa
\big[L_0,L_{nsm}\big]= n L_{nsm} \ . \nn
\eeqa\\

We finish this section mentioning that a KM algebra on $\SL/U(1)$ can
be easily deduced from the KM algebra on $\SL$. The only representations which appear in the harmonic analysis on $\SL/U(1)$ are the representations, which are chargeless with respect to $R_0$.
The only representations that survive this condition in the PB basis are the matrix elements of the
bosonic principal continuous series $\psi^0_{n i\sigma 0}$, and in the LB basis,
the elements of $W_{m0}:
\Phi_{m0k}$ (LB). More details  and the explicit brackets can be found in \cite{ram}.

\section{Soft manifolds}

This section is devoted to the construction of KM algebras on soft group manifolds, that is, on  group manifolds with a soft deformation.

\subsection{KM algebras on soft manifolds}
In Sections \ref{sec:GM} and \ref{sec:Gnc} we have considered KM algebras associated to (compact and non-compact) group manifolds. These manifolds have a large isometry group (associated to the left and right action of the group itself). Furthermore, for these manifolds the Vielbein satisfies the Maurer-Cartan
equation \eqref{eq:MC} or is a left invariant one-form. Moreover, for these manifolds the  metric tensor is naturally deduced from the Vielbein \eqref{eq:gG}.

Let G$_m$ be a group (compact or non-compact), we now we consider a smooth, `softening' deformation $G_m^{\mu }$
of the Lie group $G_m$, locally diffeomorphic to $G_m$ itself (see e.g.
\cite{cas}  and references therein).
We  further assume that the manifold $G_m^{\mu }$ has the same
parameterisation $m^{M}$ (see Section \ref{sec:GM} for G$_m=G_c$ a compact manifold and Section \ref{sec:Gnc} for G$_m=G_{nc}$ a non-compact manifold), the only difference
between $G_{m}$ and $G_{m}^{\mu }$ being at the level of the metric tensor.
 We thus assume that the Vielbein $\mu$ is an intrinsic one-form (valued in the Lie algebra $\mathfrak{g}_m$
of $G_m$)
\begin{equation*}
\mu ^{  A}(m)=\mu _{M}{}^{A}(m)\;\text{d}m^{M}
\end{equation*}%
\textit{i.e.}, it is not a Maurer-Cartan one-form (it does not
satisfy equation \eqref{eq:MC}):
\begin{equation}
\text{d}\mu +\mu \wedge \mu =R,  \label{this}
\end{equation}%
where $R$ is the curvature two-form of $\mu $.
In other words, $\mu $ is \textit{not} left-invariant
({\it i.e.}, it is a `soft', intrinsic one-form).
 It is in this sense that we
consider that the `soft' group manifold $G_{m}^{\mu }$ is a \textit{%
  deformation} of $G_m$. The metric tensor is now defined by
\begin{equation}
g_{MN}^{\mu }(m)=\mu _{M}{}^{A}(m)\mu _{N}{}^{B}(m)\;\eta _{AB} \ , \label{eq:gGmu}
\end{equation}%
where $\eta_{AB}$ is the Killing form  of the Lie algebra
$\g_m$ (positive definite for compact manifolds and undefinite for non-compact manifolds).
Taking the exterior derivative of both sides of Eq. (\ref{this}), one obtains
the Bianchi identity for the curvature of $\mu $,%
\begin{equation*}
dR+2R\wedge \mu =0\Leftrightarrow \nabla R=0,
\end{equation*}%
where the covariant derivative operator $\nabla $ on $G_{m}^{\mu }$ has been
introduced.
For instance, if $G_m= \text{ISO}(1,d-1)/\text{SO}(1,d-1)$, where SO$(1,d-1)$ (resp. ISO$(1,d-1)$)  corresponds to the Lorentz (resp. Poincar\'e) transformations in $D-$dimensions,  $G_m$ is the $D-$dimensional Minkowski  spacetime of Special Relativity, and its deformation $G^\mu_m$ the $D-$dimensional
Riemann spacetime  of General Relativity \cite{Castellani:1991et}.

The fact that the metric tensors differ for the manifold $G_m$ (see \eqref{eq:gG} or its analogue for a non-compact manifold)
and its deformation $G_m^\mu$  (see \eqref{eq:gGmu}) is not the only difference between $G_m$ and $G_m^\mu$.  Indeed, as seen previously, the manifold
$G_m$ has a large isometry group, namely $(G_m)_L \times (G_m)_R$, but in general, the isometry group of $G^\mu_m$ is reduced with respect to its undeformed analogue. Stated differently, solving the Killing equation \eqref{eq:killing} leads to less invariant vector fields. Consequently, if we follow the
construction of KM algebras in Section \ref{sec:M}, we will obtain a less richer structure.
For this reason,  we associate a KM algebra to $G_m$ following
a different strategy. We endow $G_m^\mu$ with  the scalar product
 $G_{m}^{\mu }$
\begin{equation}
(f,g)_{\mu }= \frac 1 C\int_{G_{m}}\sqrt{g^{\mu }}\;\text{d}^n m\;\overline{f(m )}\;g(m ), \label{eq:PSGmu}
\end{equation}%
where $g^{\mu }=|\det (g_{MN}^{\mu })|$ and $n$ is the dimension
of $G_m$. The coefficient $C$ depends of the manifold.
For a compact manifold $G_c$ we take $C=V$, the volume of $G_c$.
For a non-compact manifold, see \eqref{eq:spSL} for $G_m=\SL$.  Notice that, since the
parameterisation of $G_{m}^{\mu }$ and $G_{m}$ is the same, the limits of
integration are again $G_{m}$ in this case.
As a final hypothesis concerning the deformed manifold, we assume that $G_m^\mu$ is non-singular  and well defined at any point of $G_m^\mu$ (see \eqref{eq:gNS} below).\\

Let  $L^2(G_m)$ be the set of square integrable functions on $G_m$.  We would like to identify a Hilbert basis of $L^2(G^\mu_m)$ from
a Hilbert basis of $L^2(G_m)$\footnote{For $G_m=G_{nc}$
a non compact manifold, the Plancherel theorem involves normalizable and non-normalizable functions (see Section \ref{sec:Gnc}).  However the technique presented here extends to non-normalizable functions. For $G_m=G_{nc}$ we will consider two explicit examples, with an obvious generalization to any $G_{nc}$.}.
The results presented here are due to Mackey  \cite{Mc} (see also \cite{ram}).
Let $\cal M$ be a manifold  (not necessarily a group manifold) and let $L^2({\cal M})$ be the set of square integrable functions defined on $\cal M$.
Assume that the manifold $\cal M$ is  endowed with two different scalar products with measure $\text{d} \alpha$ and $\text{d} \beta$ respectively:
\begin{eqnarray}
\begin{array}{ll}
(\mathcal{M}, \text{d} \alpha): & (f,g)_\alpha= \int_{\mathcal{M}} \text{d}%
\alpha\; \overline{f(m)}\; g(m) \ ,\\[.2cm]
(\mathcal{M}, \text{d} \beta): & (f,g)_\beta= \int_{\mathcal{M}} \text{d}%
\beta\; \overline{f(m)}\; g(m) \ .
\end{array}
\notag
\end{eqnarray}
 We assume further that there exists a mapping $%
T_{\beta \alpha}$:
\begin{eqnarray}
T_{\beta \alpha}: L^2(\mathcal{M},\text{d} \beta) \to L^2(\mathcal{M} \ , \text{%
d} \alpha) \ ,  \notag
\end{eqnarray}
such that
\begin{eqnarray}
\int_{\mathcal{M}} \text{d} \alpha = \int_{\mathcal{M}} \text{d} \beta
\;T_{\beta \alpha} \ .   \notag
\end{eqnarray}
For an $n$-dimensional Riemannian manifold $\mathcal{M}$
parameterized by $m_1,\cdots,m_n$ with metric $g_\alpha$ (resp. $g_\beta$),
we have $\text{d}\alpha = \sqrt{\big|\det g_\alpha\big|}\; \text{d}^n m$
(resp. $\text{d}\beta = \sqrt{\big| \det g_\beta\big|}\; \text{d}^n m$) and
thus $T_{\beta \alpha} = \sqrt{\big|\det g_\alpha\big|/\big|\det g_\beta\big|%
}$. This means that if $\{f^\beta_i, i\in \mathbb{N}\}$ is a Hilbert basis
of $L^2(\mathcal{M},\text{d} \beta)$, then $\{f^\alpha_i = \frac{f^\beta_i }{%
\sqrt{T_{\beta \alpha}}}, i\in \mathbb{N}\}$ is a Hilbert basis for $L^2(%
\mathcal{M},\text{d} \alpha)$, and we obviously have
\begin{eqnarray}
  \label{eq:T}
(f^\beta_i,f^\beta_j)_\beta = \delta_{ij}\ \ \Longleftrightarrow \ \
(f^\alpha_i,f^\alpha_j)_\alpha = \delta_{ij} \
\end{eqnarray}
and the map $T_{\beta\alpha}$ is unitary.

Returning to our softly deformed manifold, let $g_{MN}$ be the metric of $G_m$, let  $g^\mu_{MN}$ be the metric on $G_m^\mu$ and
let $g = |\det (g_{MN})|$, $g^\mu = |\det (g^\mu_{MN})|$. Let
${\cal B}=\{\rho_I(m), I \in {\cal I}\}$ be a Hilbert basis of $L^2(G_m)$; by \eqref{eq:T}
\beqa
{\cal B}_\mu = \Big\{\rho_I^\mu(m)= \sqrt{T^\mu} \;\rho_I(m) \ , I\in {\cal I} \Big\} \ ,\nn
\eeqa
where $T=\sqrt{\frac g {g^\mu}}$
is a Hilbert basis of $L^2(G^\mu_m)$ and we have
\beqa
(\rho_I^\mu,\rho_J^\mu)_\mu  = (\rho_I,\rho_J) =\delta_{IJ} \ , \nn
\eeqa
where in the second equality we have used  the scalar product over the manifold $G_m$.
We assume that the transition function is non-degenerate, namely that
\beqa
\label{eq:gNS}
t_{\text{min}}\le T^\mu(m) \le t_{\text{max}} \ ,  \ \ \forall m \in G_m^\mu
\eeqa
with $t_{\text{min}}, t_{\text{max}} \in \mathbb R_+\setminus\{0\}$.

This construction can be adapted easily for non-normalizable functions. As an illustration, consider the Minkowski spacetime in $D$ dimensions. By the Fourier
theorem, the set of plane waves
\beqa
\label{eq:pwave}
{\cal B} = \Big\{ \Phi_{\bf p}({\bf x}) = \frac 1 {(2\pi)^{D/2}} e^{ i {\bf p}\cdot {\bf x}} \  , \ {\bf p} \in \mathbb R^{1,D-1} \Big\}
 \eeqa
 where ${\bf p}\cdot {\bf x}$ is the usual Minkowski scalar product, enables to expand any Schwartz function. Indeed, in
 this case, by Fourier transformation we have:
 \beqa
 \Psi({\bf x}) &=& \int \limits_{\mathbb R^{1,D-1}} \text{d} {\bf p}\; \widetilde{\Psi}({\bf p}) \; \Phi_{\bf p}({\bf x}) \ , \nn\\
  \widetilde{\Psi}({\bf p})&=&(\Phi_{\bf p},\Psi)=  \int \limits_{\mathbb R^{1,D-1}} \text{d} {\bf x}\; \overline{\Phi_{\bf p}({\bf x})} \; \Psi({\bf x})   \ , \nn
  \eeqa
  and
  \beqa
  (\Phi_{\bf p}, \Phi_{\bf q})=\int \limits_{\mathbb R^{1,D-1}} \text{d} {\bf p}\; \overline{\Phi_{\bf p}({\bf x})}\;  \Phi_{\bf p}({\bf x})=
  \delta({\bf p}-{\bf q}) \ . \nn
  \eeqa
  The procedure of Mackey can be extended easily in this case and
  \beqa
{\cal B}_\mu = \Big\{ \Phi^\mu_{\bf p}({\bf x}) = \frac1 {(g^\mu)^{1/4}}\Phi_{\bf p}({\bf x}) \  , \ \ {\bf p} \in \mathbb R^{1,D-1} \Big\} \nn
 \eeqa
 with
 \beqa
 (\Phi^\mu_{\bf p}, \Phi^\mu_{\bf q})_\mu=(\Phi_{\bf p}, \Phi_{\bf q})=\delta({\bf p}-{\bf q})  \ .\nn
 \eeqa
 Finally, the closure relation of the wave functions $\Phi_{\vec p}$ implies
  \beqa
  \int \text{d} {\bf p} \;\overline{\Phi_{\bf p}^\mu({\bf x})}\;\Phi_{\bf p}^\mu({\bf y})=
  \frac{\delta({\bf x}-{\bf y})}{\sqrt g^\mu} \ ,\nn
    \eeqa
  thus the set ${\cal B}_\mu$ is complete, and these
 elements enable to expand any Schwartz function of the Riemannian spacetime:
 \beqa
 \Psi({\bf x}) = \int \limits_{\mathbb R^{1,D-1}} \text{d} {\bf p}\; \widetilde{\Psi}({\bf p}) \; \Phi^\mu_{\bf p}({\bf x}) \ ,  \ \
  \widetilde{\Psi}({\bf p})=(\Phi^\mu_{\bf p},\Psi)_\mu   \ . \nn
  \eeqa
    Note however that the functions $\Phi^\mu_{\bf p}$  in ${\cal B}_\mu$ are defined by means of the  scalar product in  the Minkowski spacetime, and not the
    scalar product in the Riemann spacetime. As a second example, we now consider a deformation of $\SL$. In this case, the Mackey procedure leads to
    \beqa
    \label{eq:Hilbmu}
\psi_{n\lambda m}^\eta(\rho,\varphi_1,\varphi_2) &\to&   \psi^{\mu\eta}_{n\lambda m}(\rho,\varphi_1,\varphi_2) =
\Big(
\frac{\cosh \rho \;\sinh \rho}{ g^\mu} \Big)^{1/4}\psi_{n\lambda m}^\eta(\rho,\varphi_1,\varphi_2) \\
\psi_{ni\sigma  m}^\epsilon(\rho,\varphi_1,\varphi_2) &\to&   \psi^{\mu\epsilon}_{ni\sigma m}(\rho,\varphi_1,\varphi_2) =
\Big(
\frac{\cosh \rho \;\sinh \rho}{ g^\mu} \Big)^{1/4}\psi_{ni\sigma m}^\epsilon(\rho,\varphi_1,\varphi_2) \ , \nn
\eeqa
for respectively the matrix elements of the discrete and principal continuous series. These matrix elements satisfy
the first equation in \eqref{eq:Psi_N} and Eq.[\ref{eq:Non-norm}] with the scalar product \eqref{eq:PSGmu},  instead of the scalar product \eqref{eq:spSL}. In a similar manner, as seen for
the previous example, the set of functions in \eqref{eq:Hilbmu} is a complete set.
Of course, only the former functions are normalized, while the latter functions are not normalized. Even because of hypothesis \eqref{eq:gNS}, the functions
 $\psi^{\mu\eta}_{n\lambda m}$ are Schwartzian.
This set of functions enables us to have
a `Plancherel' decomposition on $G^\mu_{nc}$. Actually, if $f$ is a Schwartz function, we have
\beqa
&f(\rho,\varphi_1,\varphi_2) = \sum \limits_{\Lambda,m,n} \hskip -.6 truecm \bigintssss{} f^{n \Lambda m}\; \psi^\mu_{n\Lambda,m}(\rho,\varphi_1,\varphi_2)\nn\\
 &f_{\pm }^{n \lambda m}=(\psi^{\mu \pm  }_{n\lambda m},f)_\mu \ , \ \
 f^{\epsilon}_{nm}(\sigma)=(\psi^{\mu \epsilon }_{ni\sigma m},f)_\mu \ . \nn
\eeqa
with the notations of \eqref{eq:IP} and \eqref{eq:PC}.

Furthermore,
if $G_m=G_{nc}$ is non-compact, it is always possible to identify a Hilbert basis, {\it i.e.}, a complete set of orthonormal functions. The Mackey
procedure follows easily in this case. Again for $G_{nc}=\SL$, this leads to
\beqa
\label{eq:Losmu}
\Phi_{nmk}(\rho,\varphi_1,\varphi_2)\to \Phi^\mu_{nmk}(\rho,\varphi_1,\varphi_2)=\Big(
\frac{\cosh \rho \;\sinh \rho}{ g^\mu} \Big)^{1/4} \Phi_{nmk}(\rho,\varphi_1,\varphi_2) \ ,
\eeqa
which form a Hilbert basis of $L^2(\SL^\mu)$ and are also Schwartzian.

Following the notations of Section \ref{sec:Gc}, \footnote{Similar results hold to label states of non-compact
manifolds, so we take the same notations in both cases.} and in particular results due to Racah \cite{Ra}, we denote $\Psi_{LQR}$  (see \eqref{eq:HilGc},
for $G_m=G_c$)
the set of matrix elements which allows us to expand any Schwartz functions on $G_m$, and define
\beqa
\label{eq:Bmu}
{\cal B}_\mu=\Big\{\Psi^\mu_{LQR}(m) = \sqrt{T^\mu} \Psi_{LQR}(m) \ , (LQR) \in {\cal I} \Big\} \ ,
\eeqa
the corresponding set in $G^\mu_m$ through the Mackey procedure.

The next step in our construction is to decompose the product  $\Psi^\mu_{LQR}(m)\Psi^\mu_{L'Q'R'}(m)$ in the basis \eqref{eq:Bmu}. We analyze first the case when $G_m=G_c$ is a compact Lie group. As the metric tensor of the soft manifold satisfies \eqref{eq:gNS}, this in particular means that
any function $\Psi^\mu_{LQR}$ can be expanded in the Hilbert basis of $L^2(G_c)$ \eqref{eq:HilGc}. Conversely, any function $\Psi_{LQR}$
can be expanded in the Hilbert basis of $L^2(G^\mu_c)$:
\beqa
\label{eq:PsiPsiGc}
\Psi^\mu_{LQR}(m)= P_{LQR}{}^{L'Q'R'} \Psi_{L 'Q'R'}(m) \ , \ \
\Psi_{LQR}(m)= (P^{-1})_{LQR}{}^{L'Q'R'} \Psi^\mu_{L'Q'R'}(m) \ .
\eeqa
Therefore the product $\Psi^\mu_{L_1Q_1R_1}(m)\Psi^\mu_{L_2Q_2R_2}(m)$ decomposes as:
\beqa
\label{eq:PsiPsimu}
\Psi^\mu_{L_1Q_1R_1}(m)\;\Psi^\mu_{L_2Q_2R_2}(m)=
C^\mu{}_{L_1Q_1R_1;L_2Q_2R_2}^{L_3Q_3R_3}\;\Psi^\mu_{L_3Q_3R_3}(m),
\eeqa
where
\beqa
\label{eq:Cmu}
C^\mu{}_{L_1Q_1R_1;L_2Q_2R_2}^{L_3Q_3R_3}=P_{L_1Q_1R_1}{}^{L'_1Q'_1R'_1} \; P_{L_2Q_2R_2}{}^{L'_2Q'_2R'_2}
\;(P^{-1})_{L'_3Q'_3R'_3}{}^{L_3Q_3R_3} C_{L'_1Q'_1R'_1;L'_2Q'_2R'_2}^{L'_3Q'_3R'_3} \ .
\eeqa
If $G_{nc}$ is a non-compact manifold, this procedure extends naturally to any Hilbert basis such as \eqref{eq:Hilbmu} for $\SL$.
In \cite{ram3} we proceeded in a different but equivalent way to obtain the decomposition of the product  $\Psi^\mu_{L_1Q_1R_1}(m)\Psi^\mu_{L_2Q_2R_2}(m)$.
However, the proof given in \cite{ram3} does not extend to the case when $G_m=G_{nc}$ is a non-compact manifold.
We now consider  non compact-manifolds, and consider $\SL_\mu$ as an illustration.
Since the matrix element $\psi^{\eta}_{n \lambda m}$ are Schwartz functions and the metric of $\SL_\mu$ satisfies \eqref{eq:gNS}, the functions
$\psi^{\mu \eta}_{n\lambda m}$ (see \eqref{eq:Hilbmu}) are also Schwartzian, and can thus
be decomposed in the Hilbert basis of $L^2(\SL)^\perp$  \footnote{
\label{fot:1} In this analysis, in order to simplify
  the presentation we assume that $\psi^{\mu \eta}_{n \lambda m}$ decompose in
  the set of functions $\psi^\eta_{n \lambda m}$, more general situations can be encountered.
} (see Section \ref{sec:HilbSL}):
\beqa
\label{eq:ddmu}
\psi^{\mu \eta}_{n\lambda m}(\rho, \varphi_1,\varphi_2)&=&P_{n\lambda m}{}^{n' \lambda' m'} \psi^{ \eta}_{n'\lambda' m'}(\rho, \varphi_1,\varphi_2) \ , \ \nn\\
\psi^{ \eta}_{n\lambda m}(\rho, \varphi_1,\varphi_2)&=&(P^{-1})_{n\lambda m}{}^{n'\lambda' m'} \psi^{\mu \eta}_{n'\lambda' m'}(\rho, \varphi_1,\varphi_2) \ .
\eeqa
We now turn on the decomposition of  $\psi^{\mu \epsilon}_{n i\sigma  m}$. We recall that, in the case of $\SL$, the Plancherel theorem enables us to
express the matrix elements of the continuous series in terms of the Losert basis of $\LL^{\text{d}^\perp}$ and {\it vice versa} \cite{ram} (see \eqref{eq:PL} and \eqref{eq:LP}).
Thus, in the case of $\SL_\mu$ because of \eqref{eq:Hilbmu} and \eqref{eq:Losmu}, we
have exactly the same expansion, and  \eqref{eq:Hilbmu}, \eqref{eq:Losmu} reduce
to:
\beqa
 \Phi^\mu_{n  mk }(\rho, \varphi_1,\varphi_2)&=& \int \limits_0^{+\infty} \text{d} \sigma \; \sigma\tanh \pi(\sigma+i\epsilon)
 f_{nmk}(\sigma) \psi^{\mu \epsilon}_{n i\sigma  m}(\rho,\varphi_1,\varphi_2)\ , \ \ k \ge k_{\text{min}}\nn\\
 \psi^{\mu \epsilon}_{n i \sigma m}(\rho, \varphi_1,\varphi_2) &=&
 \sum \limits_{k\ge k_{\text{min}}} \overline{ f^{nmk}}(\sigma) \Phi^\mu_{n  mk}(\rho, \varphi_1,\varphi_2) \nn \ .
 \eeqa
Observe that we have only a sum over $k$ in the second equality and an integral over
 $\sigma$ in the first equality.
 We also have \footnote{As in Footnote \ref{fot:1}, we assume for simplicity
 that if $\Phi_{nmk}\in \LL^{\text{d}^\perp}$, then $\Phi_{nmk}^{\mu}\in \LL^{\text{d}^\perp}$.}
 \beqa
 \Phi^{\mu}_{nmk}(\rho, \varphi_1,\varphi_2)&=&  P_{nmk}{}^{n'm'k'} \Phi_{n'm'k'}(\rho, \varphi_1,\varphi_2) \ ,\nn\\
  \Phi_{nmk}(\rho, \varphi_1,\varphi_2)&=& (P^{-1})_{nmk}{}^{n'm'k'} \Phi^\mu_{n'm'k'}(\rho, \varphi_1,\varphi_2)\nn
  \eeqa
  where for the sum over $k'$ we have $k'\ge k_{\text{min}}$, and the sum over $n,m$ is
  unconstrained (note that if $\Phi^\mu_{nmk}\not\in
  \LL^{\text{d}^\perp}$ we have no restriction on the summation over $k'$).
   Thus
  \beqa
  \label{eq:ppmu0}
\psi^{\mu \epsilon}_{n i \sigma m}(\rho, \varphi_1,\varphi_2) &=&
\overline{ f^{nmk}}(\sigma) P_{nmk}{}^{n'm'k'}\Phi_{n'm' k'}(\rho, \varphi_1,\varphi_2) \nn \\
&=& \int \limits_0^{+\infty} \text{d} \sigma' \; \sigma'\tanh \pi(\sigma'+i\epsilon)\overline{ f^{nmk}}(\sigma) P_{nmk}{}^{n'm'k'}
f_{n'm'k'}(\sigma') \psi^{\epsilon}_{n' i\sigma'  m'}(\rho,\varphi_1,\varphi_2)\ \nn\\
&\equiv& \int \text{d} \sigma' P^\epsilon_{nm}(\sigma,\sigma') \;\psi^{\epsilon}_{n i\sigma'  m}(\rho,\varphi_1,\varphi_2) \ ,\
\eeqa
where
\beqa
 P^\epsilon_{nm}(\sigma,\sigma')= \sigma'\tanh \pi(\sigma'+i\epsilon)\overline{ f^{nmk}}(\sigma) P_{nmk}{}^{n'm'k'}
f_{n'm'k'}(\sigma') \ .\nn
\eeqa
Note that there is no summation over $n$ and $m$.
This relation can be inverted
\beqa
\label{eq:ppmu2}
\psi^{\epsilon}_{n i \sigma m}(\rho, \varphi_1,\varphi_2)
&=& \int \text{d} \sigma' (P^{-1})^\epsilon_{nm}(\sigma,\sigma')\; \psi^{\epsilon}_{n i\sigma'  m}(\rho,\varphi_1,\varphi_2) \ ,
\eeqa
with
\beqa
 (P^{-1})^\epsilon_{nm}(\sigma,\sigma')= \sigma'\tanh \pi(\sigma'+i\epsilon)\overline{ f^{nmk}}(\sigma) (P^{-1})_{nmk}{}^{n'm'k'}
f_{n'm'k'}(\sigma') \ .\nn
\eeqa
Then,  the product $\psi^{\mu}_{n_1 \bar \Lambda_1 m_1} \psi^{\mu }_{n_2 \bar \Lambda_2 m_2}$ decomposes as
\beqa
\label{eq:ppmu1}
\psi^{\mu}_{n_1 \bar \Lambda_1 m_1}(\rho, \varphi_1,\varphi_2)
\psi^{\mu }_{n_2 \bar \Lambda_2 m_2}(\rho, \varphi_1,\varphi_2)
 &=&   \sum_{\bar \Lambda} \hskip -.5 truecm \int{}\;
{ C^\mu}_{\bar \Lambda_1, \bar \Lambda_2}^{\bar \Lambda}{}_{m_1,m_2, m'_1,m'_2}
\psi^\mu_{m_1+m_2 \bar\Lambda m'_1+m'_2}(\rho,\varphi_1,\varphi_2)\nn \\
\eeqa
where the coefficients ${ C^\mu}_{\bar \Lambda_1, \bar \Lambda_2}^{\bar \Lambda}{}_{m_1,m_2, m'_1,m'_2}$ can be deduced either from
the transformations properties
\eqref{eq:ddmu}, \eqref{eq:ppmu1}, \eqref{eq:ppmu2} or from the relations \eqref{eq:matmat} involving the Clebsch-Gordan coefficients
of $\SL$.\\

To summarize, we have for any soft manifold $G_m^\mu$
\beqa
\label{eq:PPmu}
\Psi^\mu_{LQR}(m) = \sum \limits_{LQR} \hskip -.6 truecm \bigintssss{} P_{LQR}{}^{L'Q'R'} \Psi_{L'Q'R'}^\mu(m) \ , \ \
\Psi_{LQR}(m) = \sum \limits_{LQR} \hskip -.6 truecm \bigintssss{} (P^{-1})_{LQR}{}^{L'Q'R'} \Psi^{\mu}_{L'Q'R'}(m)
\eeqa
The decomposition of the product $\Psi^\mu_{LQR} \Psi^\mu_{L'Q'R'}$ follows at once by \eqref{eq:PPmu} and \eqref{eq:psipsi}.
The Lie algebra
\beqa
\g(G_m^\mu)=\Big\{T^\mu_{aLQR}(m)= T_a\psi^\mu_{LQR}(m)\ ,\ \  a=1,\cdots ,\  d,\  (L,  Q, R) \in {\cal I}\Big\} \nn
\eeqa
has Lie brackets
\beqa
\big[T^\mu_{a_1L_1Q_1R_1}(m),T^\mu_{a_2L_2Q_2R_2}(m)\big]=i f_{a_1 a_2}{}^{a_3} \;\sum
\hskip -.4 truecm \bigintssss{}\;
C^\mu{}_{L_1Q_1R_1;L_2Q_2R_2}^{L_3Q_3R_3}\;
T^\mu_{a_3L_3Q_3R_3}(m) \ , \nn
\eeqa
by \eqref{eq:LoopGc} and \eqref{eq:PPmu}.
Note that, due to relations \eqref{eq:PPmu}, the Lie algebras
$\g(G_m^\mu)$ and $\g(G_m)$ are isomorphic.\\

We now turn our attention to the identification of Hermitian  operators.
As mentioned previously,
the isometry group of the soft manifold $G_m^\mu$ is smaller than its undeformed analogue. Thus,
if we solve the Killing equation \eqref{eq:killing} for $G_m^\mu$,  we have less Killing vectors than for $G_m$.
In order to have a number of generators larger than the dimension of the isometry group of $G_m^\mu$, define \cite{ram3} (see also \cite{Hor} for
an analogous definition)
\begin{eqnarray}
  \label{eq:hermmu}
L_A^\mu=\sqrt{T^\mu} L_A \frac 1{\sqrt{T^\mu}} \ , \ \ R_A^\mu=\sqrt{T^\mu}
R_A \frac 1{\sqrt{T^\mu}} \ .
\end{eqnarray}
In general, these generators do not generate the isometry group of $G_m^\mu$, but
it follows at once that
\begin{eqnarray}
\big[L^\mu_A,L^\mu_B\big]=i c_{AB}{}^C L^\mu_C \ , \ \ \big[R^\mu_A,R^\mu_B\big]%
=i c_{AB}{}^C R^\mu_C \ , \ \ \big[L^\mu_A,R^\mu_B\big]=0 \ .  \notag
\end{eqnarray}
and
\begin{eqnarray}  \label{eq:muRep}
L^\mu_A \Psi^\mu_{LQR}(m)&=& (M^Q_A)_L{}^{L^{\prime
}}\Psi^\mu_{L^{\prime }QR}(m)  \notag \\
R^\mu_A \Psi_{LQR}(m)&=& (M^Q_{A})_R{}^{R^{\prime
}}\Psi^\mu_{LQR^{\prime }}(m)\
\end{eqnarray}
with $(M^Q_{A})_L{}^{L^{\prime}}$ the matrix elements of the representation of $G_m$ associated to the eigenvalues of the Casimir operators $Q$.
Thus $\{L_A^\mu, A=1,\cdots,n\}$ and $\{R_A^\mu,A=1,\cdots,n\}$ generate the
Lie algebra $\mathfrak{g}_m$ and $\Psi^\mu_{LQR}$ are the corresponding
matrix elements of $G_m$ (but not of $G_m^\mu$, which is not a group). This means that, locally, $G_m^\mu$ and $G_m$ are diffeomorphic. Moreover, since the operators $L_A, R_A$ are Hermitian with respect to the scalar product on $G_m$, the  operators $L^\mu_A, R^\mu_A$
will be  Hermitian with respect to the scalar product on $G^\mu_m$.
Note however, that the generators $L^\mu_A, R^\mu_A$ are {\it certainly not}
linear combinations of the generators $L_A, R_A$, simply because, in general,
the soft manifold $G_m^\mu$ has less Killing vectors than the manifold
$G_m$.

One observation is required. As we have seen,  $\psi _{LQR}^{\mu }$ are in the left and
right representation of $G_{m}$ (see \eqref{eq:muRep}). However, since the
metric tensor is deformed by the parameter $T^{\mu }$, we have to take into
account this deformation parameter when considering tensor products of
representations. In particular, if  we define
\beqa
  \Psi _{LQR}^{\mu }(m)\otimes _{\mu }\Psi _{L'Q'R'}^{\mu }(m)&\equiv& \frac{1}{\sqrt{T^{\mu }}}\Psi _{LQR}^{\mu }(m)\Psi _{L'Q'R'}^{\mu }(m)=\sqrt{T^{\mu }}\Psi _{LQR}(m)\Psi _{L'Q'R'}(m)\nn\\
  &=&
\sum  \limits_{L''Q''R''}\hskip -.8 truecm \bigintssss{}
  C_{LQRL;L'Q'R'}^{L''Q''R''}\Psi _{L''Q''R''}^{\mu }( m)\
,\nn
  \eeqa
  thus
recovering the usual results instead of \eqref{eq:PsiPsimu} (for a compact manifold).
\\

The last step in our construction is to identify relevent central extensions. First observe that
two-cocycles can be defined equivalently to the case $G_m$ by \eqref{eq:cos} or \eqref{eq:cos2}. However, because
of \eqref{eq:hermmu}, this construction is more involved. We identify relevant central extensions following
a different strategy. Indeed, we have already obtained central extensions for the algebra $\g(G_m)$ denoted
$k^i_L, k^i_R$ associated and to the two-cocycles $\omega_i^L, \omega_i^R$ (see
\eqref{eq:cosGc} for $G_m=G_c$  a compact
Lie group, and with a similar relation
for $G_m=G_{nc}$, a non-compact Lie group).

Before going further, recall some well-known properties about
two-cocycles. Let $(\{X_a, a=1,\cdots, \},[\;,\;])$ be a Lie algebra with Lie brackets
$[X_a,X_b]=i h_{ab}{}^c X_c$.
Then,  if we define a new algebra $(\{X_a, a=1,\cdots, \},[\;,\;]')$ with a new bracket
$[X_a,X_b]'= [X_a,X_b]+ \omega(X_a,X_b)$ ($\omega(X_a,X_b)$  belongs to $\mathbb R$
or $\mathbb C$ depending if we consider real or complex Lie algebras) this algebra is endowed with the structure of a Lie algebra
if $\omega$ is a two-cocycle. This is equivalent to say that the Jacobi identity is satisfied, {\it i.e.},
that we have
\beqa
      h_{bc}{}^d \;\omega(X_a,X_d) + h_{ca}{}^d\; \omega(X_b,X_d)+
    h_{ab}{}^d \;\omega(X_c,X_d) = 0 \ .\nn
    \eeqa
    Obviously, if we perform a change of basis: $X'_a=P_a{}^b X_b$, the two-cocycle is given by
    \beqa
    \label{eq:cos2mu}
\omega(X'_a,X'_b) = P_a{}^c\; P_b{}^c \omega(X_a,X_b) \ ,
    \eeqa
    and, the Jacoby identity above is satisfied with
    the new two-cocycle.\\

   Returning to $G_m^\mu$, relations \eqref{eq:PPmu} allow to  define two-cocycles of $G_m^\mu$
    from two-cocycles of $G_m$, because of \eqref{eq:cos2mu}.
    In the case of a compact Lie group
    $G_m=G_c$, \footnote{The case of a non compact Lie group is easily obtained with the
    substitution $\sum \to \sum \hskip -.25truecm\int$.} the two-cocycles take the form:
    \beqa
    \label{eq:cosmu}
    \omega_i^{\mu L}(T^\mu_{a_1L_1Q_1R_1}, T^\mu_{a_2L_2Q_2R_2})=
    k_{a_1 a_2}
    P_{L_1Q_1R_1}{}^{L_1'Q_1'R_1'}
  P_{L_2Q_2R_2}{}^{L_2'Q_2'R_2'}
     L_2'(i)
   \eta_{L'_1 Q_1'R'_1,L'_2Q_2'R'_2} \ .
    \eeqa
    (with a similar relation for $\omega_i^R$) because of \eqref{eq:PsiPsiGc}. It is important to emphasize that
    the two-cocycles above are defined with integration on the manifold $G_m$ and {\it not} on $G_m^\mu$. This observation
    is very important when we identify the series of operators which are compatible with these cocycles. It turns
    out that the operators compatible with $\omega_i^{\mu L},\omega_i^{\mu R}$ are the operators $D_i^L, D_i^R$ satisfying
    \eqref{eq:comp2} and {\it not} the operators $D_i^{\mu L}, D_i^{\mu R}$.

    The KM algebra associated to $G_c^\mu$ (with $G_c$ a compact Lie group)  is then defined by
    \beqa
\widetilde{\mathfrak{g}}(G^\mu_c) = \big\{\mathcal{T}^\mu_{aLQR},\ a=1\cdots, d,
(LQR)\in {\cal I}, k_L^i, k_R^i, i=1,\cdots \ell'\}\rtimes\{D_i^L, D_i^R, i=1,\cdots, \ell'\}\nn
\eeqa
with Lie brackets
\beqa
\label{eq:KMsoft}
\big[{\cal T}^\mu{}_{aL_1Q_1R_1},{\cal T}^\mu{}_{a_2L_2Q_2R_2}\big] &=&
i\;  f_{a_1a_2}{}^{a_3} C^\mu{}_{L_1Q_1R_1;L_2Q_2R_2}^{L_3Q_3R_3} {\cal T}^\mu{}_{a_3 L_3Q_3R_3}\nn\\&&
\hskip -3.truecm +    k_{a_1a_2}
 P_{L_1Q_1R_1}{}^{L_1'Q_1'R_1'}
  P_{L_2Q_2R_2}{}^{L_2'Q_2'R_2'}
\; \eta_{L'_1 Q_1' R'_1,L'_2Q_2'R'_2}\Big(k_L^i\;
L_2'(i)+ k_R^i\;R_2'(i)\Big)
 \ , \nn\\
 \big[D_i^L, {\cal T}_{aLRQ}\big] &=&    P_{LQR}{}^{L'Q'R'} L'(i)  {\cal T}^\mu{}_{aL'Q'R'}\ ,\\
 \big[D_i^R, {\cal T}_{aLRQ}\big] &=&   P_{LQR}{}^{L'Q'R'} R'(i)  {\cal T}^\mu{}_{aLQR'} \ , \nn
 \eeqa
  where $ C^\mu{}_{L_1Q_1R_1;L_2Q_2R_2}^{L_3Q_3R_3}$ is defined in \eqref{eq:Cmu}.
  Analogous expressions hold for $G_m=G_{nc}$ a non-compact Lie group.

  Because the algebras $\g(G_c)$ and $\g(G_c^\mu)$ are isomorphic,
  and because of \eqref{eq:cosmu}, it seems at a first glance that
  the algebras $\widetilde{\g}(G_c)$ and $\widetilde{\g}(G_c^\mu)$ are also isomorphic.
  However, for the algebra associated to $G_c$, the differential operator $D_i^{L,R}$ are
  associated to a Killing vector of $G_c$, while for the algebra $G_c^\mu$, we consider the same
  differential operators, but they {\it are not} associated to any Killing vector.
  This observation simply means that the algebras $\widetilde{\g}(G_c)$ and
  $\widetilde{\g}(G^\mu_c)$ are only diffeomorphic locally, as $G_m$ and $G_m^\mu$ are
  locally diffeomorphic.
  The case where $G_m=$U$(1)$ is very specific. Indeed, if we consider a soft manifold U$(1)^\mu$ with
  scalar product
  \beqa
(f,g)_F = \int \limits_0^{2\pi} \text{d} \theta \; F(\theta)   \;\overline{f}(\theta)\;g(\theta) \ ,\nn
  \eeqa
   a change of variables  $F(\theta) \;\text{d} \theta = \text{d} \varphi$ leads to the
  isomorphism $\widetilde{g}($U$(1)^\mu) \cong \widetilde{g}($U${}(1))$, and there do not exist
  non-trivial soft deformations of affine Lie algebras \cite{ram3}. This
  is a consequence of the property that any one-dimensional manifold without boundary is homeomorphic to the one-dimensional sphere $\mathbb{S}^{1}$.
  \\

  It is also possible in some cases to associate a Virasoro algebra of $G_m^\mu$. We briefly comment this point.
  Let $D_i^L, D_i^R$ be the set of generators of the Cartan subalgebra of $(\g_m)_L\oplus
  (\g_m)_R$ satisfying condition \eqref{eq:comp2}.
  The de Witt algebra
  (with obvious relations for $L\to R$) reads:
  \beqa
\text{Witt}_i(G_m^\mu) = \big\{\ell^\mu_{i LQR}= - \psi^\mu_{LQR} D_i{}^L \ , (LQR)\in {\cal I} \big\}
  \eeqa
  and  has Lie brackets
  \beqa
  \big[\ell^\mu_{i L_1Q_1R_1}, \ell^\mu_{i L_2Q_2R_2}\big]=  P_{L_1 Q_1 R_1}{}^{L'_1 Q'_1 R'_1}P_{L_2 Q_2 R_2}{}^{L'_2 Q'_2 R'_2}
  \Big(L_1'(i)-L'_2(i)\Big)  C_{L_1'Q_1'R_1';L_2'Q_2'R_2'}^{L''Q''R''}  \ell^\mu_{i L''Q''R''}\nn
\eeqa
because of \eqref{eq:KM-witt} and \eqref{eq:PPmu}.
On the one hand, if the corresponding operators $\ell_{i LQR}$ in  $\text{Witt}_i(G_m)$ are compatible
with the cocycle $\omega^L_i$, the operators $\ell^\mu_{i LQR}$ are compatible with the cocycle $\omega^{\mu L}_i$.
 On the other hand,  if $\text{Witt}_i(G_m)$ admits a central extension, then $\text{Witt}_i(G^\mu_m)$
 also admits a central extension, because  \eqref{eq:cos2mu} holds.\\

In this review, we will not consider any example of KM algebras associated to soft
    deformations of (compact)Lie groups, since their construction goes essentially along the same
    lines of the compact group manifolds. The interested reader may consult \cite{ram3} for more details.

 \section{Roots systems and some elements of representation theory}\label{sec:bedlewo}
In this section we would like to introduce a system of roots for the Kac-Moody
algebras considered in previous sections. In a second part we shall introduce
some elements of representations theory and in particular we shall see that
the situation is very different to the corresponding situation for affine
Lie algebras. Notwithstanding that   the results of the previous section, ({\it i.e.}, the construction of the KM algebra $\widetilde{\g}({\cal M)}$)
 can easily be extended
to any complex or real Lie algebra $\g$, in this section we {\it are assuming that   $\g$ is a compact Lie algebra}.

\subsection{Roots system}
For simplicity we  now consider   Kac-Moody algebras associated
to compact Lie groups that is ${\cal M}=G_c$ or
$G_c/H$ with $H$ a subgroup of $G_c$  \cite{mrm}. The other cases can also be considered, but
with more technical difficulties irrelevant for the analyses below --see \cite{ram} for the case where $\cal M$ is a non-compact
Lie group.

Let $\g$ be a compact Lie algebra (not to be confused with the compact
Lie algebra $\g_c$ of  Section \ref{sec:Gc}) and assume that $\g$ is of
rank $r$. Let $\{H^i, i=1,\cdots,r\}\subset \g$ be a Cartan subalgebra of $\g$,
 let $\Sigma$ be the root system of $\g$ and let $E_\alpha, \alpha \in \Sigma$ be the corresponding root-vector. To ease  the notations, we
denote $I=(LQR)$ with the notations of Section \ref{sec:Gc} then in the Cartan-Weyl basis
the Kac-Moody algebra is generated by
\beqa
\widetilde{\g}({\cal M})=\Big\{
H_I^i, E_{\alpha I}, i=1,\cdots,r, \alpha\in \Sigma, I\in {\cal I},
D_{i}, k_i, i=1\cdots, \ell'\Big\}\nn
\eeqa
and the non-vanishing Lie brackets are:
\beqa
\label{eq:CW}
\big[H^i_{I}, H^{i'}_{I '}\big] &=&  \eta_{ II'} h^{ii'}\sum _{i=1}^{\ell'} I'(i) \;k_i   \ , \nn\\
\big[H^i_{I}, E_{\alpha J} \big]&=& c_{I J}{}^K  \alpha^i \; E_{\alpha K }\  ,
\nn\\
\big[E_{\alpha_ I}, E_{\beta_ J}\big] &=&
\left\{
\begin{array}{ll}
\epsilon(\alpha,\beta) \; c_{I J }{}^K \; E_{\alpha+\beta K }\ ,
& \alpha+\beta \in \Sigma ,\\[4pt]
c_{I J }{}^K\;  \alpha\cdot H_{K}
+ \eta_{IJ}\sum\limits _{i=1}^{\ell '} J(i)\;  k_i   & \alpha + \beta=0,\\[4pt]
0,&\left\{\begin{array}{l}\alpha + \beta \ne 0\ ,\\ \alpha+\beta \not \in \Sigma , \end{array} \right.\
\end{array}
\right.\\
\big[D_i,E_{\alpha J}\big]&=& J(i) E_{\alpha_ J}\ , \nn\\
\big[D_i ,H^j_{J}]&=& J(i)  H^i_{J}\nn \ ,
\eeqa
where
\beqa
h^{ij} =\big<H^i,H^j\big>_0 \ , \ \ h_{ij}= (h^{-1})_{ij}\ , \ \ \alpha\cdot H_I = h_{ij} \;\alpha^i H^j_I  \nn
\eeqa
with the Killing form $ \big<\cdot,  \cdot\big>_0$ defined in Section
\ref{sec:Gc},
and the operators associated to roots of $\mathfrak g$  normalized as
\beqa
\big<E_\alpha, E_\beta\big>_0 = \delta_{\alpha,-\beta} \ . \nn
\eeqa
The coefficients $\epsilon(\alpha,\beta)$ depends on $\g$ and are equal to $\pm 1$ if $\g$ is simply laced.
Note the hermiticity relations:
\beqa
(H^i_I)^\dag = H^i_{I^c} \ , \ \ (E_{\alpha I})^\dag = E_{-\alpha I^c}\ , \ \ (k_i)^\dag=k_i \ , \nn
\eeqa
where $I^c$ means $I^c(i)=-I(i), i=1,\cdots,\ell'$.

In a way entirely analogous to the usual affine Lie algebra (see e.g. \cite{Goddard:1986bp}, p. 343-344)  \cite{mrm},
we can extend the Killing form to $\widetilde{ \mathfrak g}({\cal M})$ by:
\beqa
\label{eq:CSA}
\Big<{\cal T}_{aI},{\cal T}_{bJ}\Big>_1&=&\eta_{IJ} k_{ab}\ ,\nn\\
\Big<D_j,{\cal T}_{aI}\Big>_1&=&\Big<k_ j ,{\cal T}_{a I }\Big>_1 \ =\ 0\ , \\
\Big<k_i ,k_j\Big>_1 &=&\Big<D_i ,D_j\Big>_1\ =\ 0\ ,\nn\\
\Big<D_i ,k_j\Big>_1&=&\delta_{ij} \ . \nn
\eeqa

Let $\cal Q$ be the possible eigenvalues of the Casimir operators, let $q\in {\cal Q}$ and let
 ${\cal D}^q$ be the corresponding representation. Define
${\cal D}^q_0 =\{\Psi_{q,1},\cdots, \Psi_{q,n_q}\}\subset {\cal D}^q$ to be the set of vectors with zero weight.
Let $\{{\cal H}^i_{q,1},\cdots,{\cal H}_{q,n_q}^i,i=1,\cdots,r\}
\subset \{H^i_I, I, i=1,\cdots,r\in {\cal I}\}$  be the corresponding elements
in  $\widetilde{ \mathfrak g}({\cal M})$.  These elements are obviously
commuting. \footnote{For instance, if $\g=\mathfrak{su}(2)$, all bosonic, or
integer spin, representations have exactly one zero-weigh vector, and in this
case $\widetilde{\g}(\text{SU}(2))$ is an infinite rank-Lie algebra.}
 Thus from \eqref{eq:CW}, the maximal set of commuting operators
is given by
\beqa
\widetilde{\mathfrak h}=\Big\{{\cal H}^i_{n,q}\ , i=1,\cdots,r, q \in{\cal Q}, n=1,\cdots, n_q\ , D_i, k_i, i=1,\cdots,\ell'\}\ . \nn
\eeqa
This means that the Kac-Moody algebra $\widetilde{\g}({\cal M})$ is an infinite rank Lie algebra.
Due to the difficulty to deal with infinitely many commuting generators to be diagonalised
simultaneously, we define the root-space of $\widetilde{\g}({\cal M})$
considering the finite-dimensional subalgebra
$\widetilde{\mathfrak h}_0 \subset \widetilde{\mathfrak h}$ defined by:
\beqa
\widetilde{\mathfrak h}_0=\Big\{ H^i_{\bf 0}\ , i=1,\cdots,r,  D_i, k_i, i=1,\cdots,\ell'\}\ , \nn
\eeqa
(with the notations of Sect. \ref{sec:Gc} for the definition of $\ell'$)
where  $H^i_{\bf 0}$ correspond to the generators of the `loop' algebra  $\g({\cal M})$ associated to the trivial representation ${\cal D}^0$
of $\g$.
The corresponding root spaces are given by
\beqa
\label{eq:root-s}
\mathfrak g_{(\alpha, n_1,\cdots, n_{\ell'})} &=& \Big\{E_{\alpha I } \ \text{with} \ I(1) = n_1 ,\cdots , I(\ell')=n_{\ell'} \Big\} \ ,\alpha \in \Sigma,
n_1,\cdots, n_{\ell'} \in \mathbb Z \ , \nn\\
\mathfrak g_{(0, n_1,\cdots, n_{\ell'} )} &=& \Big\{H^i_{I}\  \text{with} \ I(1) = n_1 ,\cdots , I(\ell')=n_{\ell'} \Big\} \ , n_1,\cdots, n_{\ell'} \in \mathbb Z \ .
\eeqa
Since, in general, the number $\ell'$ of compatible commuting operator $D_I$ is such that $\ell'<n=\dim {\cal M}$,
unlike the usual Kac-Moody algebras, the root  spaces associated to roots are infinite dimensional and we have
\beqa
\big[\mathfrak g_{(0,{\bf n})}, \mathfrak g_{(\alpha,{\bf m})}\big]&\subset& \mathfrak  g_{(\alpha,{\bf m+n})}, \nn\\
\big[\mathfrak g_{(\alpha,{\bf m})}, \mathfrak g_{(\beta,{\bf n})}\big]&\subset& \mathfrak  g_{(\alpha+\beta,{\bf m+n})}, \ \ \alpha+ \beta \in \Sigma\nn
\eeqa
with ${\bf n } = (n_1,\cdots,n_{\ell'})$. There is one notable exception. The Kac-Moody algebra associated to U$(1)^n$ admits exactly
$\ell'=n$ compatible commuting operators \cite{mrm} (these algebra are called toroidal algebras in \cite{MRT} and
quasi-simple Lie algebras in \cite{HOEGHKROHN1990106}). Thus, in this cases  the root space $\mathfrak g_{(\alpha, n_1,\cdots, n_n)}$ is
one-dimensional.
 Finally, it is important to observe that
the Lie brackets between two elements involve not only the
root structure, but also the representation theory of $G_c$, in the form of the  Clebsch-Gordan coefficients  $c_{IJ}{}^K$ (see \eqref{eq:CW}).\\

Introduce  ${\bf 0}=(0,\cdots,0)$, then a root of $\widetilde{\g}({\cal M})$ is defined by $\tilde \alpha=(\alpha, {\bf 0}, {\bf n})$ were $\alpha \in \Sigma$
corresponds to the root of the simple compact Lie algebra $\g$, the following $\ell'$ entries correspond to the vanishing eigenvalues
of the central charges $k_1,\cdots,k_{\ell'}$ and ${\bf n} \in \mathbb Z^{\ell'}$ are the eigenvalues of $D_1,\cdots,D_{\ell'}$.

A  root  $(\alpha,{\bf 0},n_1,\cdots,n_{\ell'})$ is said to be positive if it satisfy:
\beqa
\label{eq:ord}
(\alpha,{\bf 0},n_1,\cdots,n_{\ell'}) >0 \ \ \text{if} \; \left\{
\begin{array}{cc} \text{either}&\left\{\begin{array}{l}
\exists\; k \in \{1,\cdots, \ell'\} \ \ \text{s.t.} \\
n_{\ell'}=\cdots =n_{k+1} =0\ \  \text{and} \ \ n_k >0
\end{array}\right.\\
\text{or}& n_{\ell'}=\cdots=n _1=0 \ \ \text{ and} \ \ \alpha>0 \ .
\end{array}\right.
\eeqa
 By \eqref{eq:CSA}, we can  endow the weight space with
the scalar product:
\beqa
(\alpha,c_1,\cdots,c_{\ell'},n_1,\cdots,n_{\ell'}) \cdot  (\alpha',c'_1,\cdots,c'_{\ell'},n'_1,\cdots,n'_{\ell'}) =\alpha \cdot \alpha'+
\sum_{j=1}^{\ell'}\big(n_j c'_j + n'_J c_j\big) \ , \nn
\eeqa
where $\alpha \cdot \alpha'$ is the usual scalar product in $\Sigma$.

 As happens for usual Kac-Moody algebras \cite{Kac:1990gs}, we have two types of roots.
The set of roots $(\alpha, {\bf 0}, {\bf n})$ of $\mathfrak {g}_{(\alpha, {\bf n})}$ with $\alpha \in \Sigma, {\bf n} \in \mathbb Z^{\ell'}$ satisfy
\beqa
(\alpha, {\bf 0}, {\bf n}) \cdot (\alpha, {\bf 0}, {\bf n}) = \alpha\cdot \alpha >0 \ , \nn
\eeqa
and are called {\it real roots}, whilst the set
$(0, {\bf 0}, {\bf n})$  of $\mathfrak{g}_{(0, {\bf n})}$ with $ {\bf n} \in \mathbb Z^{\ell'}$ and satisfying
\beqa
(0, {\bf 0}, {\bf n}) \cdot (0, {\bf 0}, {\bf n'}) = 0 \ , \nn
\eeqa
is called the set of  {\it  imaginary roots}. The imaginary root-space is $\ell'-$dimensional.

Recall that  $\ell'$ denotes  the number of central charges, that we call the order of centrality.
We now show that unless $\ell'=1$,  we cannot find a system of simple roots for $\widetilde{\mathfrak g}({\cal M})$.
 To this extent, introduce $\alpha_i, i=1,\cdots,r$ the simple roots of $\mathfrak g$.  If $\ell'=1$, and we denote  by $\psi$ the highest root of $\mathfrak{g}$, it is easy to see that
\beqa
\label{eq:s1}
\hat \alpha_i =(\alpha_i,0,0) \ , \ \ i=1,\cdots, r\ , \ \ \hat \alpha_{r+1}=(-\psi,0,1) \
\eeqa
is a system of simple roots of $\widetilde{\mathfrak g}({\cal M})$. Now, if we suppose that $\ell'= 2$, as the positive roots are given by
(i) $(\alpha,0,0,0,0)$ with $\alpha>0$, or (ii) $(\alpha,0,0,n_1,0)$ with $\alpha \in \Sigma, n_1>0$, or (iii)
$(\alpha,0,0,n_1,n_2), \alpha \in \Sigma, n_1\in \mathbb Z, n_2>0$ and since the roots $(\alpha,0,0,n_1,0)$ are neither bounded from below
nor from above because $n_1 \in \mathbb Z$,  we cannot define a simple root of the form $(-\psi,0,0,-n_{\mathrm{max}},1)$, where $n_{\max}$
corresponds to the highest possible value of $n_1$ (or $-n_{\max}$ the lowest possible value of $n_1$). This means that for $\ell'\ge 2$ we
can't  construct a system of simple roots.
 In other words, the only generalised Kac-Moody algebras that admit simple roots are (obviously) the usual affine algebras,
 but also the Kac-Moody algebras associated to $SU(2)/U(1)$ studied in  \cite{mrm}.

 Note that their exists an alternative generalization of affine Lie algebras  called also Kac-Moody
 algebras and introduced independently by Kac and Moody. These algebras admit a system
 of simple roots and are defined by a generalized Cartan matrix \cite{Kac:1990gs,Moody:1966gf,Mdo}.
 However both generalizations of affine Lie algebras, {\it i.e.}, Kac-Moody algebras associated to
 a Manifold ${\cal M}$ and Kac-Moody algebras associated to a generalized
 Cartan matrix exhibit fundamentally different features.
 Indeed, the former is of infinite-rank --except for the algebra $\widetilde{\g}(\text{U}(1)^n)$, see Section \ref{sec:CST}--,
but all its roots and the corresponding generators are explicitly known, whereas the latter
has finite rank, but its generators are only iteratively known in terms of the
Chevalley-Serre relations.

 \subsection{Some elements of representation theory}\label{sec:repT}
 Let $G_c$ be a compact Lie group and let $H \subset G_c$ be a subgroup. Let
 ${\cal M}=G_c$ or ${\cal M}=G_c/H$ and let $\widetilde{\g}({\cal M})$ be the
 corresponding Kac-Moody algebra.
 Representation theory of $\widetilde{\g}({\cal M})$ is very different than
 representation theory of usual affine Lie algebras when $\dim {\cal M} >1$.
 In fact we will  show that firstly  unitary
 representations of $\widetilde{\g}({\cal M})$ exist {\it iff} there is
 {\it only one non-vanishing} central charge. Further unitary representations
 forbid highest weigh representations.

 To begin, consider the Kac-Moody algebra $\widetilde{\g}(U(1)^n)$. In the Cartan-Weyl basis
 we have
 \beqa
\widetilde {\mathfrak g}(U(1)^n) = \big\{H^i_{\bf m}, E_{\alpha, \bf m}\ ,\;  \alpha \in \Sigma, \; {\bf m}\in \mathbb Z^n, \; d_i, k_i, i=1,\cdots, n \big\} \ ,\nn
\eeqa
and the Lie brackets  are given by (${\bf m}=(m_1,\cdots,m_n)$) \footnote{To be in accordance with the literature on affine Lie algebras, please note the
  change of sign in the central charges $k_i$ compared to \eqref{eq:CW}.}
\beqa
\label{eq:tor}
\big[H^i_{\bf m}, H^{i'}_{\bf m '}\big] &=&   h^{ii'} \sum\limits_{i=1}^n  m_i k_i \; \delta_{{\bf m},- {\bf m'}}  \ , \nn\\
\big[H^i_{\bf m}, E_{\alpha \bf n} \big]&=&   \alpha^i \; E_{\alpha {\bf m} + {\bf n} }\  ,
\nn\\
\big[E_{\alpha {\bf m}}, E_{\beta  {\bf n}}\big] &=&
\left\{
\begin{array}{ll}
\epsilon(\alpha,\beta) \; \; E_{\alpha+\beta {\bf m} + {\bf n} }\ ,
& \alpha+\beta \in \Sigma ,\\[4pt]
  \alpha\cdot H_{ {\bf m} + {\bf n}  }
+  \sum\limits_{i=1}^n  m_i k_i \;\delta_{{\bf m},- {\bf n}}\ ,   & \alpha + \beta=0,\\[4pt]
0,&\left\{\begin{array}{l}\alpha + \beta \ne 0\ ,\\ \alpha+\beta \not \in \Sigma , \end{array} \right.\
\end{array}
\right.\\
\big[d_i,E_{\alpha {\bf m}}\big]&=& m_i E_{\alpha \bf m}\ , \nn\\
\big[d_i ,H^j_{\bf m}]&=& m_i H^i_{\bf m}\nn \ .
\eeqa
Stress again that in this case the root-space $\g_{(\alpha,n_n,\cdots,m_n)}, \alpha \in \Sigma, m_1,\cdots,m_n \in \mathbb Z$ are   one-dimensional. Moreover, because of this last property,
the Cartan subalgebra, differently than for the generic cases ($G_c \ne U(1)^n$), {\it is not} infinite dimensional, but its dimension is equal to $\text{rk} \;\g + 2n$.

Recall that a root $(\alpha,0,{\bf m})$ is said to be positive if  it satisfies
\eqref{eq:ord}.
Assume that rk~$\g=r$ and let $\alpha_i, i=1,\cdots,r$ 
be
the simple roots  of $\g$. Introduce further  $\mu^i,i=1,\cdots, r$
the fundamental weights  of $\g$ satisfying
\beqa
2 \;\frac{ \alpha_i\cdot \mu^j}{\alpha_i\cdot \alpha_i}  = \delta_i{}^j \ . \nn
\eeqa
Since $\g \subset \widetilde{\g}(U(1)^n)$, if ${\cal D}$ is a
unitary representation of $\widetilde{\g}(U(1)^n)$,
it decomposes into ${\cal D} = \oplus {\cal D}_i$ where
where ${\cal D}_i$ are a unitary representations of $\g$. However, unitary representations of $\g$ are in one-to-one correspondence with
highest weight vectors $\mu_0 = \sum _{i=1}^r\;p_i \mu^i, p_i \in \mathbb N$.

\begin{proposition}[\cite{mrm}]
  \label{prop}
  Let $n>1$ and let $\widetilde{\g}(U(1)^n)$ be a generalized Kac-Moody algebra. Let $\mu_0= p_i \mu^i, p_i
  \in \mathbb N$ and suppose that $\big|\mu_0, {\bf c}, {\bf m_0}\big>$ is  such
  that
  \beqa
  \label{eq:HW}
  E_{\alpha, {\bf m}} \big|\mu_0, {\bf c}, {\bf m_0}\big>&=& 0 \ \ \text{if} \ \
  (\alpha, {\bf 0}, {\bf m})>0\ ,\nn \\
  H^i_{ {\bf m}} \big|\mu_0, {\bf c}, {\bf m_0}\big>&=& 0 \ \ \text{if} \ \
   (0, {\bf 0}, {\bf m})>0\ ,\nn \\
  H^i_{\bf 0}  \big|\mu_0, {\bf c}, {\bf m_0}\big>&=& \mu_0^i  \big|\mu_0, {\bf c}, {\bf m_0}\big>\ ,\\
  k_i  \big|\mu_0, {\bf c}, {\bf m_0}\big>&=&c_i  \big|\mu_0, {\bf c}, {\bf m_0}\big>\ ,\nn\\
  d_i \big|\mu_0, {\bf c}, {\bf m_0}\big>&=&(m_{0})_i  \big|\mu_0, {\bf c}, {\bf m_0}\big>\ .\nn\
  \eeqa
  Define ${\cal D}_{\mu_0, {\bf c}, {\bf m_0} }$ to be the highest weight
  representation obtained by the action of $E_{\alpha, {\bf m}}$ with
  $(\alpha, {\bf 0}, {\bf m})<0$ on the vacuum state
  $\big|\mu_0, {\bf c}, {\bf m_0}\big>$.
  \begin{enumerate}
  \item[(i)]  If  ${\cal D}_{\mu_0, {\bf c}, {\bf m_0}}$ is a unitary representation
    all central charges vanishes, but one.
  \item[(ii)]   If  $\big|\mu_0, {\bf c}, {\bf m_0}\big>$ is a highest weight
    satisfying \eqref{eq:HW} then     ${\cal D}_{\mu_0, {\bf c}, {\bf m_0}}$ is not a
    unitary representation.
    \end{enumerate}
\end{proposition}
\begin{demo}
  (i) Let $(-\alpha, {\bf 0}, {\bf m})$ be a root of $\widetilde{\g}
  (U(1)^n)$. The generators
\beqa
X^\pm_{ \alpha,   {\bf m}} = \sqrt{\frac 2 {\alpha\cdot\alpha}} E_{\mp \alpha,\pm  {\bf m}}\ ,\ \ h_\alpha =\frac 2{\alpha\cdot \alpha}
\Big(- \alpha \cdot H_{\bf 0} + \sum\limits_{i=1}^n m_i k_i \Big) \ , \nn
\eeqa
span an $\mathfrak{su}(2)-$subalgebra. Assume $(\alpha,{\bf 0}, {\bf m})>0$
thus $X_{\alpha,{\bf m}}^+ |\mu_0, {\bf c}, {\bf m_0} \rangle=0$.
 Unitarity condition $((X^{+}_{\alpha,   {\bf m}})^\dag= X^{-}_{\alpha,   {\bf m}}$)
\beqa
||X^{-}_{\alpha,   {\bf m}}\, |\mu_0, {\bf c}, {\bf m_0} \rangle||^2 &=&
\langle\mu_0, {\bf c}, {\bf m_0} \;|\big[\; X^{+}_{ \alpha,   {\bf m}}, X^{-}_{\alpha,   {\bf m}} \big]|\mu_0, {\bf c}, {\bf m_0}\rangle=
\nn\\
&=& \langle \mu_0, {\bf c}, {\bf m_0} \;|\; h_{\alpha}\;  |\; \mu_0, {\bf c}, {\bf m_0}\rangle =\frac 2{\alpha \cdot \alpha}\Big( -\alpha\cdot \mu_0 + \sum \limits _{i=1}^n c_i m_i \Big)\ge 0 \  \nn
\eeqa
implies
\beqa
\label{eq:central}
\sum \limits _{i=1}^nc_i m_i\ge \alpha\cdot \mu_0 \ .
 \eeqa
Suppose $\alpha>0$ then  $\alpha \cdot \mu_0>0$. In this case
 \eqref{eq:central} is very strong. Indeed since $(-\alpha,{\bf 0},{\bf m})>0$, this means that
${\bf m} = (m_1,\cdots,m_{k-1},m_k,0,\cdots,0)$
with $ 0<k\le n, m_k>0$ and $m_1,\cdots,m_{k-1} \in \mathbb Z$.
The  condition of \eqref{eq:central},  which must be satisfied for any $m_i \in \mathbb Z, i=1,\cdots,k-1$, is equivalent to impose that only one central charge is non-vanishing.
This proves (i).

(ii): Since only one central charge is non-vanishing,  without loss of generality we can  suppose
$c_n =c \ne 0 $ and $c_i =0, i=1,\cdots, n-1$.
Assume again $\alpha>0$ and
let $\left(-\alpha,{\bf 0},m_1,\dots ,m_{k},0,\dots 0\right)>0$ with
$ 0< k< n$, $m_k>0$ and  $(m_1,\dots ,m_{k-1})\in \mathbb{Z}^{k-1}$. The  operators (with ${\bf m}=(m_1,\cdots,m_k,0,\cdots,0)$)
\beqa
Y^\pm_{ \alpha, {\bf m}} = \sqrt{\frac 2 {\alpha\cdot\alpha}} E_{\mp \alpha,\pm  {\bf m}}\ ,\ \ h_\alpha^{\prime} =-\frac 2{\alpha\cdot \alpha}
 \alpha\cdot H_{\bf 0}   \ ,
\eeqa
also generate an $\mathfrak{su}(2)-$subalgebra. As before, the condition
$||Y^-_{\alpha, {\bf n}}\, | \mu_0, {\bf c} , {\bf m_0}\rangle||^2\geq 0$ holds, from which we deduce that
\beqa
\sum_{p=1}^{k} m_ic_i\geq \alpha\cdot\mu_0 \ .
\eeqa
However, as $c_1=\dots =c_{n-1}=0$, this cannot be satisfied if $\alpha>0$. Thus
the representation ${\cal D}_{\mu_0, {\bf c}}$ is not unitary. Which prove (ii).
\end{demo}
\begin{corollary}\label{cor}
  Let $\g$ be a simple compact Lie algebra, then
  the Kac-Moody Lie algebra $\widetilde{\g}(U(1)^n), n>1$ doesn't have any unitary highest weight representation.
\end{corollary}

These results can be extended to the case of a more general Kac-Moody algebra as $\widetilde{\g}(G_c)$ or
$\widetilde{\g}(G_c/H)$. Some elements of the proof are given in \cite{mrm}.
It is important to observe that, in absence of symmetries between the generators $D_i$, we can have $\ell'$ different possibilities given by (eventually reordering
the eigenvalues of the operators $D_i$ to define positive roots, see equation (\ref{eq:ord}))
\beqa
{\bf c} = (0,\cdots,0, c_p,0,\cdots, 0) \ , \ \ p\in\{1,\cdots,\ell'\} \ . \nn
\eeqa
\medskip

These results have been obtained in a different manner in \cite{mir}.  Let $G_c$ be a compact Lie group and
let $\g(G_c)$ and $\widetilde{\g}(G_c)$ be respectively the `loop algebra' (see \eqref{eq:LoopGc}) and the Kac-Moody
algebra (see Section \ref{sec:KMGc}). On purpose, the  authors
introduced spinors of $G_c$ ($G_c$ is  assumed to be a spin manifold). Then, they observed
 that it is easy to obtain a unitary representation
 of  the `loop-'algebra $\g(G_c)$, however this representation is {\it not } a highest weight representation. In other words
 we get a `first quantized' representation with no vacuum states annihilated by all fermion annihilation operators. They
 then showed that if we try to define a `second quantized'  or
 a highest weight representation,  some divergences do appear (see p. 380,
 \cite{mir}) and operators are ill defined. However, considering bosonic condensate fermions,
 they were able to construct a highest weight representation, but
no  invariant inner product was identified.\\

 To illustrate the situation, let us consider an explicit example.
  Let $\g$ be
a compact Lie algebra and let ${\cal D}$ be a real unitary $d-$dimensional representation. Denote
the generators of $\g$ in the representation ${\cal D}$ by  the
Hermitian  matrices $M_a, a =1\,\cdots, \dim \g$.
Introduce now  $H^i, i=1,\cdots, d$ real fermions in the representation $\cal D$.

Consider first the case where ${\cal M} = \mathbb S^1$. We have the following decomposition
(to ease the presentation we only assume here Neveu-Schwarz (NS) --and not Ramond-- fermions), {\it i.e.}, with anti-periodic boundary
conditions:
\beqa
H^i(\theta) =\sum \limits _{n \in \mathbb Z + \frac 12} b_n^i e^{-in\theta} \ . \nn
\eeqa
We further assume the quantization relations (with $\{a,b\}=ab +ba$)
\beqa
\{b_n^i,b_m^j\}=\delta^{ij} \delta_{m,-n} \ , \nn
\eeqa
and the reality condition
\beqa
(b_m^i)^\dag= b_{-m}^i \ . \nn
\eeqa
The vacuum of the representation is defined by
\beqa
b_n^i \big|0\big>=0 \ , \ \ n>0 \ ,\nn
\eeqa
in other words $b_n^i, n<0/n>0$ are creation/annihilation operators. Consider now the current
\beqa
T_a(\theta) = \frac 1 2 H^i(\theta) \;(M_a)_{ij} \;H^j(\theta) \ .\nn
\eeqa
Since $T_a$ is bilinear in the fermion fields, in order to have well defined quantities, a  normal ordering prescription must be defined. As usual
we put to the right the creation operators:
\beqa
\no \;b^i_n b^j_m\; \no = \left\{
\begin{array}{lll}
  -b^i_m b^j_n&\text{if}&n>0\\
  \phantom{-}b^i_m b^j_n&\text{if}&n<0
  \end{array}\right.\nn
\eeqa
If we decompose $T_a(\theta)=  \sum_{M \in \mathbb Z} T_{aM}\; e^{-i M \theta}$ we obtain
\beqa
\label{eq:TS1}
T_{aM} = \frac 1 2 (M_a)_{ij} \sum \limits_{m \in \mathbb Z} \no \; b_{m}^i b_{M-m}^j \;\no=
-\frac 1 2 (M_a)_{ij}  \sum \limits_{m >M} b_{M-m}^j  b_{m}^i +
\frac 1 2 (M_a)_{ij}  \sum \limits_{m <M} b_{m}^i b_{M-m}^j \ .
\eeqa
This normal ordering prescription has two advantages: (i) the vacuum is well defined $\big<0\big|T_{aM}\big|0\big> =0$
and (ii) $\|T_{aM}\big|0\big> \|$ is finite.
For the second condition (ii) observe that in  \eqref{eq:TS1}, in the second sum the term
$b_{M-m }^j$ is an annihilation operator and thus  annihilate the vacuum.
In the first sum the second  term $b_{m}^i$ is a creation  operator if $M<m<0$. Thus if $M<0$, there are
 are $-M-2$ possible values of $m$ where  $b_{M-m}^j b_{m}^i$ acts non-trivially  on $\big|0\big>$, and   consequently $\|T_{aM}\big|0\big> \|$ is finite.
So the operators $T_{aM}$ are always well defined. Furthermore, the normal ordering prescription is also of  crucial importance. Indeed,  using Wick theorem one can show that in fact the operators $T_{aM}$ generate
the affine Lie algebra $\widetilde{\g}(\mathbb S^1)$ with a  well defined central extension expressed in terms
of the quadratic Casimir operator in the representation $\cal D$
\cite{Goddard:1986bp} and thus we obtain a unitary highest weight
representation of  $\widetilde{\g}(\mathbb S^1)$.

We now analyze briefly the case where ${\cal M} = \mathbb S^1 \times \mathbb S^1$. Again considering (NS,NS) modes we get
\beqa
H^i(\theta_1,\theta_2) = \sum \limits_{m_1, m_2 \in \mathbb  Z + \frac 12} b_{m_1 m_2}^i \; e^{-i(m_1 \theta_1 + m_2 \theta_2)}  \ .
\eeqa
As for the circle, we assume the quantization relations
\beqa
\{b^i_{m_1 m_2},b^j_{n_1 n_2}\}=\delta^{ij} \delta_{m_1, -n_1}\delta_{m_2,-n_2} \ ,\nn
\eeqa
and the reality condition
\beqa
(b^i_{m_1 m_2})^\dag = b^i_{-m_1-m_2} \ .
\eeqa
To define annihilation and creation operators we introduce the following order relation
\footnote{Since we are considering only (NS,NS) fermions, $m$ and $p$ can't be equal to zero. For
R-fermions we have to account on the possibility that $m$ or $n$ equal to zero to define our order relation.}:
\beqa
(m,p)>0\ \ \  \Longleftrightarrow\ \ \   m>0  \ . \nn
\eeqa
The vacuum is defined by:
\beqa
b^i_{m_1 m_2} \big|0\big>=0\ ,\ \    m_1>0 \ , \ \ \forall  m_2 \in \mathbb Z+\frac 12  \ , \nn
\eeqa
and the normal ordering prescription by:
\beqa
\label{eq:NO}
 {}_\circ \hskip -.17truecm {}^\circ \hskip .1 truecm b^i_{mp}  b^{j}_{nq}\no = \left\{
 \begin{array}{cll}
 -  b^{j}_{n q} b^i_{mp} &\text{if~} m>0&\forall p \in \mathbb Z +\frac12\\
  \phantom{-} b^i_{mp}   b^{j}_{nq}&\text{if~} m<0&\forall p  \in  \mathbb Z +\frac12\\
\end{array}\nn
\right.
\eeqa
Decomposing as for the circle
\beqa
T_a(\theta_1,\theta_2)=\frac 1 2 (M_a)_{ij} \;\no \; H^i(\theta_1,\theta_2) \;  H^j(\theta_1,\theta_2)\; \no\nn
\eeqa
we get
\beqa
T_{aMP}&=&\frac 12 (M_{a})_{ij}  \sum \limits_{m,p \in \mathbb Z \;+\frac 12}\no b_{mp}^i b_{M-m P-p}^j \no\nn\\
&=&
-\frac 12 (M_{a})_{ij}\sum \limits_{m>M\; p \in \mathbb Z \;+(\frac 12)} b_{M-m P-p}^j b_{mp}^i +\frac 12 (M_{a})_{ij}  \sum \limits_{m<M\; p \in \mathbb Z \;+(\frac 12)} b_{mp}^i b_{M-m P-p}^j  \nn\
\eeqa
As for the circle the vacuum is well defined, but now  $\|T_{aMP}\big|0\big>\|$ diverges because when $M<0$ the first sum above involves infinitely many terms
(because of the sum  over the second indice which belongs to $\mathbb Z$).
This situation has been analyzed explicitly in \cite{Campoamor-Stursberg:2022lyx,Campoamor-Stursberg:2022ane} where
in order to have well defined generators, beyond the usual normal ordering
prescription,  a regulator was introduced,  and infinite sums  were regularized by means of Riemann
$\zeta-$function. Next, the normal ordering prescription also leads to a representation
of $\widetilde{\g}(\mathbb S^1\times \mathbb S^1)$ but with only one central extension, and again the central charge can be expressed in terms
of the quadratic Casimir operator of $\g$ in the representation ${\cal D}$ \cite{Campoamor-Stursberg:2022lyx}. Even if it was possible to construct fermions  \cite{Campoamor-Stursberg:2022lyx}
or bosons  \cite{Campoamor-Stursberg:2022ane}  representations of $\widetilde{\g}(\mathbb S^1\times \mathbb S^1)$ this procedure is not fully satisfactory because
some cut-off is needed in order to have well defined quantities. This construction is {\it not} in contradiction with Corollary \ref{cor} because we didn't consider any cut-off or regularization prescription in Proposition \ref{prop}.

\subsection{Some results on  toroidal algebras}\label{sec:CST}
In this section we would like to study more into the details the Kac-Moody algebra
$\widetilde{\g}(U(1)^n)$. We first consider the case $n=2$, having in mind that
the general case $n>2$ follows easily. Starting from the `loop-'algebra
$\g(U(1)^2)=\{T_{am_1m_2}=T_a e^{i m_1 \theta_1 + i m_2 \theta_2}\ , a =1,\cdots,\dim \g,
  m_1,m_2\in \mathbb Z\}$, its
central extensions are characterized  by closed one-forms. Since $H^1(\mathbb S^1
\times \mathbb S^1) = \mathbb Z \oplus \mathbb Z$ a general closed one-form reads:
\beqa
\gamma= i k_1\d \theta_2 -i k_2 \d \theta_1 -i\d h :=k_1 \gamma_1 +k_2 \gamma_2 + \gamma_h \ ,\nn
\eeqa
where $k_1,k_2 \in \mathbb R$ and $h$ is an arbitrary periodic function  on the two-torus. Let $h_{m_1,m_2}(\theta_1,\theta_2)=
e^{i m_1 \theta_1 + i m_2 \theta_2}, (m_1,m_2) \ne (0,0)$ and let
$\gamma_{m_1,m_2}= -k_{m_1,m_1} d h_{m_1,m_2}$ be the corresponding exact one-form.
The differential operators associated to the one-forms $\gamma_1, \gamma_2,
\gamma_{m_1,m_2}$ are respectively:
\beqa
d_1&=& -i \partial_1 \ , \nn\\
d_2&=& -i \partial_2 \ , \nn\\
d_{n_1,n_2} &=& e^{i m_1\theta_1 + i m_2 \theta_2}\big(-i m_2 \partial_1+i m_1\partial_2\big) \nn \ .
\eeqa
A generic two-cocycle is thus given by
\beqa
\omega(T_{an_1 n_2}, T_{a'm_1'n_2'}) &=&-k_1 \;\omega_{1}(T_{an_1 n_2}, T_{a'm_1'n_2'}) -
k_2 \;\omega_{2}(T_{an_1 n_2}, T_{a'm_1'n_2'}) \nn\\
&&- \hskip -.8truecm \sum \limits_{(m_1,m_2)\in \mathbb Z
  \setminus\{0,0\}} k_{m_1,m_2}\;\omega_{m_1,m_2}(T_{an_1 n_2}, T_{a'm_1'n_2'})  \nn
\eeqa
where (see Eq.[\ref{eq:cosL}])
\beqa
\label{eq:cosT2}
\omega_{1}(T_{an_1 n_2}, T_{a'm_1'n_2'}) &=&
 n_1' k_{aa'}\;\delta_{n_1,-n_1'} \;\delta_{n_2,-n_2'}\ ,\nn\\
\omega_{2}(T_{an_1 n_2}, T_{a'm_1'n_2'})
&= &n_2' k_{aa'}\;\delta_{n_1,-n_1'} \;\delta_{n_2,-n_2'}\ , \\
\omega_{m_1,m_2}(T_{an_1 n_2}, T_{a'm_1'n_2'}) &=&
k_{aa'}(m_2 n_1' - m_1 n_2')\;\delta_{n_1+n_1', -m_1} \;\delta_{n_2+n_2',-m_2} \nn\\
 &=&
k_{aa'}(n_1 n_2' - n_2 n_1')\;\delta_{n_1+n_1', -m_1} \;\delta_{n_2+n_2',-m_2} \ .\nn
\eeqa

These cocycles were also be obtained in \cite{Frappat:1989gn}.
It is direct to observe  by \eqref{eq:compat} that  the differential operators $d_1$ and $d_2$
are compatible with the cocycles $\omega_1, \omega_2, \omega_{m_1,m_2}$.

 In \cite{MRT} Moody and his collaborators    obtained, using K\"ahler differentials \cite{Ka,MY}, the universal central extension of the `loop' algebra they called toroidal algebra. It turns out,
that in fact the central extensions in \cite{MRT} coincide with the two-cocycle
above \eqref{eq:cosT2}. Indeed, the first cases (resp. second cases)  in  Eq.[3] of  \cite{MY}
correspond to the cocycle $\omega_{m_1,m_2}$ --after an appropriate rescalling--  (resp. $\omega_1,\omega_2$).
Moreover, in \cite{MY} the authors also proved that their formul\ae\ coincide
with the formul\ae \ obtained in \cite{Ka} used by Moody and collaborators.
This means that to centraly extend the `loop-'algebra $\g(U(1)^2)$ we can equivalently
use the cocycle \eqref{eq:cosT2} or
K\"ahler differentials as in \cite{MRT}.

For completeness and for further use, we briefly recall the main feature of the
Moody {\it et al} construction \cite{MRT}. Let $\mathbb C[z_1,z_1^{-1}, z_2,z_2^{-1}]$ be the Laurent polynomial ring in two variables and let $\g \otimes \mathbb C[z_1,z_1^{-1}, z_2,z_2^{-1}]$ be the `loop' algebra (see  \eqref{eq:LoopGc} in our notations). To define the universal central extension of the `loop' algebra introduce $\Omega_1$ the
set of one-forms of $\mathbb C[z_1,z_1^{-1}, z_2,z_2^{-1}]$. Finally let ${}^{\overline {\phantom {tt}}}: \Omega_1 \to \Omega_1/\text{Im}\d$. Thus for any $F \in \mathbb C[z_1,z_1^{-1}, z_2,z_2^{-1}]$ we have $\overline{\d F}=0$. The universal central extension of the `loop' algebra is given by \cite{MRT}
\beqa
\big[F\otimes x, G \otimes y]'=FG\otimes [x,y] + \overline{(\d F) G}\; \big<x,y\big>_0 \nn \ ,
\eeqa
for any $x,y \in \g, F, G \in \mathbb C[z_1,z_1^{-1}, z_2,z_2^{-1}]$ with $[\, , \, ]$ (resp. $\big<\, , \,\big>_0$)
the Lie brackets (the Killing form) of $\g$.
Since
\beqa
\overline{\d (z_1^{k_1} z_2^{k_2})} = k_1\; \overline{z_1^{k_1} z_2^{k_2} z_1^{-1} \d z_1} +k_2 \;\overline{z_1^{k_1} z_2^{k_2} z_2^{-1} \d z_2} =0 \ , \nn
\eeqa
a basis of $\Omega_1/\text{Im}\d$ is given by
\beqa
\begin{array}{lll}
  C_{k_1 k_2}=\overline{z_1^{k_1} z_2^{k_2} z_2^{-1} \d z_2}&\text{if}&k_1 \ne 0\\[4pt]
  C_{0 k_2}=\overline{ z_2^{k_2} z_1^{-1} \d z_1}&\text{if}&k_1 = 0\ ,\ \  k_2 \ne 0\\[4pt]
  c_1= \overline{ z_1^{-1} \d z_1}\ ,\ \   c_2= \overline{ z_2^{-1} \d z_2}&\text{if}& k_1=k_2=0
   \end{array}\nn
\eeqa
and
\beqa
\overline{\d(z_1^{n_1} z_2^{n_2}) z_1^{n_1'} z_2^{n_2'}}= \left\{
\begin{array}{lll}
  (n_2 n_1'-n_1 n_2')) \frac {C_{n_1+n_1' n_2 + n_2'}}{n_1 + n_1'}&\text{if}& n_1 + n_1' \ne 0\\
  (n_2 n_1'-n_1 n_2') \frac{C_{0 n_2 + n_2'}}{n_2 + n_2'}&\text{if}&n_1+n_1'=0\ ,  \ \ n_2 + n_2'\ne 0\\
  n_1 c_1 + n_2 c_2&\text{if}&n_1 +n_1'=n_2+n_2'=0
  \end{array}\right.\nn
\eeqa
which corresponds exactly to the cocycles given in \eqref{eq:cosT2}.  Thus the two constructions coincide.

Coming back to our notations, the toroidal algebra is
$T_2(\g)=\{{\cal T}_{an_1n_2}, d_1,d_2, k_1, k_2, k_{m_1m_2},$
   $a=1,\cdots,\dim \g, n_1,n_2 \in \mathbb Z, (m_1,m_2) \in \mathbb Z^2 \setminus\{0,0\}\}$ with Lie brackets (in our notations):
\beqa
\label{eq:gT2}
\big[{\cal T}_{an_1n_2}, {\cal T}_{a'n_1' n_2'}\big]&=& if_{aa'}{}^{b} {\cal T}_{b n_1 + n'_1 n_2 +n'_2} +
(k_1 n_1 + k_2 n_2) \;k_{aa'} \; \delta_{n_1,-n'_1} \;\delta_{n_2,-n'_2} \nn\\
&&+ \sum \limits_{(m_1,m_2) \in \mathbb Z^2\setminus\{0,0\}}k_{m_1 m_2} (n_2 n_1' - n_1 n_2')\;k_{aa'}\;\delta_{n_1+n_1', -m_1} \;\delta_{n_2+n_2',-m_2}\ , \nn\\
&=& if_{aa'}{}^{b} {\cal T}_{b n_1 + n'_1 n_2 +n'_2} +
(k_1 n_1 + k_2 n_2) \;k_{aa'} \; \delta_{n_1,-n'_1} \;\delta_{n_2,-n'_2} \\
&&+k_{n_1+n_1',n_2+ n_2'} (n_2 n_1' - n_1 n_2')\;k_{aa'}\ , \nn\\
\big[d_1,{\cal T}_{an_1n_2}\big]&=&n_1 {\cal T}_{an_1n_2}\ , \nn\\
\big[d_2,{\cal T}_{an_1n_2}\big]&=&n_2 {\cal T}_{an_1n_2} \ . \nn
\eeqa
 This algebra admits an interesting subalgebra
 where only two-central charges are non-vanishing, say $k_1,k_2$ and corresponds to the Kac-Moody
 algebra considered throughout  this review: $\widetilde{\g}(U(1)^2)=
 \{{\cal T}_{am_1,m_2}, d_1,d_2, k_1, k_2,
  a=1,\cdots,\dim \g, m_1,m_2 \in \mathbb Z\}$. This algebra is also named the double affine algebra in \cite{jk}.
  All the results of the two-toroidal and the double affine algebras extend naturally for $n>2$ \cite{MR2},
  {\it i.e.}, to the $n-$toroidal and $n-$affine algebras.

  Now we turn to representation theory of the double affine Lie algebra
  (or Kac-Moody algebra associated to ${\cal M}=(U(1)^2$).
In  Section \ref{sec:repT}  the results  were given, considering the root system  $\Sigma$ of $\g$. In \cite{MRT}  non-unitary representations
of $\widetilde{\g}(U(1)^2)$ were explicitly obtained considering the roots of $\widehat{\g}$, the affine extensions of $\g$.
The key observation is the following: even if $\widetilde{\g}(U(1)^2)$ doesn't admit a system of simple roots, it is possible
to have a Chevalley-Serre presentation of $\widetilde{\g}(U(1)^2)$ (and of course also of the toroidal algebra $T_2(\g)$)
in two different but equivalent manners-- (1) with the simple roots of $\g$, or (2) with the simple roots of $\widehat{\g}$.
Consider now the double-affine algebra or the Kac-Moody algebra $\widetilde{\g}(U(1)^2)$ with two central extensions $k_1,k_2$ whose Lie brackets are
given by \eqref{eq:tor} with $n=2$ (note that all results presented below extend easily for the toroidal algebra with non-vanishing
central charge $k_{m_1 m_2}$).

Let $\alpha_i, i=1,\cdots,r$ be the simple roots of $\g$ and let $A_{ij}$
be the corresponding Cartan matrix:
\beqa
A_{ij} = 2 \frac{\alpha_i\cdot \alpha_j}{\alpha_i\cdot \alpha_i}\ , \ \ 1\le i,j\le r \ .\nn
\eeqa
The matrix $A$ is non-singular, {\it i.e.},  $\det A \ne 0$. Associated to
any simple  roots of the semisimple Lie algebra $\g$, we  define the three operators for $(m_1,m_2) \in \mathbb Z^2$:
\beqa
h_{im_1 m_2} = \frac 2 {\alpha_i \cdot \alpha_i} \alpha_i \cdot H_{m_1 m_2} \ , \ \
e^{\pm }_{i m_1 m_2}= \sqrt{\frac 2 {\alpha_i \cdot \alpha_i}} E_{\pm \alpha_i m_1,m_2}\ , \ \
1 \le i \le r \ .\nn
\eeqa
The Chevalley-Serre relations  are \footnote{In \cite{MRT} the
authors were given a presentation of the toroidal (and not the double-affine Lie) algebra.}
\beqa
\label{eq:CSS}
\begin{split}
\big[k_1,h_{im_1 m_2}\big]&=\big[k_1,e^{\pm}_{i m_1m_2}\big]=
\big[k_2,h_{im_1 m_2}\big]=\big[k_2,e^{\pm }_{i m_1 m_2}\big]=0\\
\big[h_{im_1m_2}, h_{jn_1 n_2}\big]&= (k_1 m_1 + k_2 m_2)
\;\alpha^{\vee}_{i} \cdot \alpha^\vee_{j} \; \delta_{m_1,-n_1} \delta_{m_2,-n_2}\\
\big[h_{i m_1 m_2},e^{\pm }_{j n_1 n_2}\big]&= \pm A_{ij}\;
e^{\pm }_{j m_1 + n_1 m_2+n_2}\\
\big[e^\pm_{i m_1 m_2}, e^\pm_{i,n_1 n_2}\big]&=0 \nn\\
\big[e^+_{im_1 m_2},e^-_{j n_1 n_2}\big]&= \delta_{ij} \Big(h_{im_1 + n_1 m_2 +n_2}+
\frac 2 {\alpha_i \cdot \alpha_i}\;(k_1 m_1 + k_2 m_2) \; \delta_{m_1,-n_1} \delta_{m_2,-n_2}\Big)\\
{\rm  ad}^{1-A_{ij}}(e^\pm_{i mn})\cdot e^\pm_{j pq}&=0 \ , \ \ i\ne j \ ,
\end{split}
\eeqa
($\text{ad}(x)\cdot y = [y,x], \text{ad}^2(x)\cdot y= [[y,x],x]$, {\it etc.}
and $\alpha^\vee = 2 \alpha/\alpha\cdot \alpha$ is the co-root
) enable
to reproduce the whole algebra (see \eqref{eq:tor} with $n=2$) using the Serre relation, {\it i.e.}, the last
equation.

The algebra $\widetilde{\g}(U(1)^2)$ can be equivalently presented using the simple roots
of the affine Lie algebra $\widehat{\g}$ associated to $\g$. Let
$\hat\alpha_{(i)}, i=0,\cdots,r$  be the simple roots of $\widehat{\g}$:
\beqa
\label{eq:saff}
\begin{array}{llll}
\hat \alpha_{(i)}&=&(\alpha_{i},0,0)\ , & i=1,\cdots,r \ , \\
\hat \alpha_{0}&=&(-\psi,0,1) \ ,
\end{array}
\eeqa
where  $\alpha_i,\psi$ are respectively the simple roots and  the highest root of $\g$.
Let $\hat A_{ij}$ be the Cartan matrix of $\widehat{\g}$:
\beqa
\hat A_{ij}= 2 \frac{\hat{\alpha_i}\cdot \hat{\alpha_j}}{\hat{\alpha_i}\cdot \hat{\alpha_i}} \ , \ \ 0\le i,j \le r \nn
\eeqa
where the scalar product (see for example \cite{Goddard:1986bp}) is
\beqa
\hat \alpha \cdot \hat \alpha'=(\alpha,k,m)\cdot (\alpha',k',m')= \alpha \cdot \alpha' + k m' +k' m\ .  \nn
\eeqa
Note that now the Cartan matrix is singular and $\dim \;\text{Ker} A =1$, {\it i.e.}, $A$ is of corank one.
 Associated to any simple roots of $\widehat{\g}$ we define for $m \in \mathbb Z$:
 \beqa
 \label{eq:sl2aff}
\begin{array}{lll}
  \hat e^\pm_{i m} = e^\pm_{i 0m}\ , & \hat h_{im} =
 h_{i0m}\ ,& i=1\ ,\cdots,r\\
\hat e^\pm_{0m} = \sqrt{\frac 2 {\psi\cdot \psi}}\; E_{\mp \psi \pm 1  m}\ , & \hat h_{0m} =\frac 2 {\psi\cdot \psi}\;(-\psi\cdot H_{1m} +k_1)\ .&
\end{array}
\eeqa
  This determines a presentation of the algebra  as
\beqa
\label{eq:Caff}
\begin{split}
\big[k_2,\hat h_{im}\big]&=\big[k_2,\hat e^\pm_{i m}\big]=0\ ,\\
\big[\hat h_{im}, \hat h_{jn}\big]&= k_2 m  \; \hat\alpha^\vee_{i} \cdot \hat \alpha^\vee_{j} \; \delta_{m,-n}\ , \\
\big[\hat h_{im},\hat e^\pm_{j n}\big]&= \pm \hat A_{ij}\; \hat e^\pm_{j m+n} \ ,\\
\big[\hat e^+_{i m},\hat e^-_{j n}\big]&= \delta_{ij}\big( \hat h_{im+n} +
\frac 2 {\hat \alpha_i \cdot \hat\alpha_i}\;k_2 m\; \delta_{m,-n}\big)\ ,\\
\big[\hat e^\pm_{i m},\hat e^\pm_{i n}\big]&=0\ ,\nn\\
{\rm ad}^{1-\hat A_{ij}}(\hat e^\pm_{i m})\cdot \hat e^\pm_{j n}&=0\ , \ \ i\ne j \ . \\
\end{split}
\eeqa
Again the last relation, {\it i.e.}, the Serre relation enables to reproduce the whole algebra \eqref{eq:tor}.
Indeed it has been shown in \cite{rm} that the two presentations \eqref{eq:CSS} and \eqref{eq:Caff} leads to isomorphic
algebras \eqref{eq:tor}.
Similar presentations (in terms of the roots of $\g$ or $\widehat{\g}$ hold for the toroidal Lie  algebra $T_n(\g)$ and for the Kac-Moody algebra
$\widetilde{\g}(U(1)^n)$  (or $n-$affine Lie algebra) when $n>2$ \cite{MR2}.

Even if the two presentations are isomorphic they present some structural differences. Indeed, within the presentation
of $\widetilde{\g}(U(1)^n)$
with
the roots of the semisimple Lie algebra $\g$ the Cartan matrix is non-singular, whereas with the presentation with
the roots of the affine Lie algebra $\widehat{\g}$ the Cartan matrix is singular.
The presentation of torodial Lie algebras in terms of roots of affine Lie algebras has been used by several authors in
order to construct explicit representations of $T_2(\g)$ and $T_n(\g), n>2$.
In \cite{MRT, MR2}
a Vertex representation of the Toroidal Lie algebra $T_n(\g)$ (when $\g$ is simply laced)
was  obtained.  In \cite{jk} a fermion realization of $\widetilde{\g}(U(1)^2)$, {\it i.e.}, of the double affine algebra in the terminolgy
of \cite{jk}, when $\g$ is a  classical Lie algebra was explicitly constructed.
Note that representation theory of toroidal Lie algebras have been studied extensively  (see e.g. \cite{jk} and references therein).
However, since the Cartan matrix of $\widehat{\g}$ is singular, and since all these articles are based on the Chevalley-Serre
presentation of the algebra in terms of the roots of $\widehat{\g}$, all these representations are non-unitary.

\subsection{Unitary representations}
Now we turn to construct explicit unitary representations.
On purpose we assume that the only non-vanishing central charges are $k_1$ and $c_m=-m k_{0,m}, m \ne 0$. Let $\overline{\g}=\{{\cal T}_{anm},  k_1, c_k,
a=1,\cdots, \dim \g, m,n \in \mathbb Z, k \in \mathbb Z\setminus\{0\} \}$. With these notations the non-vanishing Lie brackets \eqref{eq:gT2} reduce to
\beqa
\label{eq:t1}
\big[{\cal T}_{an_1n_2}, {\cal T}_{a'n_1' n_2'}\big]
&=& if_{aa'}{}^{b} {\cal T}_{b n_1 + n'_1 n_2 +n'_2} +
n_1\Big(k_1  \;\delta_{n_2,-n'_2}
+c_{n_2+ n_2'} \Big)\;k_{aa'}\; \delta_{n_1,-n'_1}\ .
\eeqa
Consider now the mapping
\beqa
    {\cal T}_{an_1 n_2}&\mapsto&e^{in_2 \theta_2}\; {\cal T}_{an_1} \nn\\
    k_1 &\mapsto& k_1\nn\\
    c_m&\mapsto& e^{i m \theta_2} k_1 \nn
    \eeqa
    where ${\cal T}_{am}$ are the generators of the affine Lie algebra $\widehat{\g}$.
    Through this mapping \eqref{eq:t1} reduces to
    \beqa
    \label{eq:t2}
\big[ e^{i n_2 \theta_2} {\cal T}_{an_1}, e^{in'_2 \theta_2}{\cal T}_{a'n_1'} \big]=
if_{aa'}{}^{b} e^{i (n_2+n_2') \theta_2} \;{\cal T}_{b n_1 +n_1'}  + n_1 k_1e^{i (n_2+n_2') \theta_2}\; h_{aa'}\delta_{n_1,-n_1'}\ ,
\eeqa
and the two algebras \eqref{eq:t1} and \eqref{eq:t2} are isomorphic (see \cite{MRT} Proposition 2.8, or \cite{tor} Theorem 3.3).
It is immediate to observe that the latter algebra \eqref{eq:t2} is the loop algebra of the affine algebra and thus we have
the following isomorphism $\overline{\g}\cong\widehat{\g}(\mathbb S^1) \cong \widehat{\g} \otimes \mathbb S^1$.
Please note that $\big[d_2, c_{m}\big]=0$, but $\big[d_2, e^{im\theta_2} k_1\big]=m  e^{im\theta_2} k_1$.

A certain class of unitary representations
follows at once and are directly obtained from unitary representations of the affine Lie algebra $\widehat{\g}$.
Unitary highest weight representations of $\widehat{\g}$ are well  known.
The simple roots of $\widehat{\g}$ are given in \eqref{eq:saff} and   the highest weight read (recall that rk $\g=r$)
\beqa
\frac{\psi}{\psi\cdot \psi}= \sum \limits_{i=1}^r  q^i \frac{\alpha_i}{\alpha_i \cdot \alpha_i} \ ,  \ \ q^i \in \mathbb N \nn
\eeqa
The corresponding fundamental weights \cite{Goddard:1986bp} are
\beqa
\hat \mu^0 = (0,\frac12 q^0 \psi\cdot \psi,0)  \ \ \text{with} \ \ q^0 =1 \ , \
\hat \mu^i = (\mu^i,\frac12 q^i \psi\cdot \psi,0) \ , \ i=1,\cdots, r  \ .\nn
\eeqa
Let $\hat \mu_0 = \sum _{i=0}^r p_i \hat \mu^i, p_i \in \mathbb N$  and let $ |\hat \mu_0\rangle$ be a highest weight defined by
(with the notations of Section \ref{sec:CST}, Eq.[\ref{eq:sl2aff}])
\beqa
\hat e_i^+ \big|\hat \mu_0\rangle &=&0 \ ,\nn\\
\hat h_i \big|\hat \mu_0\rangle &=&p_i  \big|\hat \mu_0\rangle \ . \nn
\eeqa
Because of the expression of $h_0$ in term of the central charge $k$ (we denote $c$ its eigenvalue) a highest weight is equivalently
specified by $\mu_0 =\sum_{i=1}^r p_i \mu^i$ (the highest weight of the semisimple Lie algebra $\g$, with its associated fundamental weight $\mu^i$) and the eigenvalue of the central charge $c$  \footnote{The Cartan subalgebra of $\widehat{\g}$ is $\{h_i, i=0,\cdots,r, k,d\}$, but the eigenvalue of  $d$ for the highest state
$|\hat \mu_0\rangle$ is irrelevant. Indeed if $d|\hat \mu_0\rangle= n_0 |\hat \mu_0\rangle$, redefining $d\to
d- \frac{n_0} c k$ we have $d|\hat \mu_0\rangle=0$ and we can take  $n_0=0$ \cite{fs}.}.
Let $x=2 \frac{c}{\psi\cdot \psi}$
 be the level of the representation. The representation ${\cal D}_{\hat \mu_0}$ is unitary if
 (see {\it e.g.} \cite{Goddard:1986bp}):
\beqa
\Big(x \in \mathbb Z \ \
\text{and} \ \  c \ge \psi\cdot \mu_0 \ge 0\Big) \ \ \Leftrightarrow \ \
\Big(x= \sum\limits_{i=0}^r p_i q^i \in \mathbb Z
\ \ \text{and} \ \ x\ge \sum\limits_{i=1}^r p_i q^i\Big) \nn
\eeqa

 Consider now
\beqa
\label{eq:R}
\mathcal{R}=\left\{ |{ m}\rangle\ , \ \  m\in \mathbb{Z}\right\} \ ,
\eeqa
 the set of unitary representations of $U(1)$:
\beqa
d_2 | m\rangle = m  | m\rangle\nn \ .
\eeqa
Using  the harmonic expansion on $U(1)$
\beqa
\langle \theta_2| m\rangle = e^{i m\theta_2} \ ,  \nn
\eeqa
 unitary representations of $\overline{\mathfrak{g}}$ are given by the tensor product
\beqa
\label{eq:uni2}
\overline{\mathcal{D}}_{\hat \mu_0}=\mathcal{D}_{\hat \mu_0}\otimes \mathcal{R}
\eeqa
and correspond to a harmonic expansion of the unitary representation $\mathcal{D}_{\hat \mu_0}$
of $\widehat{\mathfrak{g}}$ on the manifold $U(1)$.
These results was anticipated in \cite{mrm}.
Note that whilst $\mathcal{D}_{\hat \mu_0}$ is a highest weight representation of $\widehat{\g}$, $\overline{\mathcal{D}}_{\hat \mu_0}$ {\it is not}
a highest weight representation of $\overline{\g}$. A similar analysis hold for the algebra $\widehat{\g}\otimes \mathbb T_{n-1}, n>1$ where
$T_{n-1}$ is the $(n-1)-$dimensional torus \cite{mrm}.

\section{Applications in physics}

The generalized KM current algebra $\mathfrak{g}(\mathcal{M})$, its
semidirect symmetry actions, and its cohomological central extensions
constitute a versatile toolkit, which can be used to investigate various
subfields of high-energy theoretical physics. This section surveys three
arenas in which the mathematical framework developed in this review finds
concrete applications in physics. Our emphasis will be on : \textit{(i)}
two-dimensional current algebra and CFT, including WZW models, Sugawara
stress tensors and the Virasoro algebra; \textit{(ii)} higher-dimensional
compactifications and spectra in KK theory, and \textit{(iii)} structures
emerging in cosmological billiards and in the hidden symmetries of
supergravity. We will attempt at keeping the presentation self-contained and
pedagogical, referring to the relevant literature for technical details and
a broader background. Across all such three contexts, the crucial inputs
have a threefold nature : geometric (e.g., choice of $\mathcal{M}$ and its
symmetry), analytic (e.g., harmonic analysis on $\mathcal{M}$), and
algebraic (e.g., compatibility of cocycles and representations).

\subsection{Two-dimensional current algebra and CFT: WZW, Sugawara, Virasoro%
}

\subsubsection*{\textit{Affine symmetry from loops and its generalization}}

On a two-dimensional worldsheet, currents $J^{a}(z)$ valued in a
finite-dimensional Lie algebra $\mathfrak{g}$ satisfy the operator product
expansions (OPEs) that encode an \textit{affine} KM algebra at a certain
level $k$. This structure emerges canonically in WZW models, where the basic
field $g(z,\bar{z})\in G$ is a group-valued map and the action includes a
Wess--Zumino topological term; see e.g.\ \cite%
{Pressley:1988qk,Goddard:1986bp,DiFrancesco:1997nk}. The holomorphic
currents realize a centrally extended loop algebra of maps $\mathbb{S}%
^{1}\!\rightarrow \!\g$, and the stress tensor $T(z)$ is obtained by the
Sugawara construction, which expresses the Virasoro generators as quadratic
combinations of KM modes, producing a central charge $c=\frac{k\,\dim
\mathfrak{g}}{k+\mathfrak{g}^{\vee }}$, where $\mathfrak{g}^{\vee }$ is the
dual Coxeter number of $\mathfrak{g}$ \cite%
{Goddard:1986bp,DiFrancesco:1997nk}.

Within the general framework presented in this review, the circle $\mathbb{S}%
^{1}$ is replaced by a higher-dimensional manifold $\mathcal{M}$, thus
yielding to Lie algebra $\mathfrak{g}(\mathcal{M})$ of $\mathfrak{g}$-valued
functions on $\mathcal{M}$, organized through a Hilbert basis adapted to the
symmetries of $\mathcal{M}$ (by exploiting the Peter-Weyl theorem on compact
groups/cosets, or the Plancherel theorem on non-compact groups/cosets).
Semidirect actions by Killing vector fields or diffeomorphism algebras may
enrich this symmetry, and compatible two-cocycles yield central extensions
that generalize the affine case. From the two-dimensional worldsheet
perspective, this construction can be viewed as a \textit{mode expansion of
currents along internal manifolds}: the worldsheet current algebra fibers
over $\mathcal{M}$, so that KM modes carry, in addition, the harmonic labels
on $\mathcal{M}$ itself. This is particularly manifest when $\mathcal{M}$ is
a \textit{compact} group manifold or a homogeneous space thereof, because in
this case the harmonic analysis is representation-theoretically well-known
and studied \cite{Pressley:1988qk,Hel,PW}. Possible issues related to normal ordering prescriptions in manifolds
of dimension $>1$
have been discussed in \cite{Campoamor-Stursberg:2022ane,Campoamor-Stursberg:2022lyx}; we also briefly  recalled them  in Sec. \ref{sec:bedlewo}.

\subsubsection*{\textit{Group manifolds and coset CFTs}}

Let $\mathcal{M}=G_{c}$ be a compact Lie group manifold. The Peter-Weyl
theorem provides an orthonormal basis $\{\Psi _{LQR}\}$ in $L^{2}(G_{c})$,
built from matrix elements of irreducible unitary representations, and the
product of two basis elements decomposes in terms of Clebsch--Gordan
coefficients. The machinery developed in this review lifts the $\mathfrak{g}$%
-valued modes $T_{a;LQR}$ to generators of a current algebra $\mathfrak{g}%
(G_{c})$ with structure constants directly expressed in terms of the
representation theory of $G_{c}$. In the CFT context, this gives a natural
language for coset models $G/H$, in which the surviving harmonics are the $H$%
-invariant components on the right (\textit{or} left), and the operator
content obeys the selection rules of the $G\rightarrow H$ branching. This
perspective clarifies how the coset primaries and their fusion rules relate
to the representation-theoretic decomposition on $G/H$, as well as to the
geometry of the coset \cite{DiFrancesco:1997nk,Harada:2020exi,Hel}.

\subsubsection*{\textit{Diffeomorphism algebras and Virasoro analogues}}

For $\mathcal{M}=\mathbb{S}^{1}$ the semidirect `partner' of the affine
algebra is the Witt algebra (and its central extension, the Virasoro
algebra). On higher spheres $\mathbb{S}^{n}$ or other higher-dimensional
manifolds $\mathcal{M}$, families of vector fields play an analogous role.
For instance, for $\mathbb{S}^{2}$ the algebra of area-preserving
diffeomorphisms is relevant, while, more generally, one can construct
subalgebras of vector fields (generated by Killing vectors or selected
modes) acting on $\mathfrak{g}(\mathcal{M})$. This leads to semidirect
products that generalize to $\mathcal{M}$ the affine--Virasoro `interplay'
on $\mathbb{S}^{1}$. Actually, this idea can be traced back to the early
literature on generalized KM algebras related to diffeomorphism groups of
closed surfaces \cite{Frappat:1989gn,HOEGHKROHN1990106}. It should also here
be recalled that the appearance of generalized (Borcherds--Kac--Moody)
algebras in CFT and moonshine phenomena may be regarded as a hint to the
depth of algebraic structures that can emerge beyond the affine case \cite%
{BORCHERDS1991330,Borcherds:1992jjg,monstrous}..

\subsection{Compactifications and KK spectra}

\subsubsection*{\textit{Mode expansions and mass towers}}

Consider a $d$-dimensional field theory on a spacetime of the form $\mathbb{R%
}^{1,d-1}\times \mathcal{M}$, where $\mathcal{M}$ is a compact internal
manifold (group manifold, coset, or deformation thereof); the expansion of
fields in an orthonormal basis of $L^{2}(\mathcal{M})$ produces KK towers
whose masses are determined by eigenvalues of Laplace-type operators on $%
\mathcal{M}$. This finds extensive application and crucial relevance in
general KK theories, as well as in dimensional compactifications in
(super)string theory or M-theory \cite%
{Salam:1981xd,Bailin:1987jd,KK1,KK3,Castellani:1991et}.

Within this framework, the generalization from $\mathbb{S}^{1}$ to $\mathcal{%
M}$ yields the following consequences: \textit{(i)} the Noether charges
associated with a compact gauge algebra $\mathfrak{g}$ become \emph{families}
of charges indexed by the harmonic labels on $\mathcal{M}$, thus realizing
the Lie algebra $\mathfrak{g}(\mathcal{M})$; \textit{(ii)} the semidirect
action of the isometries of $\mathcal{M}$ provides an interesting geometric
perspective on the selection rules and spectral degeneracies;\textit{\ (iii)}
the central extensions encode anomalies that can appear in the commutators
of KK currents when integrating over $\mathcal{M}$, with the relevant
2-cocycles in one-to-one correspondence with closed $(\dim \mathcal{M}-1)$%
-currents on $\mathcal{M}$ (as developed in this review).

\subsubsection*{\textit{Compact group manifolds and cosets}}

For $\mathcal{M}=G_{c}$, the Peter-Weyl theory organizes the KK modes of
fields into representations of $G_{c}$. The product of modes is controlled
by Clebsch-Gordan coefficients, and the associated KM algebra $\mathfrak{g}%
(G_{c})$ provides underlies the interactions and selection rules among KK
modes. For $\mathcal{M}=G_{c}/H$, the $H$-invariant modes in the right (or
left) action are retained, thus reproducing the well-known truncations of
coset compactifications and their spectra \cite%
{Hel,Castellani:1991et,Harada:2020exi}. Here we confine ourselves to adding
that this intriguingly relates to generalized Scherk--Schwarz reductions,
where group-theoretic data constrain and determine consistent truncations of
physical theories.

\subsubsection*{\textit{Toroidal and deformed (soft) manifolds}}

When the internal, higher-dimensional manifold is a torus, one recovers
familiar Abelian current algebras and their higher-rank generalizations,
while deformations of group manifolds (which go under the name of \textit{%
soft} manifolds), as used in group-geometric approaches to supergravity,
naturally fit into the same formalism \cite{cas,Castellani:1991et}. The
central extension analysis identifies which deformations admit non-trivial
2-cocycles compatible with symmetry actions, and thus when anomalous terms
can affect the reduced dynamics. Related algebraic structures, such as
toroidal Lie algebras, also arise in the analysis of currents on
higher-dimensional tori, and they have been studied in the mathematical
literature since quite a long time \cite{HOEGHKROHN1990106,MRT,MR2}.

\subsubsection*{\textit{Kac--Moody symmetries in KK theories and beyond}}

KM-like enhancements in the symmetry algebras of KK reductions and related
stringy settings have been quite extensively investigated \cite%
{dd,Castellani:1991et,Fre:2005gvm}. In this respect, we should stress that
the present framework clarifies when and how such enhancements occur : the
algebra $\mathfrak{g}(\mathcal{M})$ organizes the towers of KK currents,
thus allowing for compatible central extensions to emerge precisely when the
geometry of $\mathcal{M}$ admits closed $(\dim \mathcal{M}-1)$-currents, in
turn giving rise to the cohomological 2-cocycles. Therefore, this framework
naturally highlights the dependence on the topology and metric of $\mathcal{M%
}$, thereby providing a controlled setting to explore consistent
truncations, dualities and anomaly structures of higher-dimensional (super)
gravity theories.

\subsection{Cosmological billiards and hidden symmetries of supergravity}

\subsubsection*{\textit{BKL dynamics and billiards}}

Near spacelike singularities, the Belinsky-Khalatnikov-Lifshitz (BKL)
analysis reveals chaotic, piecewise Kasner regimes interrupted by
curvature-induced reflections - the so-called `mixmaster' behavior \cite%
{Belinsky:1970ew,mixmaster}. Remarkably, this dynamics can be geometrized as
a billiard motion in a region of hyperbolic space bounded by \emph{walls}
associated to dominant gravitational and $p$-form contributions. Somewhat
surprisingly, in many supergravity theories the billiard domain coincides
with the fundamental Weyl chamber of a hyperbolic KM algebra, thus
suggesting that an underlying infinite-dimensional symmetry controls the
near-singularity regime \cite{Henneaux:2010ys,Fre:2005gvm}.

\subsubsection*{\textit{Hidden symmetries and very-extended KM algebras}}

Dimensional reduction of supergravities give rise to large, non-compact
global symmetries, controlling the electric-magnetic duality of the
resulting theory; in string/M-theory, such a symmetry gets defined over
discrete fields, and it is named $U$-duality. In general, the target space
of scalar fields can be regarded as a manifold coordinatized by the scalars
themselves. In a wide class of cases, the target space can be modeled as a
sigma-model on cosets $G/H$, where $G$ is a non-compact Lie group and $H$ a
maximal compact subgroup. The emergence of indefinite or hyperbolic
extensions in further dimensional reductions as well as in conjectural
uplifts (`$E_{10}$/$E_{11}$'-type structures) has been widely discussed, as
it may be shown to encode `hidden' symmetries of M-theory and cosmological
dynamics (see e.g. \cite{Kleinschmidt:2006ad,Bergshoeff:2008xv,Bossard:2017wxl,Tumanov:2017whf,Bossard:2019ksx}, and references therein). In this respect, the framework presented in this
review exhibits two remarkable features : \textit{(i)} the sigma-model
structure on non-compact target spaces (e.g.\ $SL(2,\mathbb{R})$ and $SL(2,%
\mathbb{R})/U(1)$) is inherently consistent with the non-compact harmonic
analysis used to build $\mathfrak{g}(\mathcal{M})$; \textit{(ii)} the billiard/wall identifications hint at the
possibility to describe the asymptotic dynamics through KM root systems,
again consistently with the appearance of infinite-dimensional algebras in
reductions and duality webs \cite{Henneaux:2010ys,Fre:2005gvm}.

\subsubsection*{\textit{Cocycles, anomalies, and constraints}}

The analysis of the possible central extensions of the current algebras on $%
\mathcal{M}$ provides crucial insights on the existence of anomaly-like
terms in the effective dynamics, as well as on the relevant constraints
associated with conserved charges. In the aforementioned cosmological
settings, integrating currents over compact slices (or along suitable
cycles) in $\mathcal{M}$ yields to Schwinger terms controlled by the
cohomology of $L^2({\cal M})$; this essentially extends the Pressley-Segal
cocycles in the affine case to the richer $(\dim \mathcal{M}-1)$-current
data. Thus, while many open questions remain (e.g.\ precise matching between
very-extended algebras and full dynamics), the formal machinery and tools
presented in this review may allow to establish when \textit{bona fide}
infinite-dimensional symmetries are actually compatible with the underlying
geometry of the internal manifold and with the employed boundary conditions;
at the same time, potential topological and/or flux obstructions may be
detected.
 
\section{Outlook: conjectures and novel applications}

The framework developed in this review - generalized KM current algebras $%
\mathfrak{g}(\mathcal{M})$ on a manifold $\mathcal{M}$, semidirect actions
by isometries/diffeomorphisms, and cohomological classification of central
extensions - suggests several avenues for new applications in high-energy
theoretical and mathematical physics. In this final section, we will briefly
mention some potential, partly conjectural, developments in supergravity
theories, superstring/M-theory, AdS/CFT and holography. The proposals below
should be regarded as possible lines of investigation rather than
established research venues; in fact, we will stress their \textit{%
conjectural} nature, but also indicate how they could in principle be tested
and checked. We leave all this for further, future works.

\subsection{Supergravity}

\subsubsection*{\textit{KM-structured charge algebras in consistent truncations}}

Consistent truncations on group manifolds and cosets are organized in terms
of geometric data (left-invariant frames, structure constants) and by
harmonic analysis on $\mathcal{M}$ \cite{Castellani:1991et,cas}. In light of
the topics discussed in the present review, it may be conjectured that in
any truncation admitting a global orthonormal frame (e.g.\ generalized
parallelisations), the full tower of Noether charges of the truncated theory
closes into a centrally-extended current algebra of the form $\mathfrak{g}(%
\mathcal{M})$, with 2-cocycles determined by closed $(\dim \mathcal{M}-1)$%
-currents on $\mathcal{M}$ itself. This conjecture is nothing  but an
extension of the standard, affine construction on the circle $\mathbb{S}^{1}$%
, and it may potentially detect possible anomalies through the study of the
cohomology on $\mathcal{M}$. Recent progress on consistent truncations in
deformed/generalized settings provides us with a promising testing ground
\cite{Harada:2020exi,Blair:2024ofc}; in this respect, a concrete test may
consist in computing Schwinger terms for KK-reduced currents in gauged
supergravities, and match them against the cohomological classification of
2-cocycles outlined in this review.

\subsubsection*{\textit{Soft manifolds, fluxes, and anomaly constraints}}

Group-geometric approaches embed supergravity in a \textit{soft} deformation
of group manifolds, encoding fluxes and torsion as geometric data \cite%
{cas,Castellani:1991et}. In this framework, it may be conjectured that the
existence of non-trivial central extensions of $\mathfrak{g}(\mathcal{M})$
compatible with isometries imposes integrability conditions on the flux
backgrounds; such integrability conditions should turn out to be equivalent
to a subset of the Bianchi identities and tadpole/anomaly-cancellation
conditions in the dimensionally reduced, resulting physical theory. In
practice, these conditions would translate into the co-closedness of certain
$(\dim \mathcal{M}-1)$-currents, as well as into quantization constraints
for the Wess--Zumino-type terms. Remarkably, this would provide a powerful,
algebraic criterion for the consistency of flux compactifications, as well
as of their gauged supergravity limits \cite%
{Castellani:1991et,Harada:2020exi}.

\subsubsection*{\textit{Hidden symmetries beyond billiards}}

The appearance of hyperbolic KM structures in cosmological billiards
suggests a more prominent role of infinite-dimensional symmetries in
supergravity dynamics \cite{Henneaux:2010ys}. One may be led to naturally
conjecture that, away from the near-singularity regime, a subset of these
symmetries survives as an \textit{algebra of generalized currents} on
suitable slices (e.g.,\ homogeneous spatial sections) of $\mathcal{M}$, with
central terms fixed by boundary conditions and topological data. In
principle, this conjecture should be testable/falsifiable within numerical
relativity setups, or in analytic families of Bianchi cosmologies, by
constructing conserved charges associated to divergence-free vector fields
on $\mathcal{M}$ and checking their commutators against the 2-cocycles
discussed in this review. As another conjectural remark, we should not forget
to mention potential connections to duality orbits in extended
supergravities, which may further constrain the admissible cocycles \cite%
{Fre:2005gvm}.

\subsection{Superstrings and M-theory}

\subsubsection*{\textit{Worldvolume current algebras with internal labels}}

In the WZW models and related worldsheet theories, the affine symmetry
controls the spectrum and dynamics \cite{Goddard:1986bp,DiFrancesco:1997nk}.
Then, it may be conjectured a \textit{lift} of such a framework, in which
the currents acquire harmonic labels on an internal manifold $\mathcal{M}$,
thus effectively realizing $\mathfrak{g}(\mathcal{M})$ on the worldsheet.
When $\mathcal{M}$ is a group manifold or coset, Peter-Weyl theorem allows
for the explicit construction of operator bases; for non-compact $\mathcal{M}
$, Plancherel theorem should provide the relevant distributions. All in all,
this approach seemingly combines target-space geometry and worldsheet
current algebras, in a way which is compatible with string field theory, as
well as with group-geometric formulations \cite%
{Castellani:1991et,cas,Green1987}. A possible test of this conjecture would
concern heterotic compactifications on group manifolds/cosets, in which one
should compute OPE coefficients across harmonic sectors.

\subsubsection*{\textit{Brane boundaries and defect algebras}}

M2/M5 branes admit boundary/defect descriptions with current algebras on
lower-dimensi\-onal intersections. One may thus put forward the conjecture
that, in backgrounds with isometries along an internal manifold $\mathcal{M}$%
, the boundary algebra could be identified with a central extension of the
generalized KM current $\mathfrak{g}(\mathcal{M})$, with the 2-cocycle
determined by the pullback of fluxes to $(\dim \mathcal{M}-1)$-cycles.
Interestingly, this would generalize level quantization in WZW models to
higher-dimensional defects. Hints of such structures have already appeared
in early works on membranes \cite{Hop}, as well within the systematic
investigation of WZW-like terms in string/M-theory \cite{Green1987}.

\subsubsection*{\textit{Non-geometric backgrounds and doubled/exceptional geometry}}

In doubled and exceptional field theories, gauge symmetries generally mix
diffeomorphisms and $p$-form transformations. In this context, we may
conjecture that the algebra of generalized diffeomorphisms and gauge
transformations on a compact manifold $\mathcal{M}$ can be organized as a
centrally extended $\mathfrak{g}(\mathcal{M})$ built from the relevant
finite-dimensional algebra $\mathfrak{g}$ (e.g.\ $E_{d(d)}$) and a $L^{2}(%
\mathcal{M})$ basis. However, we should stress that the compatibility of the
central extensions with section/closure constraints would become a sharp
algebraic condition, possibly hard to meet/check. However, recent progress
on deformed generalized parallelisations and consistent truncations \cite%
{Blair:2024ofc} seemingly supports the idea that constructing such algebras
on explicit $\mathcal{M}$ may be feasible; in this context, flux
quantization would thus arise out as the quantization of the central charge
in the extended algebra. Among other aspects, this perspective could clarify
the representation content of KK towers in exceptional field theory, as well
as their coupling to the so-called `dual' graviton sectors \cite%
{Fre:2005gvm,KK3}.

\subsection{AdS/CFT and holography}

\subsubsection*{\textit{Asymptotic symmetries and boundary current algebras}}

In $AdS_{3}$, the Virasoro symmetry arises as an asymptotic symmetry; in
higher dimensions, boundary symmetries are more subtle. In this framework, it
could be conjectured that, whenever the bulk includes an internal compact
manifold $\mathcal{M}$ with a certain isometry algebra, the boundary CFT
exhibits a tower of conserved currents organized by $\mathfrak{g}(\mathcal{M}%
)$, whose central terms should also be sensitive to holographic counterterms
and global anomalies \cite{AdS3,AdS1,TR}. In practice, such a holographic
renormalization supplies the bilinear form used to compute 2-cocycles (in
terms of boundary two-point functions of currents), while the harmonic
analysis on $\mathcal{M}$ would determine the multiplet structure. This is
in principle testable in truncations whose consistent embeddings are known,
as well as in top-down approaches to holography, in which the manifold $%
\mathcal{M}$ is explicitly specified \cite{Blair:2024ofc}.

\subsubsection*{\textit{Coset holography and harmonic selection rules}}

For compact $\mathcal{M}=G_{c}/H$, we expect a tight match between boundary
operator algebras and the representation-theoretic decomposition on $%
\mathcal{M}$, governed by Clebsch-Gordan coefficients and $H$-invariance. We
also expect the correlators to generally obey some selection rules, which
should in turn mirror the product structure of $L^{2}(\mathcal{M})$; in this
way, central terms should therefore be predicted by the cohomology classes
of $(\dim \mathcal{M}-1)$-currents. This seems to be potentially explorable
in the holographic duals built from consistent truncations on $G_{c}/H$, and
comparable with CFT data obtained via bootstrap or integrability methods
\cite{AdS3,TR}.

\subsubsection*{\textit{Holographic anomalies from cocycles}}

One may also put forward the conjecture that certain boundary anomalies
(e.g.,\ of mixed flavor--gravitational type) could be captured by the
cohomological 2-cocycles of $\mathfrak{g}(\mathcal{M})$ computed from the
bulk : in this sense, harmonic analysis and de~Rham cohomology on
  $L^2({\cal M})$ \cite{AdS1,AdS3} would become crucial. This would be especially
remarkable, since it would provide an algebraic holographic dictionary
between fluxes on $\mathcal{M}$ and the central extensions in the boundary
current algebra.\bigskip

All in all, one conceptual consequence of the programme reviewed in this
work is the determination of a uniform algebraic language for the
description and investigation of charges, anomalies and selection rules in
string/M-theory, supergravity and holography. Even partial validations of
the several conjectures stated above would shed light onto long-standing
structural questions about hidden symmetries and consistent truncations; on
the other hand, failures would be interesting too, since they would help
establishing the true scope of current-algebraic methods beyond the circle,
at least for what concerns the aforementioned physical applications.

\bibliographystyle{utphys}
\bibliography{ref}

\end{document}